\documentclass{article}

\usepackage{amssymb,amsfonts,amsmath}
\usepackage{cite,enumerate,float}
\usepackage{color}
\usepackage{tikz}
\usetikzlibrary{arrows,snakes,backgrounds}
\usepackage{hyperref}

\usepackage{amssymb,amsfonts,amsmath,stmaryrd}
\usepackage{cite,enumerate,float,indentfirst}
\usepackage{color}
\usepackage{tikz}
\usetikzlibrary{arrows,snakes,backgrounds,calc}
\usepackage{ytableau}
\usepackage[vcentermath]{youngtab}
\usepackage{pict2e}

\def\be{\begin{eqnarray}}
\def\ee{\end{eqnarray}}
\def\nn{\nonumber}

\def\be{\begin{eqnarray}}
\def\ee{\end{eqnarray}}
\def\nn{\nonumber}

\def\p{\partial}

\def\Tr{{\rm Tr}\,}

\definecolor{red}{rgb}{1,0,0}
\definecolor{orange}{rgb}{1,0.5,0}
\definecolor{violet}{rgb}{0.7,0,1}

\hoffset= - 1.0in         

\begin{document}
\title{\vspace{1.5cm}\bf
Commutative families in $W_\infty$, integrable many-body systems and hypergeometric $\tau$-functions
}

\author{
A. Mironov$^{b,c,d,}$\footnote{mironov@lpi.ru,mironov@itep.ru},
V. Mishnyakov$^{a,b,c,e,}$\footnote{mishnyakovvv@gmail.com},
A. Morozov$^{a,c,d,}$\footnote{morozov@itep.ru},
A. Popolitov$^{a,c,d,}$\footnote{popolit@gmail.com}
}

\date{ }

\maketitle

\vspace{-6.5cm}

\begin{center}
\hfill FIAN/TD-09/23\\
\hfill IITP/TH-09/23\\
\hfill ITEP/TH-13/23\\
\hfill MIPT/TH-10/23
\end{center}

\vspace{4.5cm}

\begin{center}
$^a$ {\small {\it MIPT, Dolgoprudny, 141701, Russia}}\\
$^b$ {\small {\it Lebedev Physics Institute, Moscow 119991, Russia}}\\
$^c$ {\small {\it NRC ``Kurchatov Institute", 123182, Moscow, Russia}}\\
$^d$ {\small {\it Institute for Information Transmission Problems, Moscow 127994, Russia}}\\
$^e$ {\small{\it Institute for Theoretical and Mathematical Physics, Lomonosov Moscow State University, Moscow 119991, Russia}}
\end{center}

\vspace{.1cm}

\begin{abstract}
We explain that the set of new integrable systems,
generalizing the Calogero family and implied by the study of WLZZ models,
which was described in arXiv:2303.05273,
is only the tip of the iceberg.
We provide its wide generalization and explain that it is related to commutative subalgebras (Hamiltonians) of the $W_{1+\infty}$ algebra. We construct many such subalgebras and explain how they look in various representations. We start from the even simpler $w_\infty$ contraction, then proceed to the one-body representation in terms of differential operators on a circle, further generalizing to matrices and in their eigenvalues, in finally to the bosonic representation in terms of time-variables. Moreover, we explain that some of the subalgebras survive the $\beta$-deformation, an intermediate step from $W_{1+\infty}$  to the affine Yangian. The very explicit formulas for the corresponding Hamiltonians in these cases are provided.
Integrable many-body systems generalizing the rational Calogero model arise in the representation in terms of eigenvalues. Each element of $W_{1+\infty}$ algebra gives rise to KP/Toda $\tau$-functions. The hidden symmetry given by the families of commuting Hamiltonians is in charge of the special, (skew) hypergeometric $\tau$-functions among these.
\end{abstract}

\bigskip

\section{Introduction
\label{intro}}

At the current stage of development, one of the central objects of interest in theoretical physics
is the family of $W_\infty$ algebras and their Yangian/DIM generalizations,
which, though the later generalizations not being Lie algebras, still continue to play the role
of symmetries in physically relevant theories.
Moreover, they naturally unify in a common entity the Virasoro
and $W$-algebras.
They possess a rich set of representations,
including Fock, MacMahon and, probably, many more.

These algebras also possess a rich set of one-parametric commutative subalgebras,
of which just a small subset was revealed and described in the recent \cite{MMCal}.
These commutative subalgebras are related with each other by a rotation given by an operator $\hat O(u)$ (see sec. \ref{sec:O}). All of the $W_\infty$ algebras and their representations, that we consider are double-graded, with our prime example being the linear $W_{1+\infty}$ algebra of \cite{Pope} (see also \cite{Awata} for an extensive review)\footnote{See a non-linear version of $W_{1+\infty}$ algebra in \cite{GG}.}. We will only study elements with a well-defined grading in one of the directions. In this regard the commutative subalgebras are of two types: those lying in the positively graded part of $W_{1+\infty}$, and in its
negatively graded part.
In fact, after extending the $W_{1+\infty}$-algebra to the Kac-Radul algebra \cite{KR1} and further up to the
Ding-Iohara-Miki (DIM) algebra \cite{DI,Miki}, these commutative subalgebras from these two parts of $W_{1+\infty}$
become the two halves of Heisenberg subalgebras.
\\

The goal of this paper is to provide a fuller description of the subalgebras,
their interpretation in terms of (predominantly new) integrable systems
and of the new world of associated special functions.
Commuting 1-parametric subalgebras (infinite sets of commuting Hamiltonians)
typically arise at the level of universal enveloping algebras.
For the $W_\infty$ family, these subalgebras appear just at the level of original generators,
and there are many.
Moreover, there is a problem already to enumerate/classify these subalgebras.
In this paper, we make just a preliminary attempt, and we use a hierarchical description,
which is hardly adequate, but one should begin with something. Let us provide a brief overview first.

At the first level of this hierarchy are ``the integer rays" \cite{MMCal}, when the 1-parametric commuting families constructed as follows:
one starts from the triple of elements of the $W_{1+\infty}$ algebra\footnote{Definitions and properties of the $W_{1+\infty}$ algebra can be found in \cite{Pope,FKN2,BK,BKK,KR1,FKRN,Awata,KR2,Miki}.}: the cut-and-join operator $\hat W_0$ (corresponding to what is usually denoted by $\hat \psi_3$ in works related to the affine Yangian \cite{Tsim,Prochazka} )
and two generators $\hat E_0$, $\hat F_0$. They are used in order to construct the two families of elements of the algebra (generating operators)\footnote{In this introductory part, we omit simple numerical coefficients that we use in definitions of further sections for the sake of simplicity.}:
\begin{equation}
	\hat E_n = {\rm ad}_{\hat W_0}^n \hat E_0, \quad  \hat F_n = {\rm ad}_{\hat W_0}^n \hat F_0\,.
\end{equation}
Now one constructs the commutative elements (Hamiltonians) at any fixed $n$ as
\begin{equation}
 {\hat H}_k^{(n)} = {\rm ad}_{\hat E_{n+1}}^{k-1} \hat E_n \, , \quad \text{and} \quad  {\hat H}_{-k}^{(n)} = {\rm ad}_{\hat F_{n+1}}^{k-1} \hat F_n
\end{equation}
generated by repeated action of $\hat E_{n+1}$ (and $\hat F_{n+1}$) on operator $\hat E_n$ (and $\hat F_n$). The picture is even simpler in the so-called one-body representation of the $W_{1+\infty}$ algebra by operators $z$ and $\hat D=z{d\over dz}$, where one has \footnote{Here and further we will denote the abstract operators and their image in different representations by the same symbol, to avoid heavy notations.}
\begin{equation}
	\hat W_0={1\over 2}\hat D(\hat D+1) \, , \quad\hat E_0=z \,, \quad  \hat F_0={1\over z}
\end{equation}
and hence
\begin{equation}
\hat H_k^{(n)}=(z\hat D^n)^k
\,, \quad  \hat H_{-k}^{(n)}=(z^{-1}\hat D^n)^k .
\end{equation}
Hence, the $E$-series describe the commutative subalgebras of $W_{1+\infty}$ with positive grading, and the $F$-series, with negative grading.

The second level of hierarchy describes ``the rational rays", when one starts from another set generating operators (Hamiltonians of ``the horizontal ray"): $\hat E_0^{(q)}= {\rm ad}_{\hat E_1}^{q-1} \hat E_0 $, $\hat F_0^{(q)}={\rm ad}_{\hat F_1}^{q-1} \hat F_0$ at fixed $q$. Then, again one constructs the set of generating operators as
\begin{equation}
	\hat E_n^{(q)} = {\rm ad}_{\hat W_0}^n \hat E_0^{(q)}, \ \  \hat F_n^{(q)} = {\rm ad}_{\hat W_0}^n \hat F_0^{(q)}
\end{equation} and, further, the commutative Hamiltonians
\begin{equation}
	{\hat H}_k^{(p,q)} = {\rm ad}_{\hat E_{p+1}^{(q)}}^{k-1} \hat E_p^{(q)} \quad  \text{and}  \quad {\hat H}_{-k}^{(p)} = {\rm ad}_{\hat F_{n+1}^{(q)}}^{k-1} \hat F_n^{(q)}.
\end{equation}
In terms of the operators $z$ and $\hat D$, these operators look as $\hat E_0^{(q)}=z^q$, $\hat F_0^{(q)}=z^{-q}$ and $\hat H_k^{(p,q)}=(z^q\hat D^p)^k$, $\hat H_{-k}^{(p,q)}=(z^{-q}\hat D^p)^k$.

At last, the third, most general level of the hierarchy describes ``the cones", when the generating operators are substituted by arbitrary linear combinations in the left or right quarter-planes. In terms of the operators $z$ and $\hat D$, the cone generating operators are \begin{equation}
	\hat E^{(G,q)}=z^qG(\hat D) \,, \quad  \hat F^{(G,q)}=z^{-q}G(\hat D)
\end{equation} and the cone Hamiltonians, \begin{equation}
\hat H_k^{(G,q)}=(z^qG(\hat D))^k \,,  \quad \hat H_{-k}^{(G,q)}=(z^qG(\hat D))^k \,,
\end{equation} where $G(x)$ is a polynomial of $x$. This is the most general form of a graded commutative subfamily of $W_{1+\infty}$ algebra. The next level is a {\it functional} generalization, which requires an extension of the algebra,
we will touch this only briefly in the paper.
\\

Another hierarchy which we consider in our presentation is that of algebras and representations.
We begin with the simplest example of $w_\infty$ algebra of $2d$ area preserving diffeomorphisms, where everything is easy:
with every integer point of a half-plane, there is a well defined associated operator,
and the commutator of two operators sits at a well-defined third point.
Then, the entire pattern of commutative subalgebras is immediately seen.
However, this is only a contraction (Poisson-bracket approximation) of the far more interesting
$W_{1+\infty}$.
There we also have two integer gradings, but do not have a canonical way to prescribe an operator to a point on a half-plane,
and the commutator of two is often a combination of operators from different points
(saying more ``formally", the commutator does not preserve bigrading: the spin is not preserved under commuting).
Instead of fixing this ambiguity in the choice of basis {\it a priori}, we follow a more traditional way
of going through different representations.
We begin from the one-body representation in terms of differential operators on a circle $z$ and $\frac{d}{dz}$, where commutativity and then lift it to
matrix, eigenvalue and time-variables representations.
The commutative Hamiltonians are described by concrete multiple commutators
of $E$ or $F$ generators.

A remarkable property of $W_\infty$ is its split into a collection of commutative algebras.
In the first approximation, these are just one-dimensional rays going from the origin,
and every generator of $W_\infty$ belongs to exactly one such ray.
At the second level, it appears possible to consider linear combinations of generators
of the same grading but with arbitrary coefficients, we depict them by cones
instead of rays.
This is a property of algebra, and thus it holds in arbitrary representation,
with arbitrary central charge.

The next step the hierarchy of algebras is the $\beta$-deformation. It turns out that only the integer rays, and integer cones survive the deformation. This phenomenon has an algebraic origin, and is related to the uplift from the $W_{1+\infty}$ algebra to the affine Yangian. We will discuss this issue elsewhere \cite{MMMP2}. Here we present explicit formulas for the $\beta$-deformed operators in  eigenvalue and time-variables representations.
\\

In this paper, we proceed by going from simple representations to more complicated ones,
dedicating a special section to every one. Within every section we begin from the simplest family of commuting operators,
considered in \cite{MMCal}, and then describe step-by-step generalizations. More concretely, we start our discussion in section 2 with the simplest version, the $w_\infty$ algebra due to I. Bakas \cite{Bakas}, which is a contraction of the $W_{1+\infty}$ algebra: it already demonstrates all essential properties of commutative subfamilies. In section 3, we work with the basic definition of the $W_{1+\infty}$ algebra in terms of operators $z$, $z{d\over dz}$, and, in further sections, we discuss various representations of the $W_{1+\infty}$ algebra: the representation in terms of matrices (sec.5), in terms of their eigenvalues (secs.6-7), and in terms of infinitely many variables $p_k$ realizing the Fock representation (sec.8), i.e. the second quantization of the representation in terms of operators $z$, $z{d\over dz}$. In section 9, we discuss the $\beta$-deformation of this later. The applications of the developed formalism are discussed in secs.7,11,12: in sec.7, we explain that the commutative subalgebras are related to the Hamiltonians of integrable many-body systems of the rational Calogero type, in sec.11 we generate arbitrary (skew) hypergeometric $\tau$-functions of KP/Toda hierarchy using commutative subalgebras of extensions of the $W_{1+\infty}$ algebra, and sec.12 is devoted to finding the eigenfunctions of the commutative Hamiltonians constructed in sec.8. Further generalizations and open problems can be found in the Conclusion, and Appendix A develops the technique of obtaining manifest expressions for commuting Hamiltonians of sec.8, while Appendix B explains the technique of calculating matrix derivatives in terms of eigenvalue derivatives for needs of sec.5.

\paragraph{Notation.} We use the notation $S_R$ for the Schur function: it is a symmetric polynomial of variables $x_i$ denoted as $S_R(x_i)$, it is a graded polynomial of power sums $p_k=\sum_ix_i^k$ denoted as $S_R\{p_k\}$, and we use the notation $S_R[u]$ when all the variables $p_k=u$. Here $R$ denoted the partition (or the Young diagram): $R_1\ge R_2\ge\ldots\ge R_{l_R}$, where $l_R$ is the number of parts of $R$. We also use the notation $S_{R/Q}$ for the skew Schur functions, the notation $h_r=S_{[r]}$ for the complete homogeneous symmetric polynomials, and the notation $e_r=S_{[1^r]}$ for the elementary symmetric polynomials.
The Jack polynomials are denoted through $J_R$. All the necessary details can be found in \cite{Mac,Fulton}. We use the notation $\delta_{k|n}$ for the quantity that is 1 iff $k$ is multiple of $n$, and is zero otherwise.

\section{$w_\infty$-algebra, commutative subalgebras from rays and cones}\label{sec:diffeo}

Our simplest starting example is the $w_\infty$ algebra \cite{Bakas}, which already demonstrates all basic properties of commutative subalgebras.

\subsection{The basic example: integer rays
\label{basic}}

The  $w_\infty$ algebra of $2d$ area-preserving diffeomorphisms \cite{Bakas,MWinf}
is formed by the 2d lattice of operators $v_{n,i}$,
which satisfy the simple commutation relations
\be
\left[\hat v_{M,i}, \hat v_{N,j}\right] = \Big((M-1)j-(N-1)i\Big) \hat v_{M+N-2,i+j}
\label{winfty}
\ee
This bracket respects the grading along the horizontal (second index),
and shifts the vertical grading (first index) by two. There is also a homomorphism $\hat v_{M,i} \longrightarrow \hat v_{M+i,i}$ which does not change the commutation relations.
We fix this ambiguity imposing the condition $n\in \mathbb{Z}_{>0}$, $i\in\mathbb{Z}$.
These generators can be conveniently placed on the plane: Fig.1.

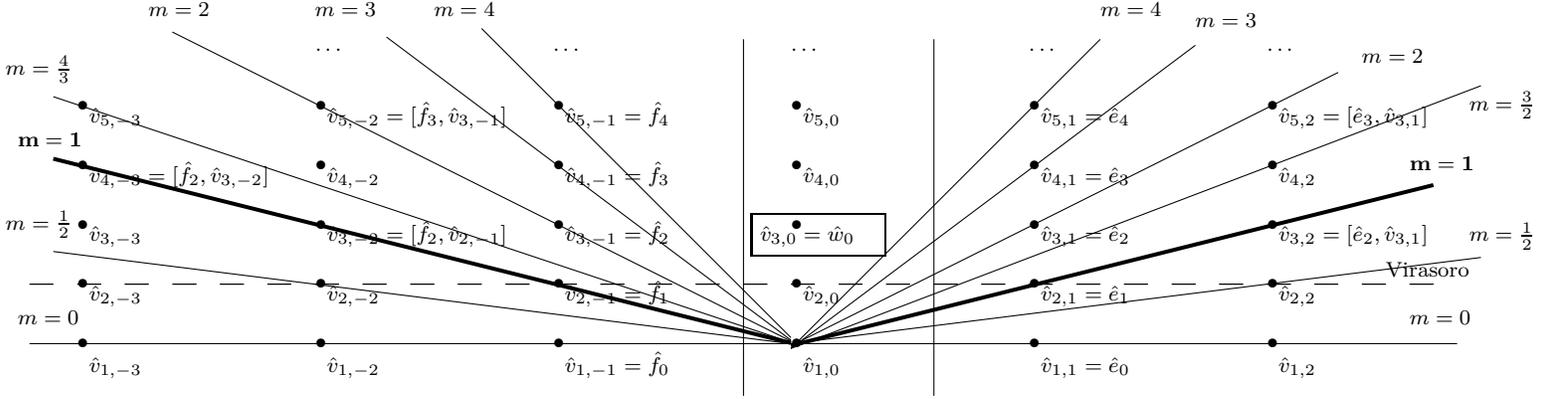
\begin{figure}[h]
\setlength{\unitlength}{.9pt}
\begin{picture}(300,170)(-285,-15)
{\footnotesize
\put(-20,-20){\line(0,1){150}}
\put(60,-20){\line(0,1){150}}

\put(0,0){\mbox{$\bullet$}}
\put(0,25){\mbox{$\bullet$}}
\put(0,50){\mbox{$\bullet$}}
\put(0,75){\mbox{$\bullet$}}
\put(0,100){\mbox{$\bullet$}}
\put(0,125){\mbox{$\ldots$}}
\put(5,-10){\mbox{$\hat v_{1,0} $}}
\put(5,20){\mbox{$\hat v_{2,0}$}}
\put(-20,45){\mbox{ $\boxed{\hat v_{3,0}=\hat w_0\phantom{5^5}}$}}
\put(5,70){\mbox{$\hat v_{4,0}$}}
\put(5,95){\mbox{$\hat v_{5,0}$}}

\put(-100,0){
\put(0,0){\mbox{$\bullet$}}
\put(0,25){\mbox{$\bullet$}}
\put(0,50){\mbox{$\bullet$}}
\put(0,75){\mbox{$\bullet$}}
\put(0,100){\mbox{$\bullet$}}
\put(0,125){\mbox{$\ldots$}}
\put(5,-10){\mbox{$\hat v_{1,-1}=\hat f_0 $}}
\put(5,20){\mbox{$\hat v_{2,-1}=\hat f_1
$}}
\put(5,45){\mbox{$\hat v_{3,-1}=\hat f_2$}}
\put(5,70){\mbox{$\hat v_{4,-1}=\hat f_3$}}
\put(5,95){\mbox{$\hat v_{5,-1}=\hat f_4$}}
}

\put(-200,0){
\put(0,0){\mbox{$\bullet$}}
\put(0,25){\mbox{$\bullet$}}
\put(0,50){\mbox{$\bullet$}}
\put(0,75){\mbox{$\bullet$}}
\put(0,100){\mbox{$\bullet$}}
\put(0,125){\mbox{$\ldots$}}
\put(5,-10){\mbox{$\hat v_{1,-2}
$}}
\put(5,20){\mbox{$\hat v_{2,-2}$}}
\put(5,45){\mbox{$\hat v_{3,-2}=[\hat f_2,\hat v_{2,-1}]$}}
\put(5,70){\mbox{$\hat v_{4,-2}$}}
\put(5,95){\mbox{$\hat v_{5,-2}=[\hat f_3, \hat v_{3,-1}]$}}
}

\put(-300,0){
\put(0,0){\mbox{$\bullet$}}
\put(0,25){\mbox{$\bullet$}}
\put(0,50){\mbox{$\bullet$}}
\put(0,75){\mbox{$\bullet$}}
\put(0,100){\mbox{$\bullet$}}
\put(5,-10){\mbox{${\footnotesize \hat v_{1,-3}
}$}}
\put(5,20){\mbox{$\hat v_{2,-3}$}}
\put(5,45){\mbox{$\hat v_{3,-3}$}}
\put(5,70){\mbox{$\hat v_{4,-3}=[\hat f_2,\hat v_{3,-2}]$}}
\put(5,95){\mbox{$\hat v_{5,-3}$}}
}

\put(100,0){
\put(0,0){\mbox{$\bullet$}}
\put(0,25){\mbox{$\bullet$}}
\put(0,50){\mbox{$\bullet$}}
\put(0,75){\mbox{$\bullet$}}
\put(0,100){\mbox{$\bullet$}}
\put(0,125){\mbox{$\ldots$}}
\put(5,-10){\mbox{$\hat v_{1,1}=\hat e_0$}}
\put(5,20){\mbox{$\hat v_{2,1}=\hat e_1$
}}
\put(5,45){\mbox{$\hat v_{3,1}=\hat e_2$}}
\put(5,70){\mbox{$\hat v_{4,1}=\hat e_3$}}
\put(5,95){\mbox{$\hat v_{5,1}=\hat e_4$}}
}

\put(200,0){
\put(0,0){\mbox{$\bullet$}}
\put(0,25){\mbox{$\bullet$}}
\put(0,50){\mbox{$\bullet$}}
\put(0,75){\mbox{$\bullet$}}
\put(0,100){\mbox{$\bullet$}}
\put(0,125){\mbox{$\ldots$}}
\put(5,-10){\mbox{$\hat v_{1,2}$
}}
\put(5,20){\mbox{$\hat v_{2,2}$}}
\put(5,45){\mbox{$\hat v_{3,2}=[\hat e_2,\hat v_{3,1}]$}}
\put(5,70){\mbox{$\hat v_{4,2}$}}
\put(5,95){\mbox{$\hat v_{5,2}=[\hat e_3, \hat v_{3,1}]$}}
}

\put(0,2){\line(-4,1){310}}
\put(0,1){\line(4,1){270}}
\put(0,3){\line(-2,1){260}}
\put(0,1){\line(2,1){230}}
\put(0,4.5){\line(-1,1){130}}
\put(0,0){\line(1,1){130}}
\put(0,0){\line(4,3){170}}
\put(0,3.4){\line(-4,3){170}}

\multiput(-320,27)(20,0){30}{\line(1,0){10}}

\put(130,140){\mbox{$m=4$}}
\put(170,135){\mbox{$m=3$}}
\put(240,120){\mbox{$m=2$}}
\put(260,75){\mbox{${\bf m=1}$}}
\put(-150,140){\mbox{$m=4$}}
\put(-200,140){\mbox{$m=3$}}
\put(-270,140){\mbox{$m=2$}}
\put(-325,85){\mbox{${\bf m=1}$}}

\put(-320,2){\line(1,0){600}}
\put(-325,10){\mbox{$m=0$}}
\put(260,10){\mbox{$m=0$}}

\multiput(-320,27)(20,0){30}{\line(1,0){10}}
\put(250,30){\mbox{Virasoro}}

\put(0,2.0){\line(-8,1){310}}
\put(0,2.0){\line(8,1){290}}
\put(0,2.5){\line(-3,1){310}}
\put(0,0.0){\line(21,8){290}}

\put(-330,50){\mbox{$m=\frac{1}{2}$}}
\put(285,45){\mbox{$m=\frac{1}{2}$}}
\put(-330,115){\mbox{$m=\frac{4}{3}$}}
\put(285,100){\mbox{$m=\frac{3}{2}$}}

\linethickness{1.5pt}
\put(0,2.2){\line(-4,1){310}}
\put(0,1.2){\line(4,1){270}}

}
\end{picture}
\caption{$2d$ integer lattice of generators of the $2d$ area-preserving diffeomorphisms.}
\label{fig:plane}
\end{figure}

\bigskip

\noindent
The commutativity properties for Hamiltonians $\hat h_k^{(m)}$ made from $\hat V$
for each ray $m$ are then the direct and simple consequences of (\ref{winfty})
\be
\boxed{
\left[\hat h_k^{(m)}, \hat h_l^{(m)}\right] = \left[\hat v_{km+1,\pm k},\hat v_{lm+1,\pm l}\right]
= \pm \Big((km)l-(lm)k\Big) \cdot \hat v_{(k+l)m,\pm k\pm l}= 0
}
\ee
We call these rays integer rays.

\subsection{Rational rays}

Only a few rays are shown in the picture, there is a ray for any
positive {\it rational} $m$,
the vertical line corresponds to the limit of $m=\pm \infty$.
One can get rid of $\pm$ in this formula by allowing negative values of $m$, but we prefer to keep
all $m$ positive for both positive and negative branches.
Note that in the standard notation (which we use) all the rays begin at point $(0,1)$,
not at the origin.
Therefore the Hamiltonians depend on the denominator and numerator of $m$ separately,
thus the better notation is $h_k^{(p,q)}$ with $p$ and $q$ co-prime.
Then the pair $(p,q)$ describes the position of the first point on the ray,
while $k$ is used to enumerate all other points on the ray $(kp,kq)$.
The rational number $m = \frac{p}{q}$ corresponds to the slope of the ray.
More explicitly,
\be
\hat h^{(p,q)}_{\pm k} := \hat v_{kp+1,\pm kq} = \dfrac{\pm1 }{q(k-1)} \left[\hat v_{p+2,\pm q},  \hat v_{ (k-1)p+1,\pm q(k-1)}      \right]
= \dfrac{(\pm 1)^{k-1}}{q (k-1)!} {\rm ad}^{k-1}_{\hat v_{p+2,\pm q}}\! \left(  \hat v_{p+1 ,\pm q }\right)
\label{hpq}
\ee
At integer $m$ (i.e. for $q=1$), we use the own notation for involved operators:
$\hat e_{m} = \hat v_{m+1,1}$ for the ``origins" $\hat h^{(m)}_1$
and $\hat e_{m+1}:=\hat v_{m+2,1}$ for the generators of the  ``negative" branches
and $\hat f_{m} = \hat v_{m+1,-1}$ and
$\hat f_{m+1}:=\hat v_{m+2,-1}$ for ``positive" ones. We call these elements $\hat e_{m}$, $\hat f_{m}$
generating operators since each this element generates with repeated commutators the whole ray.

Hamiltonians  $\hat h^{(p,\pm q)}_k$ provide an alternative labeling of points on the  lattice in Fig.1:
\be
\boxed{
\hat v_{M,i} = \hat v_{kp+1,\pm kq} = \hat h^{(p,q)}_{\pm k}
}
\ee
with co-prime $p$ and $q$.
There is a one-to-one correspondence between points and Hamiltonians,
but commuting are only Hamiltonians along the ray.

\subsection{From  rays  to cones}

In fact, (\ref{hpq}) does not exhaust the families of commuting operators.
One can take an arbitrary linear combination
\be
\hat f^{\{\alpha\}}:=\sum_m \alpha_m \hat f_m=\sum_m \alpha_m \hat v_{m+1,-1}
\ee
and a generating operator
\be
\hat {\bf f}^{\{\alpha\}}:= \sum _m \alpha_m\hat f_{m+1}=\sum_m \alpha_m \hat v_{m+2,-1}
\ee
in order to give rise a set of Hamiltonians associated with the set $\{\alpha_m\}$,
\be
\boxed{
\hat h_k^{\{\alpha\}} := {\rm ad}^{k-1}_{\hat {\bf  f}^{\{\alpha\}}}\hat f^{(\alpha)}
}
\ee
which is commutative:
\be
\left[ \hat h_k^{\{\alpha\}}, \hat h_{k'}^{\{\alpha\}}\right] = 0
\ee
For example,
\be
\hat h_2^{\{\alpha\}} = \sum_{m,m'} (m-m'+1) \alpha_m\alpha_{m'}\cdot \hat v_{m+m'+1,-2}
= \sum_{m,m'}  \alpha_m\alpha_{m'}\cdot \hat v_{m+m'+1,-2}
\nn \\
\left[ \hat h_2^{\{\alpha\}}, \hat h_{1}^{\{\alpha\}}\right]
= \sum_{m,m',m''} (m-m'+1)(m+m'-2m'')\alpha_m\alpha_{m'}\alpha_{m''}\cdot \hat v_{m+m'+m'',-3} = 0
\ee
this sum vanishes after symmetrization of $(i-i'+1)(i+i'-2i'')$ over $i,i','i''$.

This is still not the end of the story: in the same way  one can take linear combinations of $\hat v_{kp+1,\pm kq}$ describing the rational rays,
\be
\hat f^{(\alpha)}:=\sum_p \alpha_p \hat f_p=\sum_p \alpha_p \hat v_{p+1,-q}\nn\\
\hat {\bf f}^{(\alpha)}:= \sum _p \alpha_p\hat f_{p+1}=\sum_p \alpha_p \hat v_{p+2,-q}
\ee
in order to generate the Hamiltonians
\be
\boxed{
h^{\{\alpha\}}_k = \sum_{p_1,\ldots, p_k} \alpha_{p_1}\ldots\alpha_{p_k} \cdot v_{p_1+\ldots+p_k+1,-qk}
}
\label{halpha}
\ee
which are commuting. Indeed,

{\footnotesize
\be
\left[ h^{\{\alpha\}}_k, h^{\{\alpha\}}_l\right]
= \sum_{\stackrel{p'_1,\ldots,p'_l}{p_1,\ldots, p_k}}\
\underbrace{
\Big((p_1+\ldots+p_k)l-(p'_1+\ldots +p'_l)k\Big)
}_{\sum_{i=1}^k\sum_{j=1}^l (p_i-p'_j)}
\ \alpha_{p_1}\ldots \alpha_{p_k}\alpha_{p'_1}\ldots \alpha_{p'_l}\cdot
v_{(p_1+\ldots+p_k+p'_1+\ldots p'_l,-q(k+l)}
= 0
\nn
\ee
}
vanishes after  symmetrization over $\{p\}$ and $\{p'\}$

The same is true for the Hamiltonians produced by generalized $\hat{\bf e}^{(\alpha)}$
and $\hat e^{(\alpha)}$, they differ by the change of sign of $q$ in (\ref{halpha}).

One can say that instead of rays shown in Fig.1, the generalized integrable systems
are associated with ``cones" or ``bundle of rays", see Fig,2:

\begin{figure}[h]
\begin{picture}(300,170)(-250,-15)

\put(-200,0){\line(1,0){400}}
\put(0,0){\line(-5,1){200}}
\put(0,0){\line(-3,1){200}}

\put(0,0){\line(1,1){130}}
\put(0,0){\line(2,3){100}}

\put(0,0){\line(1,10){14}}
\put(0,0){\line(-2,10){28}}

\put(-195,-10){\mbox{$f$}}
\put(190,-10){\mbox{$e$}}


\put(-80,5){\line(0,1){40}}
\put(80,60){\line(0,1){90}}
\put(-82,-10){\mbox{$-q$}}
\put(78,-10){\mbox{$q$}}

\linethickness{1.5pt}
\put(0,0){\line(-4,1){200}}
\put(0,0){\line(4,5){115}}

\put(-65,90){\mbox{\rm allowed by (\ref{gencom})}}

\end{picture}

\caption{\footnotesize
One can think of the Hamiltonians (\ref{halpha}) as associated with the ``cones"
or ``bundles of rays"
instead of just rays in Fig.1,
with the profiles defined by the distribution $\alpha(\phi)$ in the angle $\phi$
(of course, only integer points of the lattice contribute).
{\it The ray} of Fig.1 is a particular distribution with $\alpha(\phi)=\delta(\phi-\phi_0)$
with a rational slope $\tan(\phi_0)$ and is shown by boldfaced lines.
There is {\bf one integrable system per cone/distribution},
and the  cones can be arbitrarily wide (or narrow).
The distribution $\alpha$ is defined at a given abscissa $\pm q$ (vertical line),
thus there is a restriction that the cone should lie entirely in the right or left quadrant
(to be fully crossed by a single vertical line at $q\ne 0$),
vertical cones are not allowed (do not describe an integrable system, see sec.\ref{novert}).
However, a slight generalization provides commuting Hamiltonians for
vertical cones as well,  see sec.\ref{novert}.
}
\end{figure}
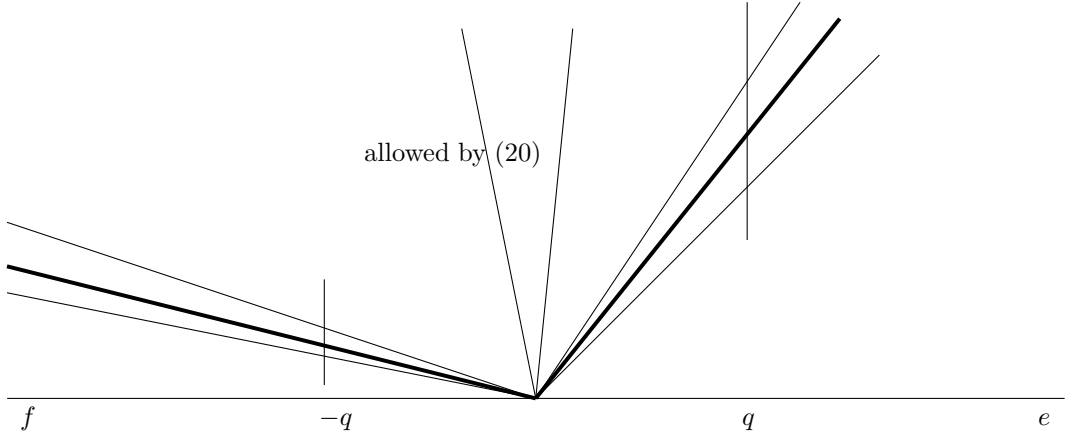

\subsection{Vertical cones are also allowed
\label{novert}}

Above construction is not directly applicable for vertical cones.
The simplest counterexample is the would be Hamiltonian $h_1\stackrel{?}{=}v_{2,1}+v_{2,-1}$.
The next one would be produced by the commutator with $v_{3,1}+v_{3,-1}$,
which is $h_2\stackrel{?}{=}-v_{3,2} + v_{3,-2}$.
Then the commutator $[h_1,h_2] = -4(v_{3,1}+v_{3,-1}) \neq 0$
is non-vanishing, the integrability property does not occur.

This is a manifestation of the universal rule: the distribution $\alpha$ should be defined at
a given value of $q$. For example,
\be
\begin{array}{ll}
{\rm if} & h_1 \stackrel{?}{=} \alpha_1 v_{m_1+1,q_1} + \alpha_2 v_{m_2+1,q_2} \nn \\ \nn \\
\text{then the Hamiltonian generator is}    &  g\stackrel{?}{=}\alpha_1 v_{m_1+2,q_1} + \alpha_2 v_{m_2+2,q_2},
\nn \\ \nn\\
\text{the second Hamiltonian} &h_2 :=[g,h_1] \stackrel{?}{=}\alpha_1^2q_1 v_{2m_1+1,2q_1}
+ \alpha_1\alpha_2(q_1+q_2)v_{m_1+m_2+1,q_1+q_2} +\alpha_2^2q_2 v_{2m_2+1,2q_2} \nn \\ \nn \\
\text{and the commutator} & [h_2,h_1]= \alpha_1\alpha_2(q_1-q_2)(m_1q_2-m_2q_1)\left(\alpha_1 v_{2m_1+m_2,2q_1+q_2}
+\alpha_2 v_{m_1+2m_2,q_1+2q_2}\right)
\end{array}
\nn \\ \nn
\ee
can not vanish when $q_1$ and $q_2$ have different signs. Moreover, if $q_1\ne q_2$, it vanishes only if ${m_1\over q_1}={m_2\over q_2}$, which means they belong to the same commutative family $v_{km+1,kq}$.

However, this is  not the end of the story.
In fact, one can construct a more general system of commuting Hamiltonians:
\be
\boxed{
h_k^{(\alpha)}=\sum_{m,n}\alpha_{mn}^{(k)} v_{m+1,n}
}
\label{gencom}
\ee
where
$\alpha^{(k)}$  are defined through powers of generating functions:
\be
\sum_{m,n}\alpha_{mn}^{(k)}x^ny^m = \left(\sum_{m,n} \alpha_{mn}^{(1)}x^ny^m\right)^k
\ee
Hamiltonians (\ref{halpha}) are a particular choice of this, but now there is no restriction on $n$,
they can be both positive and negative.
Lacking at present moment is the construction of the sequence (\ref{gencom})
by action of $w_{\infty}$ generators.

\subsection{From $w_\infty$ to $W_\infty$}

The $w_\infty$-algebra contains the $v_{2,k}=L_{-k}$ Virasoro subalgebra (dashed line in Fig.1),
but does not admit a central extension.
We will need another algebra, the $W_{1+\infty}$-algebra, which admits central extensions, and from which the $w_\infty$-algebra (\ref{winfty}) is obtained by a contraction.

In the meanwhile, though (\ref{winfty}) is just a Poisson-bracket limit of the
true commutator of the $W_{1+\infty}$ algebra, this latter algebra is formed by the same pattern of Fig.1, and, as we demonstrate below, the simple picture of rays and cones is lifted to more general representations of the algebra $W_{1+\infty}$. In fact, as we discuss in \cite{MMMP2}, this picture has an algebraic origin and, hence, is not related to a concrete representation.

In the $W_{1+\infty}$ algebra, every ray of Fig.1 is associated with commuting Hamiltonians.
For instance, at $q=1$,
\be
\boxed{
[\hat H^{(m)}_k,\hat H^{(m)}_l] =   0
}
\ee
The lift to $W_{1+\infty}$
means that this commutativity is fulfilled in any sophisticated representations
like the Fock representation in terms of a one-parametric string of variables $p_k$, which we actually use in this paper.
An analytical check is often more complicated in this case, thus a qualitative argument is desirable.
We will provide it in \cite{MMMP2}, and here make just a couple of remarks, which can be helpful in this respect.

The horizontal action of operators $\hat E_k$ and $\hat F_k$ can not take us out of the upper cone,
which is bounded by the $m=1$ rays.
To get beyond it, one obligatory needs to overpass by vertical action of $\hat W_0$ on the bottom horizontal line
(the $m=0$ ray).

Also, the simplest relations
\be
\left[H_2^{(m)}, H_1^{(m)}\right] = 0
\ee
are just the simplest Serre identities in the algebra.
The next one,
\be
\left[H_3^{(m)}, H_1^{(m)}\right] = 0
\ee
immediately reduces to it, but for higher commutation relations the story is more difficult.
We address it in a separate note \cite{MMMP2}.

To summarize, the commutative families (in particular, rays) in the $W_{1+\infty}$-algebra are the same as in $w_\infty$.
What is different is a convenient way to associate operators to integer points in the half-plane: it corresponds to linear combinations of spins (still with the fixed grading).

\section{One-body operator representation of $W_{1+\infty}$
\label{zrep}}

We consider here the most convenient way to parameterize the $W_{1+\infty}$ algebra, which is done by the one-body operators and further extending to provide a central extension. This means that we start from the algebra $w_{1+\infty}$ of differential operators on a circle. This algebra has an obvious basis given by operators $z^k{d^n\over dz^n}$, or, the more suitable for our purposes $	z^k{\hat D^n}$,
 where $\hat D:=z{d\over dz}$ ($n\in \mathbb{Z}_{\ge 0}$, $k\in\mathbb{Z}$). The basis operators are labelled by two indices, degree of $z$, $k$ is called the grading, while $n+1$ is the spin. Generic elements of the algebra are given by all possible linear combination. We will mainly consider operators of fixed grading, which may be sums of a few spins, i.e. we will deal with the operators of the form $z^kP(\hat D)$, where $P(x)$ is a polynomial.
Then the general commutation relations for such elements are
\be\label{circlecomrel}
\Big[z^nP(\hat D),z^mQ(\hat D)\Big]=z^{n+m}\left(P(\hat D+m)Q(\hat D)-P(\hat D)Q(\hat D+n)\right)
\ee
for two polynomials $P$ and $Q$. We will additionally denote the families of operators of a fixed grading $k$ and of a maximal spin $n+1$ by ${O}_{n,k}$:
\begin{equation}
	\hat{O}_{n,k} = \left\{ z^{k} \left( a_n D^n + \ldots a_0 \right) \right\}
\end{equation}
and will denote through $\sim \hat{O}_{n,k}$ the operators belonging to the family $\hat{O}_{n,k}$.
These operators can also be drawn on the half-plane on fig. \ref{fig:plane} according to the their grading and spin. We will demonstrate below that the construction of commutative families similar to that described in sec.\ref{sec:diffeo} is available.

\subsection{Integer rays}

We start with the integer rays. Consider the following identification of operators\footnote{Note that one can add to $\hat W_0$ the gradation operator $\alpha \hat D$ with an arbitrary constant $\alpha$: this just shifts $\hat D\to\hat D+\alpha$ in all formulas of this subsection.}. For the cut-and-join operator we have
\be
		\hat W_0={1\over 2}\hat D(\hat D+1)\sim\hat O_{3,0}
\ee
and for the generators of the negative branch:
\be
 \hat F_n = z^{-1}\hat D^{n}\sim\hat O_{n+1,-1},\ \ \ \ \ n\ge 0
\ee
Note, that these generators can be obtained recursively
\begin{equation}\label{Fn}
		\hat{F}_n =[F_{n-1},\hat W_0] \, , \, \hat F_0 = z^{-1}
\end{equation}
Then one obtains the commutative family similarly to \eqref{hpq}:
\be
\hat H^{(n)}_{-k} = {1\over k-1}\left[\hat F_{n+1}, \hat H^{(n)}_{-k+1}\right] =z^{-k} \prod_{i=0}^{k-1}(\hat D-i)^{n}\sim\hat O_{nk+1,-k}
\ee
with the first Hamiltonian given by the generator $\hat{F_n}$ itself:
\begin{equation}
	\hat H_{-1}^{k}= \hat F_{n}
\end{equation}
From the commutation relations \eqref{circlecomrel} it is clear that operators $\hat{F}_n$ complemented by $\hat{W}_0$ generate the whole ''positive'' part of the $w_{1+\infty}$ algebra that correspond to $\hat{O}_{n,k}$ with $k<0$. These operators are associated to what was called the positive branch of the WLZZ models \cite{China1,China2,MMsc}. In order to come to the negative WLZZ branch, one has to make \cite{MMCal} the transformation of any operator $\hat O(z)$ in accordance with
$\hat O(z)\longrightarrow -z^{-1}\hat O(z^{-1}) z$. In particular, this makes the generators of the negative branch to be:
\be\label{En}
 \hat E_n =\hat H^{(n)}_{1} =[\hat W_0,E_{n-1}]= z(\hat D+1)^{n}\sim\hat O_{n+1,1},\ \ \ \ \ n\ge 0
\ee
while the Hamiltonians are given by:
\be
\hat H^{(n)}_{k} = {1\over k-1}\left[\hat E_{n+1}, \hat H^{(n)}_{k-1}\right] = z^{k} \prod_{i=1}^{k}(\hat D+i)^n\sim\hat O_{nk+1,k}
\ee

The commutativity of the constructed subalgebras of the $w_{1+\infty}$-algebra is evident directly via commutation relations \eqref{circlecomrel}
\be
\Big[\hat H^{(n)}_{-k},\hat H^{(n)}_{-l}\Big]=z^{-k-l}\left(\prod_{i=0}^{k-1}(\hat D-i+l)^{n}\prod_{i=0}^{l-1}(\hat D-i)^n-
(k\leftrightarrow l)\right)=z^{-k-l}\left(\prod_{i=0}^{k+l-1}(\hat D-i)^n-(k\leftrightarrow l)\right)
=0\nn\\
\Big[\hat H^{(n)}_{k},\hat H^{(n)}_{l}\Big]=z^{k+l}\left(\prod_{i=1}^{k}(\hat D+i+l)^n\prod_{i=1}^{l}(\hat D+i)^n-
(k\leftrightarrow l)\right)=z^{k+l}\left(\prod_{i=1}^{k+l}(\hat D+i)^n-(k\leftrightarrow l)\right)
=0
\ee

\subsection{Rational rays}

The extended commutative sets of sec.2.2 parameterized by two integers $p$ and $q$ are obtained in these terms
by the change of variables $z\to z^q$. This corresponds to choosing $\hat F^{(q)}_0=z^{-q}$ instead of $\hat F_0=z^{-1}$ in (\ref{Fn}) and choosing $\hat E^{(q)}_0=z^{q}$ instead of $\hat E_0=z$ in (\ref{En}) with the corresponding shift ${q-1\over 2}$ in $\hat D$ in order to keep the same $\hat W_0$. This gives rise to
\be\label{Fnr}
 \hat F_n^{(q)} =\hat H^{(n,q)}_{-1} =[F_{n-1}^{(q)},\hat W_0]= z^{-q} q^n \left(\hat D-{q-1\over 2}\right)^{n}
 \sim\hat O_{n+1,-q},\ \ \ \ \ n\ge 0
\ee
\be \label{eq:h-pq-one-body}
\hat H^{(p,q)}_{- k}={1\over k-1}\left[\hat F_{p+1}^{(q)}, \hat H^{(p,q)}_{-k+1}\right] =z^{- qk}\prod_{i=0}^{k-1}
\left(\hat D-{q-1\over 2}- qi\right)^p\sim\hat O_{pk+1,-qk}
\ee
and
\be\label{Enq}
\hat E_n^{(q)} =\hat H^{(n,q)}_{1} =[\hat W_0,E_{n-1}^{(q)}]= z^q q^{n} \left(\hat D-{q-1\over 2}+1\right)^{n}\sim\hat O_{n+1,q},\ \ \ \ \ n\ge 0
\ee
\be\label{Hnq}
\hat H^{(p,q)}_{k}={1\over k-1}\left[\hat E_{p+1}^{(q)}, \hat H^{(p,q)}_{k-1}\right] =z^{ qk}\prod_{i=1}^{k}
\left(\hat D-{q-1\over 2}+qi\right)^p
\sim\hat O_{pk+1,qk}
\ee
so that the first non-trivial Hamiltonians of these sets are the generating operators $\hat F_p^{(q)}$ and $\hat E_p^{(q)}$: $\hat H^{(p,q)}_{\pm 1}\sim \hat O_{p+1,\pm q}$. Note that $\hat H^{(mp,mq)}_{\pm k}$ is not equal to $\hat H^{(p,q)}_{\pm mk}$, but their leading orders coincide: they both belong to $\hat O_{mkp+1,\pm mqk}$ and, hence, to one and the same $v_{mkp+1,\pm mqk}$ of sec.2.2.

\subsection{Cones: linear combinations}

From this consideration, it is clear that one can generate a commutative set of Hamiltonians by an arbitrary polynomial $G(x)$:
\be\label{HG}
\hat H^{(G)}_{\pm k}=z^{\pm k}\prod_{i=0}^{k-1}G(\hat D\pm i)
\ee
This class of commuting Hamiltonians can be also extended by the change of variables $z\to z^q$, which gives
\be\label{shifted}
\boxed{
\hat H^{(G,q)}_{\pm k}=z^{\pm qk}\prod_{i=0}^{k-1}G(\hat D\pm qi)}
\ee
Using
\be
G(\hat D)z^{\pm k}=z^{\pm k}G(\hat D\pm k)
\ee
one can rewrite (\ref{HG}), (\ref{shifted}) in the form
\be
\hat H^{(G)}_{\pm k}=\left(z^{\pm 1}G(\hat D)\right)^k\nn\\
\boxed{
\hat H^{(G,q)}_{\pm k}=\left(z^{\pm q}G(\hat D)\right)^k}
\ee
which makes the commutativity of Hamiltonians even more evident.

Note that the Hamiltonians $\hat H^{(G)}_{\pm k}$ are generated by the corresponding sums of $\hat F_k$ and $\hat E_k$. For instance, if the polynomial $G(x)=\sum_{i=1}^n\alpha_ix^i$, the Hamiltonian is
\be
\hat H^{(G)}_{- k}={1\over k-1}\Big[\hat{F}^{(G)},\hat H^{(G)}_{- k+1}\Big]
\ee
with the operator $\hat{F}^{(G)}$ given by
\begin{equation}
	\hat{F}^{(G)}:=\sum_{i=1}^n\alpha_i\hat F_{i+1}=z^{-1} \hat{D} \cdot G(\hat{D})
\end{equation}
and
\be
\hat H^{(G)}_{-1}=\sum_{i=1}^n\alpha_i\hat F_{i}=z^{-1} G(\hat{D})
\ee
At the same time, it is unclear how to get the most general construction of the commutative family of Hamiltonians generated by any element of $W_{1+\infty}$,
\be
H_k=\left(\sum_{n,m}\alpha_{nm}z^n\hat D^m\right)^k
\ee
by repeated commutators of generating operators. This formula is a $W_{1+\infty}$ counterpart of (\ref{gencom}).

\subsection{Central extension}
Strictly speaking, the $W_{1+\infty}$ algebra is constructed as the central extension of the $w_{1+\infty}$ algebra considered above. Denote the elements of this algebra in the basis corresponding to the differential operators on the circle as $W\Big(z^nP(\hat D)\Big)=W_n\Big(P(\hat D)\Big)$.
Then, the central extension of the algebra is
\be\label{ce}
\Big[W_n\Big(P(\hat D)\Big),W_m(Q(\hat D)\Big)\Big]=W_{n+m}\Big(P(\hat D+m)Q(\hat D)-
P(\hat D)Q(\hat D+n)\Big)+c\Psi\left(W_n\Big(P(\hat D)\Big),W_m\Big(Q(\hat D)\Big)\right)
\ee
where $c$ is the central charge and the 2-cocycle $\Psi$ is \cite{KP,Li,F,KR1}
\be
\Psi\left(W_n\Big(P(\hat D)\Big),W_m\Big(Q(\hat D)\Big)\right)=\delta_{n+m,0}\left(\sum_{j=1}^nP(-j)Q(n-j)-
\sum_{j=1}^mP(m-j)Q(-j)\right)
\ee
where the sum $\sum_{j=1}^n...=0$ at $n<1$.
\\

Our construction of commutative families treats the positive and negative branches separately (apart from the vertical cones in sec. \ref{novert}, which we do not consider further). Therefore the central charge never really emerges in any of the computations, hence we do not differentiate between the two algebras. In other words, we can think of the one-body operators as a representation of the $W_{1+\infty}$ algebra with vanishing central charge.
\\

Note also that one can choose in the completion of the $W_{1+\infty}$ algebra the basis similar to $v_{M,i}$ in the $w_\infty$-algebra of sec.2: the commutator of any two elements of the basis is a single element of the basis, and not a linear combination, this is the Kac-Radul basis \cite{KR1,Miki} with elements $u_{m,i}=W(z^mt^{i\hat D})$. The commutation relations in this basis are
\be\label{KR}
[u_{m,i},u_{n,j}]=(t^{ni}-t^{mj})u_{m+n,i+j}-c\delta_{m+n}{t^{ni}-t^{mj}\over t^{i+j}-1}
\ee
However, while respecting the grading, this basis does not respect the spin at all, and we do not use it in this paper.

\section{Beyond one-body
\label{beyondzrep}}

At this stage, a general description of the $W_\infty$-algebra ends, there are at least two conventional further directions, and they split at this point.

One option is to use {\it fermionization} and further {\it bosonisation}, which allows one to lift the $z$-representation to the ``second-quantized" level with many new parameters instead of the single $z$.
This is a canonical lifting, which gives a representation of the $W_\infty$-algebra in the Fock space, and this later can be realized
in terms of ``time" variables, an infinite set of $\{p_k\}$ which substitutes the single $z$.
We remind this standard trick in sec.8.1.
Since this is a canonical procedure, it preserves most of properties of the $z$-representation.
The only thing it can and does add to the $z$-representation is the central charge, i.e. the proper modification of commutators between
the left ($F$) and right ($E$) parts of the algebra {\it outside} the cone given by the integer rays.
Since our story of commutative subalgebras and Hamiltonians is entirely concentrated in one or another
quarter-plane, see sec.\ref{novert}, it does not depend on these ``details",
and we have a canonical lifting of its $z$-representation to times.
We discuss this direction in sec.\ref{sec:timerep}, which has two natural continuations:
to discussion of explicit formulas and a hidden structure of our Hamiltonians for
particular rays, integer \cite{MMCal} and rational, which will be shortly presented in Appendix A,
and to $\beta$-deformation,
which is done in terms of time variables (where the Schur functions are basically just substituted
by the Jack polynomials), this is briefly reminded in sec.\ref{betadef}.

Despite this is a well known direction, it has been never elaborated  in enough detail,
and even a simple question of commutative subalgebras in terms of time variables has not been raised.
More important, as  is typical for fermionization-based approach, it is severely restricted,
roughly to $N$ instead  of the $N^2$ variables, being associated with an integer central charge.
This is very much like considering the Frenkel-Kac representations of the affine Lie algebras
in terms of ${\rm rank}_G$ free fields \cite{FK,Seg} in the case of integer central charge instead of the full-fledged
Wakimoto representation of \cite{GMMOS}
in terms of ${\rm dim}_G$ fields in the case of arbitrary central charge.
Of course, it is appealing to extend representation theory of the $W_\infty$-algebra in this direction, since the Fock representation in time variables is associated with unit central charge.

Another option is to try to substitute $z$ by a matrix $\Lambda$, and this brings us into the field
of matrix representations, which we consider first, in sec.\ref{matrixrep}.
It is possible to build up explicitly arbitrary generators from the triple $\hat W$, $\hat E_0$, $\hat F_0$.
Moreover, the operators corresponding to the {\it integer} rays commute!
However, those corresponding to the rational rays breaks this: commutativity takes place only on the restricted space of
invariant (eigenvalue-dependent) polynomials.
Thus, at this stage, the success is only partial, and the full theory of commutative subalgebras
is not lifted to matrix representations but only a part of it.
The full theory on the space of eigenvalues
is considered in sec.\ref{evrep}, and it is naturally promoted to a generalization of the Calogero systems
of $N$ particles, similar to what was exposed in \cite{MMCal}, see sec.\ref{integras}.
Also, in appendix B we provide some details about going from matrix operators to the eigenvalue ones,
in a non-conventional way, avoiding direct use of eigenfunctions (as is usually done, say, in
perturbation theory in quantum mechanics).

After restriction to eigenvalues, the two approaches, those using matrices and time variables merge,
as we discuss in the short section \ref{timev}, where the two roads, which we take from now, merge again.
Merging is not full: in some sector (integer rays), the generalization to matrices works,
and one gets integrable systems of $N^2$ particles.
On the other hand, there are infinitely many time variables and only $N$ eigenvalues,
so merging implies a restriction of the time story to the Miwa locus, or, alternatively, one has to consider only the $N\to\infty$ limit.
Thus our two roads have their own advantages.

\section{Matrix operators and $N$-deformed Hamiltonians
\label{matrixrep}}

 Thus, here we give another realization, which extends one-body operators to many-body operators, but still does not admit a central extension. It can be done in terms of an $N\times N$ matrix $\Lambda$.
As compared with the one-body operators, here $z$ is substituted with $\Lambda$, $D=z\frac{\p}{\p z}$ with\footnote{
Hereafter, by the matrix derivative, we imply the derivative w.r.t. matrix elements of the transposed matrix: $\left(\frac{\partial}{\partial \Lambda}\right)_{ij}=\frac{\partial}{\partial \Lambda_{ji}}$.} $\hat{\cal D}=\Lambda\frac{\p}{\p \Lambda}=\Lambda\p_\Lambda$ and one has to take a trace of every operator.

\subsection{Integer rays}

This way, we get the following set of operators:
\be\label{W0L1}
\hat W_0={1\over 2}:\Tr \left(\Lambda{\p\over\p\Lambda}\right)^2:+N \Tr \left(\Lambda{\p\over\p\Lambda}\right)
\ = {1\over 2}\Tr \left({\p\over\p\Lambda}\Lambda^2{\p\over\p\Lambda}\right)={1\over 2}\Tr \hat{\cal D}
\Big(\hat{\cal D}+N\Big)
\ee
and the two commuting  Hamiltonian sets for the integer rays are  \cite{MMCal}:

\bigskip

\hspace{-0.65cm}\framebox{\parbox{17cm}{
\be\label{Fnm}
 \hat F_m =\hat H^{(m)}_{-1} =[F_{m-1},\hat W_0]= \Tr\Lambda^{-1}\hat {\cal D}^{m},\ \ \ \ \ m\ge 0
\ee
\be\label{Hmm}
\hat H^{(m)}_{-k} = {1\over k-1}\left[\hat F_{m+1}, \hat H^{(m)}_{-k+1}\right] =\Tr \Big(\Lambda^{-1}\hat {\cal D}^m\Big)^k
\ee
and
\be\label{Enm}
 \hat E_m =\hat H^{(m)}_{1} =[\hat W_0,E_{m-1}]= \Tr\hat{\cal D}^m\Lambda,\ \ \ \ \ m\ge 0
\ee
\be\label{Hm}
\hat H^{(m)}_{k} = {1\over k-1}\left[\hat E_{m+1}, \hat H^{(m)}_{k-1}\right] =\Tr \Big(\hat{\cal D}^m\Lambda\Big)^k
\ee
}}

\bigskip

\noindent
with different $k$ and the same $m$. Again, in accordance with the recipe of \cite{MMCal}, in order to come from the $F$-branch to the E-branch, one has to make the transformation of any operator $\hat O(z)$ in accordance with
$\hat O(\Lambda)\longrightarrow -\det^{-N}\Lambda\hat O(\lambda^{-1}) \det^N\Lambda$.

If one shifts the operator $\hat W_0$,
\be
\hat W_0^{(\alpha)} :={1\over 2}\Tr\hat{\cal D}
\Big(\hat{\cal D}+N\Big)+\alpha\Tr\hat{\cal D}
\ee
with an arbitrary constant $\alpha$, it describes the shift of operator $\hat{\cal D}\to\hat{\cal D}+\alpha$.

\subsection{Some problems with rational rays}

In order to generate rational rays, one has to choose
\be\label{E0F0m}
\hat F_0^{(q)}:={(-1)^{q-1}\over q!}{\rm ad}_{\hat F_1}^{q-1} \hat F_0={1\over q}\Tr\Lambda^{-q},\ \ \ \ \
\hat E_0^{(q)}:= {1\over q!}{\rm ad}_{\hat E_1}^{q-1} \hat E_0 ={1\over q}\Tr\Lambda^{q}
\ee
obtaining for the generating operators:
\begin{equation}
	\begin{split}
&\hat F_1^{(q)}=[\hat F_0^{(q)},\hat W_0]=
{1\over 2}\Tr\left(\Lambda^{-q}\partial_\Lambda\Lambda+\Lambda\partial_\Lambda\Lambda^{-q}\right)\\
&\hat F_2^{(q)}=[\hat F_1^{(q)},\hat W_0]={1\over 2}\sum_{{a,b=0}\atop{a+b=q-1}}\Tr
\left(\Lambda^{-a}\partial_\Lambda\Lambda^{1-b}\partial_\Lambda
+\partial_\Lambda\Lambda^{1-b}\partial_\Lambda\Lambda^{-a}\right) \\
&\hat F_3^{(q)}=[\hat F_2^{(q)},\hat W_0]=
\sum_{{a,b,c\ge 0}\atop{a+b+c=q-2}}\Big(1-{1\over 2}\delta_{b,0}\Big)\Tr\left( \Lambda^{-a}\p_\Lambda\Lambda^{1-b}\p_\Lambda\Lambda^{1-c}\p_\Lambda
+\p_\Lambda\Lambda^{1-c}\p_\Lambda\Lambda^{1-b}\p_\Lambda\Lambda^{-a}\right)
\\
&\vdots
\\
&\hat F_k^{(q)}=[\hat F_{k-1}^{(q)},\hat W_0]=
{1\over 2}\sum_{{a_i\ge 0}\atop{\sum_ia_i=q-k+1}}\xi^{(0)}_{\{a_i\}}\Tr\left( \Lambda^{-a_1}\p_\Lambda\Lambda^{1-a_2}\ldots\p_\Lambda\Lambda^{1-a_k}\p_\Lambda
+ \right. \\
& \hspace{9cm} + \left.\p_\Lambda\Lambda^{1-a_k}\p_\Lambda\Lambda^{1-a_{k-1}}\ldots\p_\Lambda\Lambda^{-a_1}\right)
\\
&\hat E_1^{(q)}=[\hat W_0,\hat E_0^{(q)}]={1\over 2}\Tr\left( \Lambda^q\p_\Lambda\Lambda
+\Lambda\p_\Lambda\Lambda^q\right)
\\
&\hat E_2^{(q)}=[\hat W_0,\hat E_1^{(q)}]=
{1\over 2}\sum_{{a,b\ge 1}\atop{a+b=q+1}}\Tr\left( \Lambda^a\p_\Lambda\Lambda^{1+b}\p_\Lambda
+\p_\Lambda\Lambda^{1+b}\p_\Lambda\Lambda^a\right)\\
&\hat E_3^{(q)}=[\hat W_0,\hat E_2^{(q)}]=\sum_{{a,b,c\ge 1}\atop{a+b+c=q+2}}\Tr\left( \Lambda^a\p_\Lambda\Big(\Lambda^{b}-{1\over 2}\delta_{b,1}\Lambda\Big)\p_\Lambda\Lambda^{c+1}\p_\Lambda
+\p_\Lambda\Lambda^{c+1}\p_\Lambda\Big(\Lambda^{b}-{1\over 2}\delta_{b,1}\Lambda\Big)\p_\Lambda\Lambda^a\right)=
\\
=&\sum_{{a,b,c\ge 1}\atop{a+b+c=q+2}}\Big(1-{1\over 2}\delta_{b,1}\Big)\Tr\left( \Lambda^a\p_\Lambda\Lambda^{b}\p_\Lambda\Lambda^{c+1}\p_\Lambda
+\p_\Lambda\Lambda^{c+1}\p_\Lambda\Lambda^{b}\p_\Lambda\Lambda^a\right)
\\
& \vdots
\\
&\hat E_k^{(q)}=[\hat W_0,\hat F_{k-1}^{(q)}]=
{1\over 2}\sum_{{a_i\ge 1}\atop{\sum_ia_i=q+k-1}}\xi^{(1)}_{\{a_i\}}\Tr\left( \Lambda^{a_1}\p_\Lambda\Lambda^{a_2}\ldots\Lambda^{a_{k-1}}\p_\Lambda\Lambda^{a_k+1}\p_\Lambda
+ \right. \\
& \hspace{9cm} + \left.\p_\Lambda\Lambda^{a_k+1}\p_\Lambda\Lambda^{a_{k-1}}\ldots\p_\Lambda\Lambda^{a_1}\right)
	\end{split}
\end{equation}
where the strange coefficients $\xi_{\{a_i\}}$ are non-trivial only at the boundary of summation range (otherwise, they are just $k!$) and depends only on $a_2,\ldots,a_{k-1}$:
\be
\xi^{(p)}_{\emptyset}=1\nn\\
\xi^{(p)}_{a_2}=2+\delta_{2}\nn\\
\xi^{(p)}_{a_2,a_3}=6+3\sum_{i=2}^3\delta_{i}+\delta_{2}\delta_{3}\nn\\
\ldots\nn
\ee
\be
\xi^{(p)}_{a_2,a_3,a_4,a_5,a_6}=720+360\sum_{i=2}^6\delta_{i}+180\sum_{i<j+1}\delta_i\delta_j
+120\sum_{i=2}^5\delta_i\delta_{i+1}+90\delta_2\delta_4\delta_6
+60\Big(\delta_2\delta_3\delta_5+\delta_2\delta_4\delta_5
+\delta_2\delta_3\delta_6+\nn\\
+\delta_3\delta_4\delta_6+\delta_2\delta_5\delta_6+\delta_3\delta_5\delta_6\Big)
+30\sum_{i=2}^4\delta_i\delta_{i+1}\delta_{i+2}+\left({20\over\delta_4}+{15\over\delta_3} +{15\over\delta_5}
+{6\over\delta_2}+{6\over\delta_6}\right)\delta_2\delta_3\delta_4\delta_5\delta_6
+\delta_2\delta_3\delta_4\delta_5\delta_6
\ee
$$
\ldots
$$
where we denoted $\delta_i=-\delta_{a_i,p}$.
The coefficients are the multinomial coefficients of expansion $(\sum_{i=1}^{k+1}x_i)^{k}$, for instance,
\be
\Big(\sum_{i=1}^{7}x_i\Big)^6=720m_{[1,1,1,1,1,1]}+360m_{[2,1,1,1,1]}+180m_{[2,2,1,1]}+90m_{[2,2,2]}
+120m_{[3,1,1,1]}+60m_{[3,2,1]}+\nn\\
+30m_{[4,1,1]}+20m_{[3,3]}+15m_{[4,2]}+6m_{[5,1]}+m_{[6]}
\nn\\
\hbox{etc}
\ee
Here $m_R$ are the monomial symmetric functions. The coefficient in front of $m_R$ with $|R|=k$ is equal to
\be
{k!\over R_1!\ldots R_{k-1}!R_k!}
\ee
and some parts $R_i$ of the partition $R$ can be zero. This coefficient is also equal to the number of ways to label the partition $R$ of size $|R|=k$ with numbers 1 to $k$.

Now what is the correspondence between concrete terms in $\xi^{(p)}$ and partitions? The rule of thumb is as follows: one has to remove the leftmost column of the Young diagram corresponding to the partition. Then, each cycle of length $r$ of the remaining Young diagram corresponds to $r$ successive $\delta_i$'s. For instance, look at $\xi^{(p)}_{a_2,a_3,a_4,a_5,a_6}$. The first partition consists of just one column so that, after removing the leftmost column, nothing remains: hence, no $\delta_i$. Similarly, the second partition becomes just $[1]$, which means it corresponds to single $\delta_i$'s. The third partition becomes $[1,1]$, which means only two non-neighbour $\delta_i$'s contribute, while the next partition becomes $[2]$, i.e. the cycles of length 2 contribute, they are just $\delta_i\delta_{i+1}$, etc.

These generating operators give rise to the Hamiltonians, the simplest of them being
\be
H_2^{(1,q)}=q[E_2^{(q)},E_1^{(q)}]=\sum_{{a,b,c\ge 1}\atop{a+b+c=2q+2}}\eta_b\Tr\Lambda^a\p_\Lambda\Lambda^b\p_\Lambda\Lambda^c
\ee
where
\be
\eta_{q+1+ k}=2q-3|k|-\delta_{k,0}+{q+1\over 2}\delta_{k+q,0},\ \ \ \ \ -q\le k\le q-1
\ee
However, the Hamiltonians constructed by the generating operators $\hat F^{(q)}_k$, $\hat E^{(q)}_k$ {\bf do not commute} for exception of the integer ray series (\ref{Hmm}), (\ref{Hm}) associated with $q=1$. For instance, in the case of $N=2$,
\be\label{ex}
[E_1^{(q)},H_2^{(1,q)}]=\hbox{Pol}^{(q)}\Big(\Tr \Lambda^k,\det \Lambda\Big)\cdot\underbrace{\Tr\left[2\Lambda\p^2_\Lambda\Lambda+
N^2\p^2_\Lambda\Lambda^2+N^2\Lambda^2\p^2_\Lambda-(N^2+1)\left(\Lambda\p_\Lambda\Lambda\p_\Lambda+
\p_\Lambda\Lambda\p_\Lambda\Lambda\right)\right]}_{\hat{\mathrm{Diff}}}
\ee
where $\hbox{Pol}^{(q)}\Big(\Tr \Lambda^k,\det \Lambda\Big)$ is an invariant polynomial of $\Lambda$ depending on $q$ such that $\hbox{Pol}^{(1)}=0$, while the differential part $\hat{\mathrm{Diff}}$ does not depend on $q$.

Nevertheless, the Hamiltonians {\bf do commute} when acting on the invariant polynomials of matrices, i.e. of\footnote{In fact, the Hamiltonians generated by the $E$-operators commute also on polynomials of $\Tr\Lambda^{-k}$, but those generated by the $F$-operators do not commute.}  $p_k=\Tr\Lambda^k$ (in particular, they commute at $N=1$, when describe the one-body operators). In particular, $\hat{\cal D}$ in (\ref{ex}) vanishes on invariant polynomials of $\Lambda$.

Hence, in order to construct a representation of $W_{1+\infty}$ in terms of matrices, one has additionally to project matrix operators realizing this representation onto the subspace of invariant polynomials of matrices.
This allows us to construct a representation of $W_{1+\infty}$ and the commuting Hamiltonians in invariant variables $p_k=\Tr\Lambda^k$. This is much similar to constructing the $\beta$-deformation of the Hamiltonians in terms of matrix eigenvalues in \cite{MMCal} via the Dunkl operators: they also commute only when acting on symmetric functions of the eigenvalues.

In particular, this means that the rational rays can not be used for constructing Calogero like Hamiltonians in matrix terms (only in terms of eigenvalues, see the next section), though the integer rays can. This is what has been done in \cite{MMCal} even for the $\beta$-deformed case, corresponding to a non-trivial Calogero coupling constant: the commutativity of the integer ray Hamiltonians survives the $\beta$-deformation.

\section{Eigenvalue representation
\label{evrep}}

Instead of constructing a matrix realization, as we did in the previous section, one may equally well use the eigenvalue terms, an advantage of such a representation being that the operators act on the space where there is no angular matrix variables from the very beginning. In fact, just this eigenvalue representation is related to the many-body integrable systems of the Calogero type, as we discuss in the next section.

\subsection{Integer rays}

We start with the following set of operators:
\be
\hat W_0&=&{1\over 2}\sum_i\lambda_i^2{\p^2\over\p\lambda_i^2}+\sum_{i\ne j}{\lambda_i^2\over \lambda_i-\lambda_j}{\p\over\p\lambda_i}+\sum_i\lambda_i{\p\over\p\lambda_i}
\ee
which is equivalent to (\ref{W0L1}) in terms of eigenvalues, and two generating operators
\be\label{F0E0ev}
\hat F_0=\sum_i\lambda^{-1}_i,\ \ \ \ \ \hat E_0=\sum_i\lambda_i,\ \ \ \ \
\ee
Now, with our standard procedure, we obtain
\be
\hat F_1&=&\hat H_{-1}^{(1)}=[\hat F_0,\hat W_0]=\sum_i{\p\over\p\lambda_i}\nn\\
\hat F_2&=&\hat H_{-1}^{(2)}=[\hat F_1,\hat W_0]=\sum_i\lambda_i{\p^2\over\p\lambda_i^2}
+2\sum_{i\ne j}{\lambda_i\over \lambda_i-\lambda_j}{\p\over\p\lambda_i}+\sum_i{\p\over\p\lambda_i}\nn\\
\ldots
\ee
and the first series of Hamiltonians is \cite{MMCal}
\be
\hat H_{-k}^{(1)}=\Tr{\p^k\over\p\Lambda^k}=\sum_i\sum_{{I\subset [1,\dots,N]}\atop{|I|=k-1}}\prod_{j\in I}{1\over\lambda_i-\lambda_j}
{\p\over\p\lambda_i}
\ee
Note that the sum includes the terms with poles at $i=j$, which are resolved by the L'H\^ospital's rule.

One can also realize these Hamiltonians in terms of the operators $\hat {\mathfrak{D}}_i=\lambda_i\hat {\mathfrak{d}}_i$ with
\be
\hat {\mathfrak{d}}_i={\p\over\p\lambda_i}+\sum_{j\ne i}{1\over \lambda_i-\lambda_j}
\ee
Indeed,
\be
\hat H_{-k}^{(1)}=\sum_i\left(\lambda_i^{-1}\hat {\mathfrak{D}}_i\right)^k
\ee
Similarly, the higher commuting families of the Hamiltonians are
\be\label{H-km}
\boxed{
\hat H_{-k}^{(m)} = \sum_i \left(\lambda_i^{-1}\hat{\mathfrak{D}}_i^m
\right)^k
}
\ee

We discuss here only the $F$-generating operators , since, in order to construct the $E$-generating operators, one just has to make in all operators of this section the substitution $\lambda_i\to\lambda_i^{-1}$, and replace any operator $\hat A(\lambda_i)\to \left(\prod_{i=1}^N\lambda_i\right)^{-N}\cdot
\hat A(\lambda_i^{-1})\cdot\left(\prod_{i=1}^N\lambda_i\right)^{N}$. Indeed, one can check that this works for the first generating operators: this leaves the cut-and-join operator $\hat W_0$ intact while making $\hat E_0^{(q)}$ from $\hat F_0^{(q)}$.  These operators define all others, hence, the procedure is correct.

\subsection{Rational rays}

In the case of rational rays, we repeat the chain of commutators as in the previous section,
\be\label{E0F0mr}
\hat F_0^{(q)}:={(-1)^{q-1}\over q!}{\rm ad}_{\hat F_1}^{q-1} \hat F_0={1\over q}\sum_i\lambda_i^{-q},\ \ \ \ \
\hat E_0^{(q)}:= {1\over q!}{\rm ad}_{\hat E_1}^{q-1} \hat E_0 ={1\over q}\sum_i\lambda_i^{q}
\ee
obtaining for the generating operators:
\begin{equation}
\begin{split}
&\hat F_1^{(q)}=[\hat F_0^{(q)},\hat W_0]=
\sum_i\lambda_i^{1-q\over 2}{\p\over\p\lambda_i}\lambda_i^{1-q\over 2}+\sum_{i\ne j}{\lambda_i^{1-q}\over\lambda_i-\lambda_j}
\\
&\hat F_2^{(q)}={1\over q}[\hat F_1^{(q)},\hat W_0]=
\sum_i\lambda_i^{1-q\over 2}{\p\over\p\lambda_i}\lambda_i{\p\over\p\lambda_i}\lambda_i^{1-q\over 2}
+2 \sum_{i\ne j}{\lambda_i^{2-q\over 2}\over \lambda_i-\lambda_j}{\p\over\p\lambda_i}\lambda_i^{2-q\over 2}
+\sum_{i\ne j\ne k}{\lambda_i^{2-q}\over(\lambda_i-\lambda_j)(\lambda_i-\lambda_k)}
\end{split}
\end{equation}
As one can see the formulas become more involved than in the integer rays case, and the first Hamiltonians are
\begin{equation}
	\begin{split}
&\hat H_{-1}^{(1,q)}=\hat F_1^{(q)}=
\sum_i\lambda_i^{1-q\over 2}{\p\over\p\lambda_i}\lambda_i^{1-q\over 2}+\sum_{i\ne j}{\lambda_i^{1-q}\over\lambda_i-\lambda_j}
\\
&\hat H_{-2}^{(1,q)}={1\over q}[\hat F_1^{(q)},\hat F_2^{(q)}]=
\sum_i\lambda_i^{1-q\over 2}{\p\over\p\lambda_i}\lambda_i^{1-q}{\p\over\p\lambda_i}\lambda_i^{1-q\over 2}
+2 \sum_{i\ne j}{\lambda_i^{1-q}\over \lambda_i-\lambda_j}{\p\over\p\lambda_i}\lambda_i^{1-q}
+\sum_{i\ne j\ne k}{\lambda_i^{2-2q}\over(\lambda_i-\lambda_j)(\lambda_i-\lambda_k)}
	\end{split}
\end{equation}
However, the answer for the Hamiltonians can be again rewritten in terms of the operator ${\mathfrak{D}}_i$:
\be\label{H-kpq}
\boxed{
\hat H_{-k}^{(p,q)} = q^{pk}\sum_i \left(\lambda_i^{1-q\over 2}\cdot\lambda_i^{-1}\hat{\mathfrak{D}}_i^p\cdot\lambda_i^{1-q\over 2}
\right)^k
}
\ee

These Hamiltonians are commutative, and the eigenvalue representation provides a representation of the $W_\infty$-algebra, which is not surprising, since all these formulas can be obtained from those of the previous section when acting on the invariant polynomials of matrices. This representation, however, does not possess a central charge because of (\ref{E0F0mr}): $\hat E_0^{(q)}$ and $\hat F_0^{(q)}$ commute with each other, while their commutator, in accordance with (\ref{ce}), is proportional to the central charge.

\section{Many-body systems of Calogero type
\label{integras}}

\subsection{Many-body integrable systems}

As we discussed in \cite{MMCal}, one can naturally interpret the Hamiltonians of the previous section at integer rays as Hamiltonians of an integrable system, with variables $\lambda_i$ or maybe their degrees playing the role of the coordinates in the system. In fact, the Hamiltonians (\ref{H-km}) at any $m$ give rise to a trivial integrable system: one can make the transformation
\be\label{van}
\hat {\mathfrak H}_{-k}^{(m)}=\Delta(\lambda)\cdot\hat H_{-k}^{(m)}\cdot\Delta(\lambda)^{-1}
\ee
with $\Delta(\lambda):=\prod_{i<j}(\lambda_i-\lambda_j)$, and the new Hamiltonians $\hat {\mathfrak H}_{-k}^{(m)}$ are sums of Hamiltonians of non-interacting particles. In particular, in the case of $\hat {\mathfrak H}_{-k}^{(1)}$, the Hamiltonians describe just free particles, and they are associated with the so called free fermion point of the Calogero system.

This property persists for rational rays: the same transformation (\ref{van}) of Hamiltonians (\ref{H-kpq}) at any $p$ and $q$ again leads to new Hamiltonians that are sums of Hamiltonians of non-interacting particles.

In order to generate non-trivial interacting Hamiltonians, one has to consider the $\beta$-deformation of the Hamiltonians, and a counterpart of (\ref{van}) is, after the deformation,
\be\label{vanb}
\hat {\mathfrak H}_{-k}^{(m)}=\Delta(\lambda)^\beta\cdot\hat H_{-k}^{(m)}(\beta)\cdot\Delta(\lambda)^{-\beta}
\ee
so that $\hat {\mathfrak H}_{-k}^{(1)}$ become the (interacting) rational Calogero Hamiltonians.

\subsection{$\beta$-deformation of integer rays}

In order to construct the $\beta$-deformation, we consider a deformation of the $\hat W_0$-operator:
\be
\hat W_0&=&{1\over 2}\sum_i\lambda_i^2{\p^2\over\p\lambda_i^2}+\beta\sum_{i\ne j}{\lambda_i^2\over \lambda_i-\lambda_j}{\p\over\p\lambda_i}+\sum_i\lambda_i{\p\over\p\lambda_i}
\ee
while the generating operators $\hat F_0$, $\hat E_0$ in (\ref{F0E0ev}) remain intact.

Then, in the case of integer rays, introducing a deformation of the operator $\hat {\mathfrak{d}}_i$ (which is the Dunkle operator)
\be
\hat {\mathfrak{d}}_i^\beta={\p\over\p\lambda_i}+\beta\sum_{j\ne i}{1\over \lambda_i-\lambda_j}(1-P_{ij})
\ee
where $P_{ij}$ is the operator permuting $i$ and $j$, one can construct the Hamiltonians \cite{MMCal}
\be\label{H-kmD}
\boxed{
\hat H_{-k}^{(m)} = \sum_i \left(\lambda_i^{-1}\hat{\mathfrak{D}}_i^m
\right)^k\Big|_{symm}
}
\ee
which form a commutative family at any $m$ if acting on symmetric functions of $\lambda_i$.
Hence, we manifestly indicated projecting onto symmetric functions of $\lambda_i$

This projection onto symmetric functions is not required only in the case of $m=1$, when the Hamiltonians, upon the transformation (\ref{vanb}), describe the rational Calogero Hamiltonians.

In fact, even more is correct: one can construct cones as before introducing the generating operator
\begin{equation}
	\hat{F}^{(G)}:=\sum_{i=1}^p\alpha_i\hat F_{i+1}
\end{equation}
with arbitrary coefficients $\alpha_i$ so that the Hamiltonians are
\be
\hat H^{(G)}_{- k}={1\over k-1}\Big[\hat{F}^{(G)},\hat H^{(G)}_{- k+1}\Big]
\ee
In particular,
\be
\hat H^{(G)}_{-1}=\sum_{i=1}^p\alpha_i\hat F_{i}
\ee
These Hamiltonians for any concrete set of $\{\alpha_i\}$ form a commutative family, and, hence, a new non-trivial integrable many-body system of Calogero type.

In order to construct the $\beta$-deformed Hamiltonians $H^{(m)}_{k}$, one note that $H^{(m)}_{k}$ are related with $H^{(m)}_{-k}$ by the substitution $H^{(m)}_{k}(\lambda_i)=\prod_i\lambda_i^{-N\beta+\beta-1}\cdot H^{(m)}_{k}(\lambda_i^{-1})\cdot\prod_i\lambda_i^{N\beta-\beta+1}$ \cite{MMCal}.

\subsection{$\beta$-deformation of rational rays}

In the case of rational rays, we repeat the chain of commutators as in the previous section:
\be\label{E0F0}
\hat F_0^{(q)}:={(-1)^{q-1}\over q!}{\rm ad}_{\hat F_1}^{q-1} \hat F_0={1\over q}\sum_i\lambda_i^{-q},\ \ \ \ \
\hat E_0^{(q)}:= {1\over q!}{\rm ad}_{\hat E_1}^{q-1} \hat E_0 ={1\over q}\sum_i\lambda_i^{q}
\ee
obtaining for the generating operators:
\begin{equation}
	\begin{split}
\hat F_1^{(q)}=&[\hat F_0^{(q)},\hat W_0]=
\sum_i\lambda_i^{1-q\over 2}{\p\over\p\lambda_i}\lambda_i^{1-q\over 2}
+\beta\sum_{i\ne j}{\lambda_i^{1-q}\over\lambda_i-\lambda_j}
\\
\hat F_2^{(q)}=&{1\over q}[\hat F_1^{(q)},\hat W_0]=
\sum_i\lambda_i^{1-q\over 2}{\p\over\p\lambda_i}\lambda_i{\p\over\p\lambda_i}\lambda_i^{1-q\over 2}
+2\beta \sum_{i\ne j}{\lambda_i^{3-\beta-q\over 2}\over \lambda_i-\lambda_j}{\p\over\p\lambda_i}\lambda_i^{1+\beta-q\over 2}
+\\&
+\beta^2\sum_{i\ne j\ne k}{\lambda_i^{2-q}\over(\lambda_i-\lambda_j)(\lambda_i-\lambda_k)}+{\beta(\beta-1)\over 2q}\sum_{i\ne j}\sum_{a=1}^{q-1}a(q-a)\lambda_i^{-a}\lambda_j^{a-q}
	\end{split}
\end{equation}
\be
\dots
\ee
Now one constructs the first Hamiltonians, they look as
\begin{equation}
	\begin{split}
\hat H_{-1}^{(1,q)}=&\hat F_1^{(q)}=
\sum_i\lambda_i^{1-q\over 2}{\p\over\p\lambda_i}\lambda_i^{1-q\over 2}
+\beta\sum_{i\ne j}{\lambda_i^{1-q}\over\lambda_i-\lambda_j}
\\
\hat H_{-2}^{(1,q)}=&{1\over q}[\hat F_1^{(q)},\hat F_2^{(q)}]=
\sum_i\lambda_i^{1-q\over 2}{\p\over\p\lambda_i}\lambda_i^{1-q}{\p\over\p\lambda_i}\lambda_i^{1-q\over 2}
+2\beta \sum_{i\ne j}{\lambda_i^{1-q}\over \lambda_i-\lambda_j}{\p\over\p\lambda_i}\lambda_i^{1-q}
+\\&+ \beta^2\sum_{i\ne j\ne k}{\lambda_i^{2-q}\over(\lambda_i-\lambda_j)(\lambda_i-\lambda_k)}
+{\beta(\beta-1)\over 2q^2}\sum_{i\ne j}\sum_{a=1}^{2q-1}((a+1)q^2+(q-a)a^2+|q-a|^3)\lambda_i^{-a}\lambda_j^{a-2q}
	\end{split}
\end{equation}
These Hamiltonians {\bf do not commute} unless $\beta=1$.

\subsection{Rational Calogero system as affine Yangian}

Thus, we realize that the $\beta$-deformation of our construction gives rise to the commutative integer ray families and to the non-commutative rational ray ones. In fact, this is not surprising: one can check that the constructed $\beta$-deformation describes the deformation of the $W_{1+\infty}$ algebra to the affine Yangian of $\mathfrak{gl}_1$ \cite{Tsim,Prochazka}. In particular, the defining affine Yangian relations \cite[Eqs.(Y0)-(Y6)]{Tsim} are fulfilled upon the identification $f_i=(-1)^i\hat F_i$, $e_i=\hat E_i$, $\psi_3-\beta(\beta-1)\psi_2=6\hat W_0$, $h_1=1$, $h_2=-\beta$, $h_3=\beta-1$ (i.e. $\sigma_2=-1-\beta(\beta-1)$, $\sigma_3=-\beta(\beta-1)$).

On the other hand, as we explain in \cite{MMMP2}, the $\beta$-deformation of $W_{1+\infty}$ to the affine Yangian preserves commutativity only of the integer rays. Therefore, only integer rays provide commutative families of non-trivial many-body Hamiltonians, while the rational ones admit only Hamiltonians associated with non-interacting particles. At the same time, the integer rays give us a clear interpretation of the rational Calogero model with a non-trivial coupling constant and its generalizations in terms of the affine Yangian.

{\section{Operators in $p$-variables \label{sec:timerep}}

In order to realize the central extension at the operator level, one has to come from the one-body operators to second quantization \cite{FKN2}. Technically, this is done by introducing to the game differential operators in variables $p_k$ \cite{FKN2,Awata}. This can be done in two different ways.

\paragraph{From matrix operators to time variables.} As we already explained in section 5, in variance with one-body operators, the matrix operators do not give rise to a representation of the $W_{1+\infty}$ algebra. However, when acting on the subspace of invariant polynomials of matrices, i.e. on the space of matrix traces, $p_k=\Tr\Lambda^k$, the matrix operators {\bf do realize} a $W_{1+\infty}$ algebra representation (see Appendix B for technical details). In fact, in practice this gives rise to the representation that has no central charge, which is clear already from the fact that the operators $\hat E_0^{(q)}$ and $\hat F_0^{(q)}$ commute with each other. Because of this, coming to variables $p_k=\Tr\Lambda^k$ (and bringing $N$ to $\infty$, see \cite{V} for an accurate description of this procedure) can provide us only with integer ray operators in terms of $p_k$, since they do not feel the central charge (see \cite{MMCal}).
In fact, there is another option: one can properly construct the Borel subalgebra of $W_{1+\infty}$ algebra associated with the $E$-branch (the right part of the Figure), it does not feel the central charge either. In order to construct the whole algebra representation, one should properly construct also the $F$-branch. This is done by the change $F_0^{(q)}={1\over q}\Tr\Lambda^{-q}\to {\p\over\p p_q}$. The justification of this procedure comes from the second approach.

\paragraph{Second quantization.}
The second possibility of constructing a $W_{1+\infty}$ algebra representation with a non-trivial central charge is the second quantization of the one-body operators: one realizes this representation in terms of the $\widehat{U(1)}$-currents, or uses just their mode expansion. In particular, one has to substitute the one-body operator $F_0^{(q)}={z^{-k}\over q}$ with the second quantized operator $F_0^{(q)}={k\over q}{\p\over\p p_k}$.

In general, one proceeds as follows \cite{FKN2,Awata}: given a differential operator $z^n G(\hat D)$, one constructs the second quantized operator that acts in the fermion Fock space.
Basic building blocks are free fermion currents
\be
\psi(z)=\sum_{n\in\mathbb{Z}}\psi_nz^{-n},\ \ \ \ \ \psi^\ast(z)=\sum_{n\in\mathbb{Z}}\psi^\ast_nz^{-n-1}
\ee
where
\be
\{\psi^\ast_m,\psi_n\}=\delta_{n+m,0}
\ee
and all other anticommutators are vanishing. Now, introducing the normal ordering $\ddag\psi^\ast _n\psi_m\ddag$ understood as $\psi^\ast_n\psi_m$ if $n\le -1$ and $-\psi_m\psi^\ast_n$ otherwise, one constructs the second quantized operator corresponding to the one-body operator in accordance with the rule
\be
z^n G(\hat D) \longrightarrow W\left(z^n G(\hat D)\right)
= \oint \frac{dz}{2 \pi i} \ddag\psi^*(z) z^n G(\hat D) \psi(z) \ddag
\label{Ofromz}
\ee
One can further use the boson-fermion correspondence
\be\label{bos}
\psi^\ast(z)=:e^{\phi(z)}:,\ \ \ \ \ \psi(z)=:e^{-\phi(z)}:
\ee
where the scalar field is defined as
\be\label{freeboson}
\phi(z)=\sum_{k\ge 1} \left({\hat a_k^\dag\over k} z^{-k}-z^{k}\hat a_k\right)+\hat a_0+\log(z) \hat a_0^\dag
\ee
the normal ordering $:\ldots:$ means that all annihilation operators are put to the right, and
\be
[\hat a_n,\hat a_m^\dag]=\delta_{n+m,0}
\ee
Now one can use
\be
\psi^\ast(z)\psi(w)=\ddag\psi^\ast(z)\psi(w)\ddag+{1\over z-w}=:e^{\phi(z)}::e^{-\phi(w)}:={1\over z-w}:e^{\phi(z)-\phi(w)}:
\ee
etc.

Thus, one gets
\be\label{WH}
\boxed{
W\left(z^nG(\hat D)\right)
= \oint \frac{dz}{2 \pi i} \ddag\psi^*(z) z^n G(\hat D)\psi(z) \ddag=\oint \frac{dz}{2 \pi i} z^n\lim_{w\to z}G(\hat D_w)\left( {1\over z-w}:e^{\phi(z)-\phi(w)}:
-{1\over z-w}\right)}
\ee
In particular,
\be\label{WE}
W\left(z^n\right)
= \oint \frac{dz}{2 \pi i} \ddag\psi^*(z) z^n \psi(z) \ddag=\oint \frac{dz}{2 \pi i} z^n\lim_{w\to z}\left( {1\over z-w}:e^{\phi(z)-\phi(w)}:
-{1\over z-w}\right)=\oint \frac{dz}{2 \pi i} z^nJ(z)
\ee
where $J(z)$ is the $\widehat{U(1)}$-current
\be
J(z)=\p_z\phi(z)=-\sum_{k\ge 1} \left(\hat a_k^\dag z^{-k-1}+kz^{k-1}\hat a_k\right)+ {\hat a_0^\dag\over z}
\ee
From (\ref{WE}) one obtains that, at $n>0$,
\be\label{Wzn}
W(z^n)=-\hat a_n^\dag,\ \ \ \ W(z^{-n})=-n\hat a_n,\ \ \ \ \ W(1)=\hat a_0^\dag
\ee
and we realize the creation and annihilation operators in terms of differential operators in variables $p_k$:
\be\label{ap}
\hat a_n^\dag=-p_n,\ \ \hat a_n=-{\p\over\p p_n}
\ee
and consider the representation where $ \hat a_0^\dag$ acts as multiplication by an arbitrary parameter $u$, which we denote by introducing
\begin{equation}
	p_0=u\,.
\end{equation}
To shorten the formulas, it is convenient to use the notation:
\begin{equation}
	p_{-k}=k \frac{\partial}{\partial p_{k}} \, .
\end{equation}.

\subsection{The basic example}
Using either of the two methods described in the previous subsection, one obtains
\begin{equation}
	\begin{split}
		\hat W_0 =  W&\left( \dfrac{\hat D(\hat D-1)}{2} + u  \hat D \right) =
		\\
		&=\frac{1}{2}\sum_{a,b=1} \left(abp_{a+b}\frac{\p^2}{\p p_a\p p_b} + (a+b)p_ap_b\frac{\p}{\p p_{a+b}}\right)
		+ u\sum_{a=1} ap_a\frac{\p}{\p p_a}+{u^3\over 6}={1\over 6}\sum_{a,b,c\in\mathbb{Z}}^{a+b+c=0}:p_ap_bp_c:
	\end{split}
\end{equation}
where the normal ordering $:\ldots :$ puts all derivatives on the right. An arbitrary parameter $u$ is called spectral flow in \cite{Awata}.
\\

As follows from (\ref{Wzn})-(\ref{ap}),
\be
\hat E_0 = p_1,\ \ \ \ \ \hat F_0 = \frac{\p}{\p p_1}
\ee
and one obtains further
\be
\hat E_{m+1} = [\hat W_0, \hat E_{m}], \ \ \ \  \hat F_{m+1} = [\hat F_m,\hat W_0]
\ee
In particular,
\be\label{EF1p}
\hat E_1 = \sum_a ap_{a+1}\frac{\p}{\p p_a}+up_1={1\over 2}\sum_{a,b\in\mathbb{Z}}^{a+b=1}:p_ap_b:\ \ \ \ \ \ \ \
\hat F_1 = \sum_{a=1} (a+1)p_a\frac{\p}{\p p_{a+1}}+u{\p\over\p p_1}={1\over 2}\sum_{a,b\in\mathbb{Z}}^{a+b=-q}:p_ap_b:
\ee
\be\label{EF2p}
\hat E_2= {1\over 3}\sum_{a,b\in\mathbb{Z}}^{a+b+c=1}:p_ap_bp_c:\ \ \ \ \ \ \ \ \ \ \
\hat F_2= {1\over 3}\sum_{a,b\in\mathbb{Z}}^{a+b+c=-1}:p_ap_bp_c:
\ee

Then the Hamiltonians
\begin{equation}
	\begin{split}
		\hat H^{(m)}_k &= {\rm ad}_{\hat F_{m+1}}^{k-1} \hat F_m =
		\left[\hat F_{m+1}, \ldots \left[\hat F_{m+1},[\hat F_{m+1},\hat F_m]\right]\ldots\right]
		\\
		\hat H^{(-m)}_k &= {\rm ad}_{\hat E_{m+1}}^{k-1} \hat E_m
	\end{split}
\end{equation}
commute for a given $m\in \mathbb{Z}$ and arbitrary $k,k'\in \mathbb{Z}_{>0}$:
\be
\left[\hat H^{(m)}_{k},\hat H^{(m)}_{k'}\right] = 0
\ee

\subsection{Rational rays}

In order to deal with the rational rays, one starts as usual with (see (\ref{Wzn})-(\ref{ap}))
\be
\hat E_0^{(q)} := \frac{1}{q!}{\rm ad}^{q-1}_{\hat E_1}\!(\hat E_0)  = {p_q\over q},\ \ \ \ \
\hat F_0^{(q)} := \frac{1}{q!}{\rm ad}^{q-1}_{\hat F_1}\!(\hat F_0) = {p_{-q}\over q}
\ee
and then use our standard recursive definitions
\be
\hat E_{k+1}^{(q)} := [\hat W_0, \hat E_{k}^{(q)}], \ \ \ \  \hat F_{k+1}^{(q)} := [\hat F_k^{(q)},\hat W_0]
\ee
in order to generate the commutative set of Hamiltonians $\hat H^{(G,q)}_{\pm k}$, which are a counterpart of (\ref{shifted}).

In particular,
\begin{equation}
	\begin{split}
		&\hat E_1^{(q)}= \sum_{a\ge 1}ap_{a+q}{\p\over\p p_a}+{1\over 2}\sum_{a,b\ge 1}^{a+b=q}p_ap_b+up_q
		={1\over 2}\sum_{a,b\in\mathbb{Z}}^{a+b=q}:p_ap_b:
		\\&
		\hat F_1^{(q)}=\sum_{a\ge 1}(a+q)p_a{\p\over\p p_{a+q}}+{1\over 2}\sum_{a,b\ge 1}^{a+b=q}ab{\p^2\over\p p_a\p p_b} +u{\p\over\p p_q}={1\over 2}\sum_{a,b\in\mathbb{Z}}^{a+b=-q}:p_ap_b:
	\end{split}
\end{equation}
\begin{equation}
	\begin{split}
		&\hat E_2^{(q)}= {q\over 3}\sum_{a,b\in\mathbb{Z}}^{a+b+c=q}:p_ap_bp_c:+{q(q^2-1)\over 12}p_q\\
&\hat F_2^{(q)}= {q\over 3}\sum_{a,b\in\mathbb{Z}}^{a+b+c=-q}:p_ap_bp_c:+{q(q^2-1)\over 12}p_{-q}
	\end{split}
\end{equation}
The generators $\hat E_1^{(q)}$ are associated with the generators $L_{-q}$ of the Virasoro algebra, $\hat F_1^{(q)}$, with the generators $L_{q}$, and $\hat E_k^{(q)}/q^{k-1}$, $\hat F_k^{(q)}/q^{k-1}$, with generators $W_{\pm q}^{(k+1)}$ of the $W^{(k+1)}$-algebra. Commuting these operators, one can observe that the central charge is equal to 1:
\be
\phantom{.}[L_n,L_{-n}]=2nL_0+{n(n^2-1)\over 12},\nn\\
L_n={1\over 2}\sum_{a,b\in\mathbb{Z}}^{a+b=n}:p_ap_b:
\ee

\subsection{Explicit formulas for the generating operators}
Contrary to the one-body representation, the iterative commutation of second-quantized operators in $p_k$ variables gets increasingly more complicated, and it is not clear whether general formulas can be derived or proven this way. On the other hand for the operators in question is possible to explicitly perform second quantization in terms of generating function, i.e. the fermionization-bosonization construction \eqref{Ofromz} allows one
to obtain general formulas for the centrally-extended operators.
\\

Namely, $\hat{F}_m^{(q)} = H^{(m,q)}_{-1}$ in this representation is equal to
(the formula for $\hat{E}_m^{(q)} = H^{(m,q)}_{1}$ is analogous,
see Appendix A.2 for the derivation)
\begin{align}
\!\!\!\!  \hat{F}_m^{(q)} = & \ m! \,q^m \sum_{l=0}^m \sum_{i=0}^l
    \frac{B_{l-i}\left(\frac{1-q}{2}\right)}{(l-i)! (m+1-l)!}
    \oint \frac{d z}{2 \pi i z^{1+q}}
    \sum_{k_1+\dots+k_{m+1-l}=i}
    :\!\! \prod_{i=1}^{m+1-l}
    \frac{(-1)^{k_i}}{(k_i+1)!}
    \sum_{a_i \in \mathbb{Z}} a_i^{k_i} z^{-a_i} p_{a_i}
    : \ =
 \end{align}
\vspace{-0.3cm}
\be
\boxed{
\ \ =  q^m \sum_{l=0}^m
    \frac{m!}{(m+1-l)!}
    \sum_{i=0}^l
    \frac{(-1)^iB_{l-i}\left(\frac{1-q}{2}\right)}{(l-i)!}
    \!\!\!\! \sum_{a_1+\dots+a_{m+1-l} = -q}  \!\!\!\!\!\!\!\!
    \ \ \ \
    \left (
    \sum_{k_1+\dots+k_{m+1-l}=i}
     \prod_{j=1}^{m+1-l}
    \frac{1}{(k_j+1)!}
    a_j^{k_j} \right )
     :\!\! \prod_{j=1}^{m+1-l} p_{a_j} :}
\nn
\ee
where $B_m(x)$ are the Bernoulli polynomials.

The expression can be used to generate any number of terms in the operator quite fast. However, it does not reveal the deeper combinatoric structure that can be seen experimentally. Mainly the restricted sum over $k_i$ actually produces comprehensible dependence on $a_j$. For example, it is not obvious that the sum vanishes for odd $l$(???). More interestingly, what is not immediately obvious from these general formulas is that, in fact, the dependence on
$q$ is \textbf{very simple and universal}.
Namely, one can experimentally (or, by exercising simple combinatorics) obtain that
\begin{align} \label{eq:f-gen}
  \frac{\hat{F}_m^{(\textbf{q})}}{\textbf{q}^{m-1}}=
  & \ \sum_{a_1+\dots+a_{m+1}=-\textbf{q}}\frac{:\prod_{j=1}^{m+1} p_{a_j}:}{m+1}
  + \sum_{a_1+\dots+a_{m-1}=-\textbf{q}}\frac{m!}{(m-2)!}\left[\frac{1}{24} r_{[1],m-1} - \frac{1}{24}\right]\frac{:\prod_{j=1}^{m-1} p_{a_j}:}{m-1}
  \\ \notag
  + & \
  \sum_{a_1+\dots+a_{m-3}=-\textbf{q}}\frac{m!}{(m-4)!}
  \left[\frac{(-1)}{2880} r_{[2],m-3} + \frac{1}{1152} r_{[1,1],m-3}
  + \frac{(-1)}{576} r_{[1],m-3} + \frac{7}{5760}
  \right]
  \frac{:\prod_{j=1}^{m-3} p_{a_j}:}{m-3}
  \\ \notag
  + & \
  \sum_{a_1+\dots+a_{m-5}=-\textbf{q}}\frac{m!}{(m-6)!}
  \Bigg[
  \frac{1}{181440} r_{[3],m-5} + \frac{(-1)}{69120} r_{[2,1],m-5}
  + \frac{1}{82944} r_{[1,1,1],m-5}
  + \frac{1}{69120} r_{[2],m-5}
  \\ \notag
  & \quad \quad \quad \quad \quad \quad \quad \quad
  + \frac{(-1)}{27648} r_{[1,1],m-5} + \frac{7}{138240} r_{[1],m-5} + \frac{(-31)}{967680}
  \Bigg]
  \frac{:\prod_{j=1}^{m-5} p_{a_j}:}{m-5}
  \\ \notag
  + & \ \cdots
\label{Hpq1}
\end{align}
where we have put in boldface all the (few) places where $q$ enters.
The polynomials $r_{[d_1,\dots,d_m],l}$ are just the power sums
in variables $a_i$ and are equal to
\begin{align}
  r_{[d_1,\dots,d_m],l} = \prod_{j=1}^m r_{d_j, l} \ \ \ \ \ \
  r_{d, l} = \left ( \sum_{i=1}^l a_i^{2 d} \right )
\end{align}

\noindent Furthermore, polynomials in brackets are obtained from the simple generating function
\begin{align}
\label{BerG}
\mathcal{G}(u) = & \ \frac{u}{e^{u/2} - e^{-u/2}}
  \cdot \exp \left ( \sum_{j=1}^\infty r_j \frac{B_{2 j} u^{2 j}}{(2 j!) (2 j)} \right )
  \\ \notag
  = & \ 1 + u^2 \frac{1}{24}(r_1 - 1)
  + u^4 \left (\frac{7}{5760} - \frac{1}{576} r_1
  + \frac{1}{1152} r_1^2
  - \frac{1}{2880} r_2 \right ) + o(u^6)
\end{align}

\subsection{Explicit formulas for Hamiltonians}
Similar formulas can be also derived for the families of Hamiltonians.

\subsubsection{Calogero ray}
\label{sec:one-body-boso}
For the first (Calogero, $m=1$) series the fermionization-bosonization
construction can be applied directly, and the corresponding general formula for the Hamiltonians reads (see Appendix A.1 for the derivation):
\begin{equation}\label{genformbinom}
\boxed{
	\hat{H}^{(1)}_{-k} = - k! \cdot \sum_{n=1}^k \sum_{ \sum_{i=1}^n a_i = -k}
	:\prod_{j=1}^n \dfrac{p_{a_j}}{a_j} : \left( \sum_{\substack{ k_1, \ldots, k_n  > 0 \\ \sum_{i=1}^n k_i = k+1 } } \prod_{i=1}^{n} \binom{-a_i}{k_i} \right)}
\end{equation}
where the sum is over all tuples $a=\{a_1,a_2\ldots a_{n} \}$ and $\{k_1,\ldots k_n \}$ with the respective conditions.
\\

On the other hand, one can use the fact that the second quantization prescription
\eqref{Ofromz} is \textit{linear} and assemble expression for $H$ from expression
for $F$ (Eq.\eqref{eq:f-gen}). Recall that the one-body formulas for $H^{(1,1)}_{-k}$ and $F^{(q)}_m$ are
  (see Eqs.~\eqref{eq:h-pq-one-body}~and~\eqref{Fnr}, respectively)
  \begin{align}
    H^{(1,1)}_{-k} = & \ z^{-k} \prod_{i=0}^{k-1}
    \left (
    \hat{D} - i
    \right )
    \\ \notag
    \tilde{F}^{(q)}_m = & \ F^{(q)}_m q^{-m} = z^{-q}
    \left ( \hat D + \frac{(1-q)}{2} \right )^m
  \end{align}
A re-expansion of the one-body Hamiltonians $H^{(1,1)}_{-k}$ in the basis
  of the one-body operators $F^{(q)}_m$ is
  \begin{align}
    H^{(1,1)}_{-k} = & \ \sum_{j=0}^k C_j F^{(k)}_j
    = 1 \cdot F^{(k)}_k + \frac{(k+1)!}{(k-2)!}\frac{(-1)}{4!} F^{(k)}_{k-2}
    + \frac{(k+1)!}{(k-4)!}\frac{(5 k + 7)}{6! 8} F^{(k)}_{k-4} + ...
    \\ \notag
    = & \ \sum_{l=0}^\infty \frac{(k+1)!}{(k - 2 l)!}
    \left ( \oint\frac{du}{u^{2l+1}} \mathcal{G}(u,k) \right ) F^{(k)}_{k - 2 l},
  \end{align}
  where $\mathcal{G}(u,k)$ is
  \begin{align}
    \mathcal{G}(u,k) = & \ \frac{1}{k+1} + y + \frac{k}{2} y^2 + \frac{k(k-1)}{3!} y^3 + ...
    = \frac{1}{(k+1)} \left ( 1 + y \right )^{k+1},
    \\ \notag
    \text{with} & \ y = \frac{u}{e^{u/2} - e^{-u/2}} - 1
  \end{align}
Thus, one finally obtains
  \begin{align}
    \boxed{
      H^{(1,1)}_{-k} = \sum_{l=0}^\infty \frac{k!}{(k - 2 l)!}
      \left ( \oint\frac{du}{u^{2l+1}} \left ( \frac{u}{e^{u/2} - e^{-u/2}} \right )^{k+1}
      \right ) F^{(k)}_{k - 2 l},
    }
  \end{align}

  Again, the generating function for the coefficients turns out to be closely
  related to the Bernoulli numbers/polynomials, since
  \begin{align}
    \frac{u}{e^{u/2} - e^{-u/2}} = \sum_{n=0}^\infty B_n\left(\frac{1}{2}\right)
    \frac{u^n}{n!}
  \end{align}

  \subsubsection{Generic $(p,q)$ ray}

  Generic Hamiltonians $H^{(p,q)}_{-k}$ can be expanded into $F^{(q)}_m$ basis
as well. Indeed
\begin{align} \label{eq:hpq-in-f}
  H^{(p,q)}_{-k} = & \ z^{-q k} \prod_{i=0}^{k-1} \left (
  \hat{D} + \frac{(1 - q)}{2} - q i
  \right )^p  \\ \notag
  = & \ \sum_{m=0}^{p k} C_{p,m} (k) \cdot \frac{F^{(q k)}_{p k - m}}{(q k)^{p k - m}},
\end{align}
where the coefficients $C_{p,m} (k)$ are clearly equal to the following simple residue
\begin{align}
  C_{p,m} (k) = & \ \oint_{z=0} \frac{d z}{2 \pi i z} \frac{1}{z^{p z - m}}
  \prod_{a=0}^{k-1}\left(z + \frac{q(k-1) - 2 a}{2} \right)^p
  = \oint_{\substack{w=0\\w = 1/z}} \frac{d w}{2 \pi i w}
  w^{-m} \prod_{a=0}^{k-1}\left(1 + \frac{q(k-1) - 2 a}{2} w \right)^p
  \\ \notag
  = & \ \oint_{w=0} \frac{d w}{2 \pi i w} w^{-m} \exp \left (
  - \sum_{l=1}^\infty \left (\frac{-1}{2}\right)^l \frac{1}{l} w^l q^l p
  \sum_{a=0}^{k-1} \left (
  k-1-2a
  \right )^l
  \right )
  = 1\frac{(-q)^m}{2^m}
  s_{[m]} \left \{
  p_l = -p \sum_{a=0}^{k-1} (k-1 - 2 a)^l
  \right \}
\end{align}
In fact, the ``peculiar'' harmonic sums in the arguments of symmetric Schur functions
can be evaluated
\begin{align}
  \sum_{a=0}^{k-1} (k-1 - 2 a)^l = & \ \left [\sum_{a=0}^{\infty} - \sum_{a=k}^\infty \right ](k-1 - 2 a)^l = (-2)^l\left (\zeta\left(-l, \frac{1-k}{2}\right ) - \zeta\left(-l, \frac{1+k}{2}\right ) \right ) \\ \notag
  = & \ \frac{1}{l+1} \left( B_{l+1}\Big(\frac{1 + k}{2}\Big) - B_{l+1}\Big(\frac{1 - k}{2}\Big) \right),
\end{align}
where we have used in the last equality the well-known relation between the Hurwitz
zeta-function $\zeta(-l,a)$ at negative integral values of the first argument
and the Bernoulli polynomial $B_{l+1}(a)$.

Thus, in the end, we have
\begin{align} \label{eq:c-coeffs}
  \boxed{
    C_{p,m} (k) = q^m
    s_{[m]} \left\{
    p_l = -p \frac{1}{(l+1)}
   \left( B_{l+1}\Big(\frac{1 + k}{2}\Big) - B_{l+1}\Big(\frac{1 - k}{2}\Big) \right)\right\}
  }
\end{align}
and it is easy to check that only even coefficients, that is $C_{p, 2 l} (k)$,
are non-vanishing.

Combining \eqref{eq:c-coeffs} with \eqref{eq:f-gen} one finally obtains
\begin{align}
\label{eq:h-formula-explicit1}
  H^{(p,q)}_{-k} = & \ \sum_{l=0}^{\frac{pq+1}{2} }
\left(\sum_{a_1+\cdots+a_{p k + 1 - 2 l} = -q k}  G_{p,q,k,l}\{a\}\ :p_{a_1} \cdots p_{a_{p k + 1 - 2 l}}:\right)
\end{align}
or, in more detail
 \be
 \boxed{
  H^{(p,q)}_{-k} =  \ \sum_{l=0}^\infty
  \left(\sum_{a_1+\cdots+a_{p k + 1 - 2 l} = -q k}\!\!\!\!\!\!\!
  \frac{:p_{a_1} \cdots p_{a_{p k + 1 - 2 l}}:}{(p k + 1 - 2 l)} \times
  \left [\sum_{m=0}^l
    \frac{(p k - 2 l + 2 m)!}{(p k - 2 l)!} C_{p, 2(l-m)}(k) \oint \frac{du}{2 \pi i u^{2 m + 1}} \mathcal{G}(u)
    \right ]\right)}
\nn
 \ee
 \begin{align} \label{eq:h-formula-explicit}
  = &
  \sum_{a_1+\cdots+a_{p k + 1} = -q k}
  \frac{:p_{a_1} \cdots p_{a_{p k + 1}}:}{(p k + 1)} \ +
  \\ \notag
  & \ \ \ \ + \sum_{a_1+\cdots+a_{p k - 1} = -q k} \!\!\!\!\!\!\!
  \frac{:p_{a_1} \cdots p_{a_{p k - 1}}:}{(p k - 1)}
  \left [
  \frac{1}{24} r_{[1],p k - 1} \cdot k p (k p - 1)
  -\frac{1}{24} \cdot k p \cdot \left(k^2 q^2 + k p - q^2 - 1\right)
  \right ] \ +
  \\ \notag  \\ \notag
   & \ \ \ \ + \ \  \text{smaller sets }\{a\}
\end{align}
which correctly reproduces, in particular, the Calogero (1,1) ray
\be
  H^{(1,1)}_{-k} =
 \!\!\!\!\!\!\!\! \sum_{a_1+\cdots+a_{k + 1} = - k}\!\!\!\!\!\!\!\!\!
  \frac{:p_{a_1} \cdots p_{a_{k + 1}}:}{(k + 1)}
  + \!\!\!\!\!\!\!\!\!\sum_{a_1+\cdots+a_{k - 1} = -k}\!\!\!\!\!\!\!\!\!
  \frac{:p_{a_1} \cdots p_{a_{k - 1}}:}{(k - 1)}
  \left [
  \frac{1}{24} r_{[1],k - 1} \cdot k (k - 1)
  -\frac{1}{24} \cdot k (k-1)(k+2)
  \right ]
  \ee
The Hamiltonians of the next (2,1) ray are
  \be
  H^{(2,1)}_{-k} =
  \sum_{a_1+\cdots+a_{2 k + 1} = - k}
  \frac{:p_{a_1} \cdots p_{a_{2 k + 1}}:}{(2 k + 1)}
  + \sum_{a_1+\cdots+a_{2 k - 1} = -k}
  \frac{:p_{a_1} \cdots p_{a_{2 k - 1}}:}{(2 k - 1)}
  \left [
  \frac{1}{12} r_{[1],2 k - 1} \cdot k (2 k - 1)
  -\frac{1}{12} \cdot k (k^2 + 2 k - 2)
  \right ]
\ee

An important feature of formula \eqref{eq:h-formula-explicit} is
that the dependence on $k$ is contained in \eqref{eq:c-coeffs}
while the peculiar numeric coefficients are made from \eqref{BerG},
and, in this sense, an apparent complexity is split into two manageable
generating functions.

\subsection{Cones}

One can again consider a more general case
\be
\hat H^{(G)}_{- k}={1\over k-1}\Big[\hat{F}^{(G)},\hat H^{(G)}_{- k+1}\Big]
\ee
with the operator $\hat{F}^{(G)}$ given by
\begin{equation}
	\hat{F}^{(G)}:=\sum_{i=1}^p\alpha_i\hat F_{i+1}
\end{equation}
and
\be
\hat H^{(G)}_{-1}=\sum_{i=1}^p\alpha_i\hat F_{i}
\ee
with arbitrary coefficients $\alpha_i$.

Including the rational rays provides us with the most general commutative subsets of the $W_{1+\infty}$ algebra that preserve the grading:
\be
\hat H^{(G,q)}_{\pm k}={1\over k-1}\Big[\hat{F}^{(G,\pm q)},\hat H^{(G,q)}_{\pm (k-1)}\Big]
\ee
with the operator $\hat{F}^{(G,\pm q)}$ given by
\begin{equation}
	\hat{F}^{(G,\pm q)}:=\sum_{i=1}^p\alpha_i{\rm ad}_{\hat W_0^{i+1}}p_{\pm q}
\end{equation}
and
\be
\hat H^{(G,q)}_{\pm 1}=\sum_{i=1}^p\alpha_i{\rm ad}_{\hat W_0^{i}}p_{\pm q}
\ee
}

Note that introducing the parameter $u$ in the formulas of the present section can be considered as a particular simple example of the cone case.

\section{$\beta$-deformation
\label{betadef}}

\subsection{Integer rays}

As we already observed in sec.7, the $\beta$-deformation does not provide commutativity of the rational rays, however, the integer ray families are still commutative in this case. As in sec.7, in order to make this deformation, one considers a slight modification of $\hat W_0$, while the generating operators $\hat F_0$, $\hat E_0$ remain intact. By abuse of notations be will denote the $\beta$-deformed operators of this section by the same symbols:
\begin{align}\label{W0b}
\hat W_0 :=  & \ \frac{1}{2}\sum_{a,b=1} \left(abp_{a+b}\frac{\p^2}{\p p_a\p p_b} + \beta (a+b)p_ap_b\frac{\p}{\p p_{a+b}}\right)
+ \beta u_\beta\sum_{a=1} ap_a\frac{\p}{\p p_a}+{\beta u_\beta^3\over 6}
+{1-\beta\over 2}\sum_aa(a-1)p_a{\p\over\p p_a}= \\ \notag
= & \ {\beta^2\over 6}\sum_{a,b,c\in\mathbb{Z}}^{a+b+c=0}:p_ap_bp_c:+ {\beta(1 - \beta)\over 4}\sum_{a,b\in\mathbb{Z}}^{a+b=0} \left(|a| - 1\right) :p_ap_{b}:
\end{align}
where\footnote{We choose the value of $p_0$ in the most convenient way. One can definitely arbitrarily shift it by an arbitrary constant $\alpha$ with the operator $\exp\left(\alpha{\p\over\p u}\right)$.}  $p_0=u_\beta=u+(\beta-1)/(2\beta)$ and $p_{-k}=\beta^{-1}k{\p\over\p p_k}$. Then, the first few generating operators are
\begin{align}
\hat F_1&= \beta \sum_{b=0} p_b p_{-b-1} \\ \notag
\hat{F}_2 & = [\hat F_1,\hat W_0] =
\frac{\beta^2}{3} \sum_{a+b+c = -1} : p_a p_b p_c : + \beta(1-\beta) \sum_b b \cdot p_b p_{-b-1}
\end{align}

\begin{align}
\hat F_3 = [\hat F_2,\hat W_0] = & \
\frac{\beta^4}{4} \sum_{a+b+c+d=-1}^\infty : p_a p_b p_c p_d :
\\ \notag
+ & \frac{3}{2} \beta^2(1-\beta)
\left [
  \sum_{a,b=0}^\infty (a+b) p_a p_b p_{-a-b-1}
  + \sum_{a,b=0}^\infty \left(a + b - \frac{4}{3} \right)
p_{a+b-1} p_{-a} p_{-b}
  \right ]
\\ \notag
+ & \sum_{a=0}^\infty \left(\frac{\beta^2}{2} a + \beta(1 - \frac{3 \beta}{2}
+ \beta^2)a^2  \right)
p_a p_{-a-1}
\\ \notag
\end{align}
and
\begin{align}
   \hat E_1 = & \ [\hat W_0, p_1] = \frac{\beta}{2} \sum_{a+b=1} :p_{a} p_{b}:
    \\ \notag
   \hat E_2 = & \ [\hat W_0,\hat E_1] = \frac{\beta^2}{3} \sum_{a+b+c=1}^\infty :p_a p_b p_c:
   + \beta (1 - \beta) \sum_{k=0}^\infty k p_{k+1} p_{-k}
        \\ \notag
\hat  E_3 = [\hat W_0,\hat E_2] = & \
\frac{\beta^3}{4} \sum_{a+b+c+d=1}^\infty :p_a p_b p_c p_d:
+ \frac{3 \beta^2(1-\beta)}{2} \left [
  \sum_{a,b=0}^\infty \left( a + b - \frac{4}{3}\right) p_a p_b p_{-a-b+1}
  \sum_{a,b=0}^\infty \left( a + b \right) p_{a+b+1} p_{-a} p_{-b}
  \right ] \\ \notag
+ & \sum_{a=0}^\infty
\left ((a - 1)^2 \beta (1 + \beta^2) - \beta^2 \frac{(a-1) (3 (a - 1) - 1)}{2} \right)
p_a p_{-a + 1}
\end{align}

We see that, as compared to the $\beta = 1$ case (see Eqs.\eqref{EF1p} and \eqref{EF2p})
the structures get more complicated. The polynomials in index variables can now
have both odd and even powers, and the symmetry between different choices of
positive/negative indices that allowed one to write everything in terms of normal
ordered products is lost (at the level of $\hat{F}_3$ and $\hat{E}_3$,
this is seen most clearly).

\bigskip

The Hamiltonians in this case are commutative, the first of them being
\be
\hat H_{-1}^{(1)}&=&\hat F_1= \frac{\beta}{2} \sum_{a+b=-1} :p_a p_b: \nn\\
\hat{H}_{-2}^{(1)}&=&[\hat H_1^{(1)},\hat F_2]= \frac{\beta^2}{3}
\sum_{a+b+c=-2} : p_a p_b p_c : +\beta(1-\beta) \sum_{b=0} (b+1)p_b p_{-b-2} \nn\\
\hat H_{-3}^{(1)}&=&{1\over 2}[\hat H_2^{(1)},\hat F_2]=
\frac{\beta^3}{4}
\sum_{a+b+c+d=-3} : p_a p_b p_c p_d :+\nn\\
&+&{3\beta^2(1-\beta)\over 2}\sum_{a,b=0}\left[(a+b-2)p_{a+b-3} p_{-a} p_{-b} + (a+b+2)p_a p_b
p_{-a-b-3} \right]+\nn\\
&+&
{2\beta^3-3\beta^2+2 \beta \over 2}\sum_{a=0}(a-1)(a-2)p_{a-3} p_{-a} \nn\\
   \hat H_1^{(1)} &= &\ =\hat E_1= \beta \sum_{a=0} p_{a+1} p_{-a} \nn\\
\hat  H_{2}^{(1)} &= & \ [E_2, E_1] =
\frac{\beta^2}{3} \sum_{a+b+c=2}^\infty :p_a p_b p_c:
+ \beta(1-\beta) \sum_{k=0}^\infty (k+1) p_{k+2} p_{-k}
\ee
for the Calogero series, and
\begin{align}
\hat{H}_{-1}^{(2)} = &\hat F_2 =
\frac{\beta^2}{3} \sum_{a+b+c = -1} : p_a p_b p_c : + \beta(1-\beta) \sum_b b \cdot p_b p_{-b-1}
\end{align}

\begin{align}
  \hat H_{-2}^{(2)} = [\hat F_2,\hat F_3] = & \
  \frac{\beta^4}{5} \sum_{a+b+c+d+e=-2} :p_a p_b p_c p_d p_e:  +\ 2 (1 - \beta)\beta^2 \sum_{a,b,c} (1 + a + b + c) p_a p_b p_c \partial_{a+b+c+2}
\\ \notag
    & \ \ \ + \sum_{a_1, a_2}
    { \scriptstyle \beta \left( 1 - \beta + \beta^2 + a_1^2 (2 - 3 \beta + 2 \beta^2) +
    a_2^2 (2 - 3 \beta + 2 \beta^2) + a_2 (3 - 4 \beta + 3 \beta^2) +
    a_1 (3 - 4 \beta + 3 \beta^2 + a_2 (3 - 5 \beta + 3 \beta^2)) \right )
    }
    p_{a_1} p_{a_2} \partial_{a_1 + a_2 + 2}
\\ \notag
    & \ \ \ + \sum_{a_1} \frac{1}{6} a_1 (1 + a_1) (1 - \beta) (6 - \beta + 6 \beta^2
    + a_1 (6 - 5 \beta + 6 \beta^2)) p_{a_1} \partial_{a_1 + 2}
\\ \notag
    & \ \ \ + \sum_{a_1 + a_2 + 2 = b_1 + b_2}
    {\scriptstyle \left (
    \frac{1}{2}(5 b_1 + 5 b_2 - 8) (1 - \beta) \beta
    + {\color{red} \theta \left ( b_1 - a_1 - 1\right ) } 2 (b_1 - a_1 - 1) (1 - \beta) \beta
    \right ) } p_{a_1} p_{a_2} \partial_{b_1} \partial_{b_2}
\\ \notag
    & \ \ \ + \sum_{b_1,b_2}
       {\scriptstyle (5 - 9 \beta + 5 \beta^2 + b_1^2 (2 - 3 \beta + 2 \beta^2) +
    b_2^2 (2 - 3 \beta + 2 \beta^2)
    + b_1 (-2 + b_2) (3 - 5 \beta + 3 \beta^2) -
    2 b_2 (3 - 5 \beta + 3 \beta^2))}
    p_{b_1 + b_2 - 2} \partial_{b_1} \partial_{b_2}
\\ \notag
+ & 2 (1 - \beta) \sum_{b_1, b_2, b_3} (b_1 + b_2 + b_3 - 2)
    p_{b_1 + b_2 + b_3 - 2} \partial_{b_1} \partial_{b_2} \partial_{b_3}
\end{align}
for the (2,1) series. We see that expressions at the moment get quickly very
involved, and some parts of the structure (the Heaviside $\theta$-functions,
highlighted in red), at the moment, appear to be purely \textit{ad hoc}.
We are planning to return to this point elsewhere.

\subsection{Rational rays}

The $\beta$-deformation (\ref{W0b}) determines all operators for the rational rays too.
In particular, we have
\begin{align}
  \hat{F}^{(q)}_0 := \beta^{-1} \frac{\partial}{\partial p_q}
  \sim ad_{\hat{F}_1}^{q-1} \hat{F}_0;
  \ \ \ \hat{E}^{(q)}_0 := \frac{p_q}{q} \sim ad_{\hat{E}_1}^{q-1} \hat{E}_0
\end{align}
and
\begin{align}
  \hat{F}^{(q)}_1 = [F_0^{(q)}, \hat{W}_0]
  = & \frac{\beta}{2} \sum_{a+b=-q} : p_a p_b : + \frac{(1-\beta)}{2} (q-1) p_{-q}
  \\ \notag
  \hat{E}^{(q)}_1 = [\hat{W}_0, E_0^{(q)}]
  = & \frac{\beta}{2} \sum_{a+b=q} : p_a p_b : + \frac{(1-\beta)}{2} (q-1) p_{q}
\end{align}

\noindent Note that these $\hat{F}^{(q)}_1$ and $\hat{E}^{(q)}_1$ do not form the Virasoro algebra.
One can arrive at the Virasoro algebra only by shifting $\hat{F}^{(q)}_1$ with a proper component of the $U(1)$-current (note that one can also add $p_q$ to $L_{-q}$ with an arbitrary coefficient):
\be
L_{-q}={\beta \over 2}\sum_{a,b\in\mathbb{Z}}^{a+b=q}:p_ap_b:+{(1-\beta)\over 2}(q-1) p_q
\ee
This Virasoro algebra has the central charge
\be
c=1-3\Big(\sqrt{\beta}-{1\over\sqrt{\beta}}\Big)^2
\ee

Similarly, utilizing the normal ordering relation $p_a p_b = \beta^{-1} \delta_{a+b,0} \theta(b) b \ \ + :p_a p_b:$, one obtains
\begin{align}
\hat F_2^{(q)}= & \frac{\beta^2}{3} q \sum_{a,b,c \in \mathbb{Z}}^{a+b+c=-q}
: p_a p_b p_c :
- \frac{\beta(1-\beta)}{2} \mathop{\sum}_{a,b\in\mathbb{Z}}^{a+b=-q} b (|b|-1) :p_a p_b:
+ \frac{\beta}{6} q(q^2-1) p_{-q}
+ \frac{(1-\beta)q(q-1)}{2} \hat{F}_1^{(q)}
\\ \notag
\hat E_2^{(q)}= &
\frac{\beta^2}{3} q \sum_{a,b,c \in \mathbb{Z}}^{a+b+c=q}
: p_a p_b p_c :
- \frac{\beta(1-\beta)}{2} \mathop{\sum}_{a,b\in\mathbb{Z}}^{a+b=q} b (|b|-1) :p_a p_b:
- \frac{\beta}{6} q(q^2-1) p_{q}
+ \frac{(1-\beta)q(q-1)}{2} \hat{E}_1^{(q)}
\end{align}
Now one can calculate, for instance, $\hat H_{-2}^{(1,q)}={1\over q}[\hat F_1^{(q)},\hat F_2^{(q)}]$ and observe that it does not commute with $\hat F_1^{(q)}$.
In fact, already action of
$[[\hat F_1^{(q)},\hat F_2^{(q)}], \hat F_1^{(q)}]$ on $p_{3 q}$ is non-vanishing
and proportional to $(1-\beta)$.
This confirms the claim of sec.7 that the $\beta$-deformation destroys the commutativity of the rational rays while preserving it for the integer rays.

\section{Equivalence of times and eigenvalues
\label{timev}}

Now we can unify two different attempts which we made:
representations in terms of matrices/eigenvalues and times.
The matrix representation is very interesting,
however,
not all subalgebras that correspond to $W_{1+\infty}$ commutative subalgebras are actually commutative.
It is not fully clear what it means, whether this {\it proves} that there is no
matrix representation of the full-fledged $W_{1+\infty}$ or we should just look better
or maybe consider a slight modification of the algebras.
If we just build a $\Lambda$-algebra as we did with $z$, i.e. consider the algebra of generators
$\Tr \Lambda^m \frac{\p}{\p \Lambda}^n$, commutators will necessarily contain multi-trace operators,
i.e. the algebra increases dramatically.
As we saw, in some cases (like integer rays) this can be avoided,
but in generic situation of rational rays we have not succeeded.

Anyhow,  representation theory and {\it all} commutative subalgebras are well preserved,
if we restrict the action of operators to functions of matrix eigenvalues.
But this is equivalent to description in terms of times, provided times are constrained
to lie on the Miwa locus $p_k=\Tr \Lambda^k=\sum\lambda_i^k$: this is a restriction because for $N\times N$
matrices this is just an $N$-dimensional space.
Restriction on the time-side is not immediate, because operators contain derivatives.
Still one can demonstrate the equivalence of our formulas
in secs.6 and 8.

\section{Origin of the hypergeometric $\tau$-functions in commutative families}

\subsection{(Skew) hypergeometric $\tau$-functions}

The hypergeometric $\tau$-functions of the KP/Toda hierarchy \cite{GKM2,OS,AMMN1,AMMN2} are given explicitly by the expression\footnote{Note that the usual choice of time variables in integrable hierarchies is $t_k=p_k/k$.}
\be\label{htau}
\tau\{p_k;\bar p_k\}=\sum_R\left(\prod_{i,j\in R}f(j-i)\right)S_R\{p_k\}S_R\{\bar p_k\}
\ee
where $f(x)$ is an arbitrary function of $x$ without poles at integer points.
The sum (\ref{htau}) is a KP $\tau$-function w.r.t. the both sets of times, $p_k$ and $\bar p_k$, and the dependence on the Toda zeroth (discrete) time requires further specification (see \cite{Taka}).

Similarly, the skew hypergeometric $\tau$-functions of the KP/Toda hierarchy \cite{Ch1,Ch2} are given explicitly by the expression
\be\label{shtau}
\tau\{g_k;p_k|\bar p_k\}=\sum_{Q,R}\left(\prod_{i,j\in R/Q}f(j-i)\right)S_{R/Q}\{\bar p_k\}S_R\{g_k\}S_Q\{p_k\}
\ee
This sum is a KP $\tau$-function w.r.t. the sets of times $g_k$ and $p_k$, while $\bar p_k$ are just parameters.

These $\tau$-functions are given by explicit expressions, but, in this section, we are going to explain how the (skew) hypergeometric $\tau$-functions can be invariantly described as generated by the commutative subalgebras of the $W_{1+\infty}$ algebra. In fact, it is known \cite{Orlov,TT} that a $\tau$-function of the KP/Toda hierarchy typically has a $W$-representation with the $W$-operator being an element of the $W_\infty$ algebra. Here we explain this claim in the case of hypergeometric $\tau$-functions, when the $W$-operator acts on a trivial state \cite{OS}, and in the case of skew hypergeometric functions: in the both cases, the $\tau$-functions are generated by commutative subfamilies of the $W_\infty$ algebra.

\subsection{Examples of partition functions: integer rays}

First of all, as an example we consider the case of integer rays. In this case, the $\tau$-functions generated by the commutative Hamiltonians are \cite{Ch1,Ch2}
\be\label{Zn}
\tau\{p_k;\bar p_k\}=\exp\left( \sum_{k} \dfrac{\bar{p}_k \hat H^{(m)}_k}{k} \right) \cdot 1=
\sum_R\left(\prod_{i,j\in R}(u+j-i)^m\right)S_R\{p_k\}S_R\{\bar p_k\}
\ee
and
\be
\tau\{g_k;p_k|\bar p_k\}=\exp\left( \sum_{k} \dfrac{\bar{p}_n \hat H^{(m)}_{-k}}{k} \right) \cdot \exp\left(\sum_k{g_kp_k\over k}
\right)=\sum_{Q,R}\left(\prod_{i,j\in R/Q}(u+j-i)^m\right)S_{R/Q}\{\bar p_k\}S_R\{g_k\}S_Q\{p_k\}
\ee
These sums coincide with (\ref{htau}), (\ref{shtau}) upon choosing the function $f(x)=u+x$.

\subsection{Action of $W_{1+\infty}$ algebra on characters}

In order to deal with these character expansion systematically, let us formulate the general rule in terms of the $W_{1+\infty}$ algebra generators. Using the approach of sec.8.1, one can obtain that action of the $W_{1+\infty}$ algebra operator $z^{-n}\prod_{i=1}^{n-1} G(D-i)$ formulated in the second quantized variables $p_k$ on the Schur function $S_R$ is given by the formula
\be\label{OnSchur}
    W[G]_{n}S_R:=W\left( z^{n}\prod_{i=1}^{n} G(D+i) \right) S_R = \sum_{Q} \prod_{(i,j) \in Q/R}G(j-i) \langle p_n S_R \big| S_Q \rangle S_{Q}\label{FG+}\\
    W[G]_{-n}S_R:=W\left(z^{-n}\prod_{i=0}^{n-1} G(D-i) \right) S_R = \sum_{Q} \prod_{(i,j) \in Q/R}G(j-i) \Big\langle
    n{\p S_R\over\p p_n} \Big| S_Q \Big\rangle S_{Q}\label{FG-}
\ee
For all $n$, each of these two families is commutative because the one-body operators commute, therefore one can take the sum of them to construct the character expansion:
\begin{equation}\label{mainG}
\boxed{
\exp\left( \sum_{k} \dfrac{\bar{p}_k W[G]_{k}}{k} \right) \cdot 1 = \sum_R \left(\prod_{(i,j) \in R} G(j-i) \right) S_R(\bar{p}) S_{R}(p)
}
\end{equation}
Thus, we generate an expression for the hypergeometric $\tau$-function (\ref{htau}) with $G=f$, i.e. associated with the cone commutative subfamilies of the $W_{1+\infty}$ algebra. Note that, in accordance with the rule (\ref{FG+}), in order to determine action of the operator $z^nF(\hat D)$ on the Schur functions, one has to find such a function $G$ that
\be\label{FG}
F(\hat D)=\prod_{i=1}^{n} G(D+i)
\ee

The skew counterpart of formula (\ref{mainG}) looks as
\be\label{mainG2}
\boxed{
\exp\left( \sum_{k} \dfrac{\bar{p}_k W[G]_{-k}}{k} \right) \cdot \exp\left(\sum_k{g_kp_k\over k} \right)
= \sum_{Q,R}\left(\prod_{i,j\in R/Q}G(j-i)\right)S_{R/Q}\{\bar p_k\}S_R\{g_k\}S_Q\{p_k\}
}
\ee

\subsection{Rational rays: (1,2) rays}

First of all, let us note that the function $G$ is a polynomial, while $f$ in (\ref{htau}) is just a function without poles at integer points. This means that, strictly speaking, one has to extend the definition of the $W_{1+\infty}$ algebra in order to include not only polynomials but power series in $\hat D$ as well\footnote{Note that the Kac-Radul variant of the algebra (\ref{KR}) \cite{KR1,Miki} is a completion of the $W_{1+\infty}$ algebra of sec.3.}. Consider now an example of the simplest rational ray (1,2) and the generating operator $\hat E_1^{(2)}$,  (\ref{Enq}):
    \begin{equation}\label{DE1}
    W(\hat E_1^{(2)})=W\left(2z^{2}\left(\hat D+u+{1\over 2}\right)\right)=\sum_{a}a p_{a+2} \dfrac{\partial}{\partial p_a} + {p_1^2\over 2}+up_2
    \end{equation}
For the sake of simplicity, we put at the moment $u=0$.

The operator (\ref{DE1}) commutes with all Hamiltonians (\ref{Hnq}). However, one can find more Hamiltonians: in accordance with (\ref{FG}), one can construct such a function $G$ that
\begin{equation}\label{factorization}
    2\hat D+1=\prod_{i=1}^{2} G(\hat D+i)=G(\hat D+1)G(\hat D+2)
\end{equation}
and then the whole family $z^n\prod_{i=1}^{n} G(\hat D+i)$ will be commutative. In this concrete case, this means the following. This equation (\ref{factorization}) has the solution
\be
G(x)=2{\Gamma\left({x\over 2}+{3\over 4}\right)\over\Gamma\left({x\over 2}+{1\over 4}\right)}
\ee
and we obtain a commutative set
\be\label{Hk}
\hat{\cal H}_k^{(1,2)}=(2z)^k\prod_{i=1}^k{\Gamma\left({\hat D\over 2}+{3\over 4}+i\right)\over\Gamma\left({\hat D\over 2}+{1\over 4}+i\right)}
\ee
the even members of this set being $\hat H_k^{(1,2)}$.

Using (\ref{FG+}), one obtains for the new Hamiltonian $\hat{\cal H}_1^{(1,2)}$
\begin{equation}
    \begin{split}
        \hat{\cal H}_1^{(1,2)}=&W\left(   z^{-1} \frac{2 \Gamma \left(\frac{D}{2}+\frac{3}{4}\right)}{\Gamma
   \left(\frac{D}{2}+\frac{1}{4}\right)} \right)=
2\dfrac{\Gamma(3/4)}{\Gamma(1/4)}p_1 + \frac{\pi}{\sqrt{2} \Gamma(3/4)^2 } p_2 \dfrac{\partial}{\partial p_1} - 2\dfrac{\Gamma(3/4)}{\Gamma(1/4)} p_1^2 \dfrac{\partial}{\partial p_1} +
   \\
   &+ \left(\frac{p_1^3 \Gamma \left(-\frac{3}{4}\right) \Gamma
   \left(\frac{5}{4}\right)}{\sqrt{2} \pi }+\frac{\sqrt{2} \pi  p_2
   p_1}{\Gamma \left(\frac{1}{4}\right) \Gamma
   \left(\frac{5}{4}\right)}-\frac{p_3 \Gamma \left(-\frac{3}{4}\right)
   \Gamma \left(\frac{5}{4}\right)}{\sqrt{2} \pi } \right) \dfrac{\partial}{\partial p_2}+
   \\
   &+\left(\frac{4 p_1^3 \Gamma \left(\frac{3}{4}\right)}{\Gamma
   \left(\frac{1}{4}\right)}-\frac{\sqrt{2} \pi  p_2 p_1}{\Gamma
   \left(\frac{3}{4}\right)^2}+\frac{2 \sqrt{2} p_3 \Gamma
   \left(\frac{3}{4}\right)^2}{\pi } \right) \dfrac{\partial^2}{\partial p_1^2} +\ldots
   \end{split}
    \end{equation}

\subsection{Rational rays: general case}

Now using (\ref{mainG}), one obtains that
\begin{equation}
	\begin{split}
		\exp\left( \dfrac{\hat{E}^{(2)}_1}{2} \right) \cdot 1 =
\sum_R \prod_{(i,j) \in R} \left(  \frac{2 \Gamma \left(\frac{(j-i)}{2}+\frac{3}{4}\right)}{\Gamma
			\left(\frac{(j-i)}{2}+\frac{1}{4}\right)} \right) S_R(\delta_{k,2})S_R(p) =
\ \ \ \ \ \ \ \ \ \ \ \ \ \ \ \ \ \ \ \ \
		\\
		=
		1+ \frac{p_1^2}{2} + \frac{1}{8} \left(p_1^4+4 p_3 p_1\right) + \frac{1}{48} \left(p_1^6+12 p_3 p_1^3+24 p_5 p_1+8 p_3^2\right) + \frac{1}{48} \left(p_1^6+12 p_3 p_1^3+24 p_5 p_1+8 p_3^2\right) + \ldots
	\end{split}
\end{equation}
It evidently does not contain odd $p_k$. Similarly, for arbitrary $q$, this partition function does not depend on $p_{qk}$. Hence, it can be expanded in the Hall-Littlewood polynomials taken at $t=\exp\Big({2\pi i\over q}\Big)$ \cite{MMGKM}.

Upon switching on $u$, the formula becomes
\begin{equation}
	\begin{split}
		\exp\left( \dfrac{\hat{E}^{(2)}_1}{2} \right) \cdot 1 =	\exp\left( \dfrac{\hat{H}^{(1,2)}_1}{2} \right) \cdot 1 =
\sum_R 2^{|R|}\prod_{(i,j) \in R} \left(  \frac{\Gamma \left(\frac{(j-i)}{2}+\frac{3}{4}+{u\over 2}\right)}{\Gamma
			\left(\frac{(j-i)}{2}+\frac{1}{4}+{u\over 2}\right)} \right) S_R(\delta_{k,2})S_R(p)
	\end{split}
\end{equation}
and, generally at arbitrary $q$,
\begin{equation}
	\begin{split}
		\exp\left( \dfrac{\hat{E}^{(q)}_1}{q^{q-1}} \right) \cdot 1 =	\exp\left( \dfrac{\hat{H}^{(1,q)}_1}{q^{q-1}} \right) \cdot 1 =
\sum_R q^{|R|}\prod_{(i,j) \in R} \left(  \frac{\Gamma \left(\frac{(j-i)}{q}+\frac{q+1}{2q}+{u\over q}\right)}{\Gamma
			\left(\frac{(j-i)}{q}+\frac{q-1}{2q}+{u\over q}\right)} \right) S_R(\delta_{k,q})S_R(p)
	\end{split}
\end{equation}
At last, the partition function at general $p$ and $q$ is
\begin{equation}\label{Hrr+}
	\begin{split}
	\exp\left( \sum_k{\bar p_{qk}\over k}\dfrac{\hat{H}^{(p,q)}_k}{q^{q-1}} \right) \cdot 1 =
\sum_R \prod_{(i,j) \in R} \left(  \frac{\Gamma \left(\frac{(j-i)}{q}+\frac{q+1}{2q}+{u\over q}\right)}{\Gamma
			\left(\frac{(j-i)}{q}+\frac{q-1}{2q}+{u\over q}\right)} \right)^p S_R(\bar p)S_R(p)
	\end{split}
\end{equation}
Note that we have to request that the $\Gamma$-functions in the numerator of this expression have no poles, i.e. that ${q-1\over 2}+u\ne\mathbb{Z}$. Then, one can include in the exponential the Hamiltonians given by $z^n\prod_{i=1}^nG(\hat D+i)$ at all $n$, similarly to considering $\hat{\cal H}_k^{(1,2)}$ at all (not only even) $k$ in (\ref{Hk}), and finally obtain
\begin{equation}
	\begin{split}
	\exp\left( \sum_k{\bar p_{k}\over k}\dfrac{\hat{\cal H}^{(p,q)}_k}{q^{q-1}} \right) \cdot 1 =
\sum_R \prod_{(i,j) \in R} \left(  \frac{\Gamma \left(\frac{(j-i)}{q}+\frac{q+1}{2q}+{u\over q}\right)}{\Gamma
			\left(\frac{(j-i)}{q}+\frac{q-1}{2q}+{u\over q}\right)} \right)^p S_R(\bar p)S_R(p)
	\end{split}
\end{equation}
This is correct for generic $u$. If, however, $u$ is such that ${q-1\over 2}+u$ is a negative integer $-m$, there can happen poles in the $\Gamma$-functions, and one has to leave in the sum in this expression only variables $\bar p_{qk}$. In this case, the Schur functions are non-zero only if $|R|$ is a multiple of $q$ and $R$ can be tiled with border strips of length $q$ so that all potential poles at $i-j$ are cancelled with the corresponding zero at $i-j-1$ due to the denominator. For instance, at $q=2$ the Young diagram that can not be tiled with border strips of length 2 is $R = [3,2,1]$, and $S_{[3,2,1]}$ vanishes when all odd times are zero.

Similarly,
\begin{equation}\label{Hrr-}
	\begin{split}
	\exp\left( \sum_k{\bar p_{k}\over k}\dfrac{\hat{\cal H}^{(p,q)}_{-k}}{q^{q-1}} \right) \cdot
\exp\left( \sum_k{g_kp_k\over k} \right)=
\sum_R \prod_{(i,j) \in R/Q} \left(  \frac{\Gamma \left(\frac{(j-i)}{q}+\frac{q+1}{2q}+{u\over q}\right)}{\Gamma
			\left(\frac{(j-i)}{q}+\frac{q-1}{2q}+{u\over q}\right)} \right)^p S_{R/Q}(\bar p)S_Q(p)S_R(g)
	\end{split}
\end{equation}

\section{Common eigenfunctions of commuting Hamiltonians
\label{commonev}}

Commuting Hamiltonians have common eigenfunctions.
Of course, in such basic systems as we discuss in this paper,
these eigenfunctions should be also something special.
We begin in sec.\ref{calef} with a sketch of a more traditional story, where eigenfunctions
are in terms of matrices or of eigenvalues, and conjugation is made with the help of Vandermonde factors,
and then in sec.\ref{timefCal} and further
proceed to a far more interesting case of eigenfunctions in terms  of time-variables.

\subsection{Calogero systems
\label{calef}}

As we discussed throughout the paper, the commutative families of $W_{1+\infty}$, i.e. Hamiltonians can be realized in terms of matrices when acting on invariant polynomials (sec.5), or in terms of their eigenvalues (sec.6-7). Therefore, it is natural to start with discussing eigenfunctions of the Hamiltonians in these terms. In fact, it is quite well-investigated problem. For instance, it is known \cite{MMN} that the eigenfunctions of the ``vertical" line, i.e. of operators $\hat O_{k,0}$ (with $\hat O_{3,0}=\hat W_0$) are the Schur functions. This is consistent with the fact that this line describes the trigonometric Calogero-Sutherland system in the free fermion point: the eigenfunctions of the Hamiltonians of the trigonometric Calogero-Sutherland system are well-known \cite{Jack,St,Mac,Turb,GP} to be the Jack polynomials multiplied by the Vandermonde factor due to (\ref{vanb}), which become the Schur functions at the free fermion point $\beta=1$. The commutative family $\hat H^{(1)}_{-k}$ is nothing but the rational Calogero Hamiltonians, and their eigenfunctions can not be polynomial, since the Hamiltonians, similarly to the trigonometric case, are scaling invariant, but have a non-zero scaling dimension in variance with the trigonometric Calogero-Sutherland Hamiltonians. They can be certainly ``regularized" by adding dimensionful oscillator terms, and then the polynomial eigenfunctions exist and are called Hi-Jack polynomials \cite{UW}.

In the next subsections, we construct non-polynomial eigenfunctions of the rational Calogero Hamiltonians given as a formal power series instead of polynomials. However, as was first noticed in \cite{Per,Turb}, since the Calogero eigenfunctions are symmetric functions of coordinates, it is usually more effective to work not in the coordinate variables $\lambda_i$, but in the time variables $p_k$. The only drawback of working in these coordinates is that the eigenfunctions of the Hamiltonians $\hat H^{(*)}_{-k}$ and $\hat H^{(*)}_{k}$ in coordinates are related by a simple transformation $\lambda_i\to\lambda_i^{-1}$ (up to a simple multiplier), while in variables $p_k$ the formulas are much more involved.

\subsection{Integer rays}

\subsubsection{Calogero ray
\label{timefCal}}

We start with working in variables $p_k$.  The Hamiltonian $\hat H^{(1)}_{-1} = \sum_{a=0}^\infty (a+1)p_a\frac{\p}{\p p_{a+1}}$ has a set of obvious eigenfunctions:
\be
\psi_\lambda = \sum_{i=0}^\infty \frac{\lambda^i p_i}{i!}, \ \ \ \
\hat H^{(1)}_{-1}\psi_\lambda = \lambda\psi_\lambda
\ee
They are, however, linear in times and are not the eigenfunctions of $\hat H^{(1)}_{-k}$ with $k\geq 2$.
Instead,  $\psi^{(\vec k)}_\lambda :=\prod_a \hat H_{-k_a}^{(1)} \psi_\lambda$
are the new eigenfunctions of  $\hat H^{(1)}_{-1}$
with the same eigenvalue $\lambda$:
\be
\hat H^{(1)}_{-1}\psi^{(\vec k)}_\lambda = \lambda\psi^{(\vec k)}_\lambda
\ee
We can search for the common eigenfunctions among linear combinations of $\psi^{(\vec k)}_\lambda$.
They are already multilinear in time variables $p_k$.

Also, since  $\hat H^{(1)}_{-1}$ is of the first-order in all $p$-derivatives,
all powers of $\psi$ are  its eigenfunctions.
Moreover, one can take a linear combination of the products $\prod_{a=1}^n \psi_{\lambda_a}$
that have the same eigenvalue $\lambda=\sum_{a=1}^n \lambda_a$.

Another option is to use the fact that the Hamiltonians $\hat H^{(1)}_{-k}$ act nicely on complete homogeneous symmetric polynomials:
\be\label{actiononsymmetric}
\hat H^{(1)}_{-k}  h_r = \frac{(k+u-1)!}{(u-1)!} h_{r-k}
\ee
Then
\be
\Psi_\lambda = \sum_r \frac{\lambda^r h_r}{(u+r-1)!}
\label{psisym}
\ee
is a common eigenfunction of the entire Calogero ray:
\be
\hat H^{(1)}_{-k} \Psi_\lambda = \lambda^k \Psi_\lambda
\ee
A more general eigenfunction is
\be
\boxed{
\Psi\{p,\bar p\} = \sum_R \frac{
  S_R\{p\}S_R\{\bar p\}S_R\{\delta_{k,1}\} }{ S_R[u]}
}
\label{efH11}
\ee
where the free parameters are the dual time variables $\bar p_k$.
Then Hamiltonians act on $p$-variables, while $\bar p$ define the eigenvalues:
\be\label{88}
\hat H^{(1)}_{-k}\{p\} \Psi\{p,\bar p\}  =
\bar p_k \Psi\{p,\bar p\}
\ee
Similarly to \eqref{actiononsymmetric}, the structure of the eigenfunction is due to a nice action of the Hamiltonians on the Schur functions, which is a special case of \eqref{OnSchur}:
\begin{equation}
	H_{-k}^{(1)} S_R \left\{ p \right\} = \sum_Q \left( \prod_{(i,j) \in Q/R} (u+j-i)  \right) \Big\langle \dfrac{\partial}{\partial p_k }S_R \Big | S_Q \Big\rangle S_Q\left\{ p \right\}
\end{equation}
For example, let us see how the eigenfunction equation works for the simple Hamiltonian $H_{-1}^{(1)}= \sum\limits_a (a+1) p_a \dfrac{\partial}{\partial p_{a+1}} + u \dfrac{\partial}{\partial p_1}$ :
\begin{equation}
	\begin{split}
		H_{-1}^{(1)}& \left( 1 + p_1\bar p_1 u^{-1} +\frac{(p_1^2+p_2)(\bar p_1^2+\bar p_2) }{2u(u+1)}
		+\frac{ (p_1^2-p_2)(\bar p_1^2-\bar p_2)}{2 u(u-1)}  + \ldots \right) =
		\\
		= &\left( \bar{p}_1 + \dfrac{\bar{p}_1^2+ \bar{p}_2}{u} + \dfrac{\bar{p}_1^2 -  \bar{p}_2}{u} \ldots  \right) = \bar{p}_1 \left(  1 + p_1 \bar{p}_1 u^{-1}  \ldots  \right)
	\end{split}
\end{equation}
Eq.(\ref{efH11}) is surprisingly similar to the superintegrability formulas
for Gaussian matrix models \cite{MMsi},
but the time-independent factor is inverted.
Eq.(\ref{psisym}) corresponds to the choice of all $\bar p_k=1$, when $S\{\bar p_k\}=S_R[1]$
is unity for symmetric $R=[r]$ and zero for all other $R$.

In fact, equation (\ref{88}) is rather evident: note that (\ref{efH11}) is nothing \cite{IZce} but the Harish-Chandra-Itzykson-Zuber integral \cite{IZ} being realized in terms of matrices $p_k=\Tr\Lambda^k$ and $\bar p_k=\Tr\bar\Lambda^k$:
\be\label{IZ}
\Psi\{p,\bar p\} =\int dU\exp\left(\Tr U\Lambda U^\dag\bar\Lambda\right)
\ee
while, in these terms (see (\ref{Hmm})),
\be\label{HIZ}
\hat H^{(1)}_{-k}=\Tr \left({\p\over\p\Lambda}\right)^k
\ee
For (\ref{IZ}) and (\ref{HIZ}), (\ref{88}) is evident.

\subsubsection{Hamiltonians $\hat H^{(m)}_{-k}$}

A similar formula for the eigenfunctions for an arbitrary integer ray is

\bigskip

\hspace{-0.65cm}\framebox{\parbox{17cm}{
\be\label{hr0}
\hat H^{(m)}_{-k}\{p\} \Psi^{(m)}\{p,\bar p\}& =& \bar p_k \Psi^{(m)}\{p,\bar p\}\nn\\
\Psi^{(m)}\{p,\bar p\}& =& \sum_R S_R\{p\}S_R\{\bar p\}\left(\frac{S_R\{\delta_{k,1}\} }{ S_R[u]}\right)^m
\ee
}}

\bigskip

\noindent
In fact, these formulas follow from the relation
\be
n{\p\over\p p_n}\exp\left(\sum_{k=1}{p_k\bar p_k\over k}\right)=\bar p_n\exp\left(\sum_{k=1}{p_k\bar p_k\over k}\right)
\ee
The point is that the Hamiltonians $\hat H^{(m)}_{-k}$ can be presented \cite{Ch1} in the form
\be\label{r1}
\hat H^{(m)}_{-k}=\hat O(u)^{-m}\cdot k{\p\over\p p_k}\cdot\hat O(u)^m
\ee
where the operator $\hat O(u)$ constructed in \cite{AMMN2} (see also \cite[Eq.(14)]{Ch2}) acts on the Schur functions in accordance with
\begin{equation}
    \hat {O}(u) \cdot S_R\{p_k\} = \dfrac{S_R(u)}{S_R(\delta_{k,1})} S_R\{p_k\}
\end{equation}
Now using the Cauchy identity
\be
\exp\left(\sum_{k=1}{p_k\bar p_k\over k}\right)=\sum_RS_R\{p_k\}S_R\{\bar p_k\}
\ee
one immediately obtains (\ref{88}) and (\ref{hr0}).

These formulas for the eigenfunctions, (\ref{88}) and (\ref{hr0}) can be definitely rewritten in terms of coordinates with the substitution $p_k=\sum_i\lambda_i^k$. The corresponding Hamiltonians in terms of $\lambda_i$ can be found in sec.6.

\subsubsection{Hamiltonians $\hat H^{(m)}_{k}$}

Now note that, in accordance with \cite{Ch1},
\be\label{r2}
\hat H^{(m)}_k=\hat O(u)^m\cdot p_k\cdot\hat O(u)^{-m}
\ee
Hence, from
\be
p_n\exp\left(\sum_{k=1}{p_k\bar p_k\over k}\right)=n{\p\over\p \bar p_n}\exp\left(\sum_{k=1}{p_k\bar p_k\over k}\right)
\ee
it immediately follows that formula \eqref{88} is dual to a similar relation for the positive ray:
\begin{equation}\label{dH}
	H^{(m)}_{k} \{ p \} \Phi^{(m)} \left\{ p, \bar{p} \right\} =k\dfrac{\partial}{\partial \bar{p}_k} \Phi^{(m)} \left\{ p, \bar{p} \right\}
\end{equation}
with
\begin{equation}\label{def}
	\Phi^{(m)} \left\{p,\bar{p} \right\} = \sum_R S_R\{p\}S_R\{\bar p\} \left({S_R[u] \over S_R\{\delta_{k,1}\} }\right)^m
\end{equation}
which is nothing but formula (\ref{Zn}),
\be
\Phi^{(m)} \left\{p,\bar{p} \right\}=\exp\left( \sum_{k} \dfrac{\bar{p}_k \hat H^{(m)}_k}{k} \right) \cdot 1
\ee
Again, to demonstrate how this works consider the example $H_{1}^{(1)}= \sum\limits_a a p_{a+1} \dfrac{\partial}{\partial p_a} + u p_1$:
\begin{equation}
	\begin{split}
		H_{1}^{(1)}& \left( 1 + p_1\bar p_1 u +\frac{(p_1^2+p_2)(\bar p_1^2+\bar p_2) }{2}u(u+1)
		+\frac{ (p_1^2-p_2)(\bar p_1^2-\bar p_2)}{2 }u(u-1)  + \ldots \right) =
		\\
		= &\left( p_1 u +   \bar{p}_1\left[(p_1^2+p_2)u(u+1)
		+(p_1^2-p_2)u(u-1)  \right]   \ldots  \right) =
		\\
		 = &  \dfrac{\partial}{\partial \bar{p}_1} \left( 1 + p_1\bar p_1 u +\frac{(p_1^2+p_2)(\bar p_1^2+\bar p_2) }{2}u(u+1)
		+\frac{ (p_1^2-p_2)(\bar p_1^2-\bar p_2)}{2 }u(u-1)  + \ldots \right)
	\end{split}
\end{equation}
Equation (\ref{dH}) represents the statement that commuting Hamiltonians are behind superintegrability in a simple form: the flow in $\bar{p}_m$ direction is exactly given by the corresponding commutative Hamiltonians of the Calogero type. In order to generate the eigenvalue equation instead of (\ref{dH}), one can make the Fourier transform of $\Phi^{(m)} \left\{p,\bar{p} \right\}$ in variables $\bar p_k$, however, it is rather singular. Instead, one can return back to the coordinate variables $\lambda_i$ and note that $H^{(m)}_{k}$ are related with $H^{(m)}_{-k}$ by the substitution $H^{(m)}_{k}(\lambda_i)=\prod_i\lambda_i^{-N}\cdot H^{(m)}_{k}(\lambda_i^{-1})\cdot\prod_i\lambda_i^N$ \cite{MMCal}. Now, using this substitution, formula (\ref{hr0}) and formula
\be
S_R(\lambda_i^{-1})=\left(\prod_{i=1}^N\lambda_i\right)^{-R_1}S_{\bar R}(\lambda_i)
\ee
where $\bar R$ is the conjugate representation, one comes to the eigenfunctions of the Hamiltonians $H^{(m)}_{k}$:

\bigskip

\hspace{-0.65cm}\framebox{\parbox{17cm}{
\be
\hat H^{(m)}_k\{p\} \bar\Psi^{(m)}\{p,\bar p\} &=& \bar p_k \bar\Psi^{(m)}\{p,\bar p\}\nn\\
\bar\Psi^{(m)}\{p,\bar p\} &=& \sum_R \left(\prod_{i=1}^N\lambda_i\right)^{-N-R_1}
S_{\bar R}\{p\}S_R\{\bar p\}\left(\frac{S_R\{\delta_{k,1}\} }{ S_R[u]}\right)^m
\ee
}}

\subsection{$\beta$-deformation}

The $\beta$-deformation of all these formulas is absolutely immediate: one just has to substitute the Schur functions with the Jack polynomials. More exactly,
\be
{n\over\beta}{\p\over\p p_n}\exp\left(\sum_{k=1}{\beta p_k\bar p_k\over k}\right)=\bar p_n\exp\left(\sum_{k=1}{\beta p_k\bar p_k\over k}\right)
\ee
and the $\beta$-deformed Hamiltonians $\hat H^{(m)}_{-k}$ can be presented \cite{Ch2} in the form
\be\label{rb}
\hat H^{(m)}_{-k}=\hat O^\beta(u)^{-m}\cdot {k\over\beta}{\p\over\p p_k}\cdot\hat O^\beta(u)^m
\ee
where the operator $\hat O^\beta(u)$ acts on the Jack polynomials \cite{China1,China2,MO,Ch2} in accordance with
\begin{equation}
    \hat {O}^\beta(u) \cdot J_R\{p_k\} = \dfrac{J_R(u)}{J_R(\delta_{k,1})} J_R\{p_k\}
\end{equation}
Now using the Cauchy identity
\be
\exp\left(\sum_{k=1}{\beta p_k\bar p_k\over k}\right)=\sum_R{J_R\{p_k\}J_R\{\bar p_k\}\over ||J_R||}
\ee
where $||J_R||$ is the norm square of the Jack polynomial,
\be
||J_R||:={\overline{G}^\beta_{R^\vee R}(0)\over G^\beta_{RR^\vee}(0)}\beta^{|R|}\ \ \ \ \ \ \
G_{R'R''}^\beta(x):=\prod_{(i,j)\in R'}\Big(x+R'_i-j+\beta(R''_j- i+1)\Big)
\ee
with the bar over the functions denoting the substitution $\beta\to\beta^{-1}$,
one immediately obtains

\bigskip

\hspace{-0.65cm}\framebox{\parbox{17cm}{
\be
\hat H^{(m)}_{-k}\{p\} \Psi^{(m)}\{p,\bar p\}& =& \bar p_k \Psi^{(m)}\{p,\bar p\}\nn\\
\Psi^{(m)}\{p,\bar p\}& =& \sum_R {J_R\{p\}J_R\{\bar p\}\over ||J_R||}\left(\frac{J_R\{\delta_{k,1}\} }{ J_R[u]}\right)^m
\ee
}}

\bigskip

\noindent
Again, in order to construct the eigenfunctions of the $\beta$-deformed Hamiltonians $H^{(m)}_{k}$, one returns back to the coordinate variables $\lambda_i$ and note that $H^{(m)}_{k}$ are related with $H^{(m)}_{-k}$ by the substitution $H^{(m)}_{k}(\lambda_i)=\prod_i\lambda_i^{-N\beta+\beta-1}\cdot H^{(m)}_{k}(\lambda_i^{-1})\cdot \prod_i\lambda_i^{N\beta-\beta+1}$ \cite{MMCal}. Now, using this substitution, formula (\ref{hr0}) and formula
\be
J_R(\lambda_i^{-1})=\left(\prod_{i=1}^N\lambda_i\right)^{-R_1}J_{\bar R}(\lambda_i)
\ee
one comes to the eigenfunctions of the Hamiltonians $H^{(m)}_{k}$:

\bigskip

\hspace{-0.65cm}\framebox{\parbox{17cm}{
\be
\hat H^{(m)}_k\{p\} \bar\Psi^{(m)}\{p,\bar p\} &=& \bar p_k \bar\Psi^{(m)}\{p,\bar p\}\nn\\
\bar\Psi^{(m)}\{p,\bar p\} &=& \sum_R \left(\prod_{i=1}^N\lambda_i\right)^{-\beta N+\beta-1-R_1}
{J_{\bar R}\{p\}J_R\{\bar p\}\over ||J_R||}\left(\frac{J_R\{\delta_{k,1}\} }{ J_R[u]}\right)^m
\ee
}}

\subsection{Rational rays}

In order to come to the rational rays, let us note that, as implied by results of sec.12.2, the Hamiltonians corresponding to the rational rays are generated by
\be\label{rq}
\hat{H}^{(p,q)}_{k}=\hat O_q(u)^p\cdot p_{qk}\cdot\hat O_q(u)^{-p}\\
\hat H^{(p,q)}_{-k}=\hat O_q(u)^{-p}\cdot qk{\p\over\p p_{qk}}\cdot\hat O_q(u)^p
\ee
where the operator $\hat O_q(u)$ acts on the Schur function in accordance with
\be
 \hat {\cal O}_q(u)S_R\{p_k\}=\prod_{(i,j) \in R} \frac{\Gamma \left(\frac{(j-i)}{q}+\frac{q+1}{2q}+{u\over q}\right)}{\Gamma
			\left(\frac{(j-i)}{q}+\frac{q-1}{2q}+{u\over q}\right)} S_R\{p_k\}=:\xi_R^{(q)}(u)S_R\{p_k\}
\ee
Hence, repeating our previous arguments, one obtains

\bigskip

\hspace{-0.65cm}\framebox{\parbox{17cm}{
\be\label{hr}
\hat H^{(p,q)}_{-k}\{p\} \Psi^{(p,q)}\{p,\bar p\} &=& \bar p_{qk} \Psi^{(p,q)}\{p,\bar p\}\nn\\
\Psi^{(p,q)}\{p,\bar p\}& =& \sum_R S_R\{p\}S_R\{\bar p\}\left(\xi_R^{(q)}(u)\right)^{-p}
\ee
}}

\bigskip

\noindent
and
\begin{equation}
\hat H^{(p,q)}_{k} \{ p \} \Phi^{(p,q)} \left\{ p, \bar{p} \right\} =qk\dfrac{\partial}{\partial \bar{p}_{qk}} \Phi^{(p,q)} \left\{ p, \bar{p} \right\}
\end{equation}
with
\begin{equation}
	\Phi^{(p,q)} \left\{p,\bar{p} \right\} = \sum_R S_R\{p\}S_R\{\bar p\}\left(\xi_R^{(q)}(u)\right)^{p}
\end{equation}

In fact, these relations are immediately extended to the set of Hamiltonians $\hat{\cal H}_k$ of the previous section:
\be
\hat{\cal H}^{(p,q)}_{k}=\hat O_q(u)^p\cdot p_{k}\cdot\hat O_q(u)^{-p}\\
\hat{\cal H}^{(p,q)}_{-k}=\hat O_q(u)^{-p}\cdot k{\p\over\p p_{k}}\cdot\hat O_q(u)^p
\ee
and
\be
\boxed{\hat{\cal H}^{(p,q)}_k\{p\} \Psi^{(p,q)}\{p,\bar p\} = \bar p_{k} \Psi^{(p,q)}\{p,\bar p\}}\\
\hat{\cal H}^{(p,q)}_{k} \{ p \} \Phi^{(p,q)} \left\{ p, \bar{p} \right\} =k\dfrac{\partial}{\partial \bar{p}_{qk}} \Phi^{(p,q)} \left\{ p, \bar{p}\right\}
\ee

\section{Constructing operator $\hat{O}$}\label{sec:O}

As we discussed in the previous section, the Hamiltonians at all rays can be constructed from the horizontal line $H_{\pm k}^{(0)}$ by the operator $\hat {O}(u)$ with the rotation (see (\ref{r1})-(\ref{r2}))
\begin{equation}
	H^{(m)}_{\pm k} = \left(\hat{O}(u)\right)^{\pm m} H^{(0)}_{\pm k} \left(\hat{O}(u)\right)^{\mp m}
\end{equation}
i.e. $\hat O$ rotates the integer ray $m$ to the integer ray $m+1$.

This operator can be manifestly constructed in all representations that we discussed in this paper and has an especially simple form when acting on characters. Since the character representation is directly related to the form of the operator in the one-body formalism we expect a simple formula in terms of one-body operators. Indeed, one has \cite{AMMN2}:
\begin{equation}\label{Oob}
	\hat{O}(u) = \dfrac{\Gamma(\hat D+u)}{\Gamma(u)}
\end{equation}
which can be immediately checked (where we put $u=1$ for the sake of simplicity):
\begin{equation}
	\begin{split}
		 \hat H_{-k}^{(0)}\cdot \hat O^m ={1\over z^{k}}  \left(\Gamma(\hat D+ 1)\right)^m={1\over z^k} &\left(\Gamma(\hat D-k+1)\right)^m\prod_{i=0}^{k-1}(\hat D-i)^m=
		\\
		&=
		\left(\Gamma(\hat D+ 1)\right)^m{1\over z^k}\prod_{i=0}^{k-1}(\hat D-i)^m=\hat O^m\cdot \hat H_{-k}^{(m)}
	\end{split}
\end{equation}
and similarly for the integer rays $H_k^{(m)}$.

One can easily write down this operator for the rational series (\ref{rq}):
\be\boxed{
\hat{O}_q(u) = \dfrac{\Gamma\Big({\hat D\over q}+u\Big)}{\Gamma(u)}
}
\ee
This explains the emergence of the corresponding $\Gamma$-functions in formulas of sec.11.

This operator can be also constructed in terms of matrix variables via the cut-and-join operators $W_\Delta$ \cite{MMN} in accordance with \cite[sec.2.1]{Ch2}. In order to realize it in terms of time variables, one can either evaluate it from the matrix representation, or  presented it in the form convenient for the second quantization:
\begin{equation}\label{O}
	\hat{O}(u) = \dfrac{\Gamma(u+\hat D)}{\Gamma(u)} = u^{W_{0}^{(2)}} \exp\left( \sum_{m=1}^{\infty} \dfrac{(-1)^{m+1} \left(\hat W_{0}^{(m+2)}-\hat D^{m}\right)}{m u^{m}} \right)
\end{equation}
where $\hat W_{0}^{(m)}$ form the commutative family of the spin $m$ operators lying on the vertical line of Figure 1:
\begin{equation}
	\hat W_{0}^{(m)} :=  \rho_{m-2}(\hat D)
\end{equation}
with the degree $m$ polynomial $\rho_m(k)$ of $k$ defined as the generalized harmonic number
\be
\rho_m(k):=\sum_{i=1}^{k} i^m
\ee
which can be analytically continued to give:
\begin{equation}\label{toD}
	\rho_k(\hat{D}) =\sum_{j=1}^{k-1} \binom{k}{j} \dfrac{B_{k-j+1}}{k-j+1} \hat{D}^{j} + \dfrac{\hat D^k}{2}  + \dfrac{\hat D^{k+1}}{k+1} \,,\quad k \geq 1
\end{equation}
with $B_n$ being the Bernoulli numbers.
In particular, the operator $\hat W_0$ throughout the paper is $\hat W_0^{(3)}$, and the first examples are
\be
\hat W_0^{(2)}=\rho_{0}(\hat D)=\hat D\nn\\
\hat W_0^{(3)}=\rho_{1}(\hat D)={1\over 2}\hat D(\hat D+1)
\ee
These operators $\hat W_{0}^{(m)}$ can be also associated with the cut-and-join operators $W_{[m-1]}$ \cite{MMN}. Now one can deal with (\ref{O}) in accordance with the second-quantization formula (\ref{WH}), using, for example, the expansion \eqref{toD} into simple basis elements.

Note that, in the particular case of $u=1$, the expansion has the form \cite{GR}
\begin{equation}
	\hat{O}(u=1) = \Gamma(1+\hat D)= \exp\left(-\gamma\hat D+ \sum_{m=2}^{\infty} \dfrac{(-1)^{m} \zeta(m)}{m}\hat D^{m} \right)
\end{equation}
where $\zeta(m)$ is the Riemann $\zeta$-function, and $\gamma$ is the Euler constant.

Note also that the combination $\hat W_{0}^{(m+2)}-\hat D^{m}$ can be rewritten in terms of the operators $\psi_i=(\hat D+1)^i-\hat D^i$ from \cite{Tsim,Prochazka} as
\be
\psi_i-\psi_1=2\sum_{i=1}^k\Big(2^{k-i}-1\Big)\binom{k}{i-1}\left(\hat W_{0}^{(m+2)}-\hat D^{m}\right)
\ee
This representation allows one to realize the operator $\hat O$ in purely algebraic terms, see a discussion in \cite[sec.4.3]{MMMP2}.

A manifest construction of the operator $\hat O^\beta$ from sec.12.3 is more involved, since there is no one-body formulation in the $\beta$-deformed case, and hence formulas similar to (\ref{Oob}) are not available. It can be obtained from a rather involved construction of the operator $\hat O^{(p,q)}$ in the $(q,t)$-deformed case \cite[Sec.3]{Ch3} in the limit of $t=q^\beta$, $q\to 1$.

\section{Algebraic summary
\label{alsum}}

The 1-loop Kac-Moody extension $\widehat G$ of a simple Lie algebra $G$ adds a new index to the generators.
Further 2-loop extension to the Yangian/DIM family $\widehat{\!\widehat {G}}$ adds two:
the generators are additionally labeled by integer points of the half-plane,
and the would be positive roots, by points of the quarter-plane.
In fact, the DIM procedure can be applied even to the $\widehat G$, thus providing a {\it pagoda}
algebra \ $\widehat{\widehat{\!\widehat {G}}}$ of \cite{MMZ}.
The structure in this quarter-plane is highly non-trivial by itself
and in certain sense is exhaustive \cite{GLM}, at least from the point of view of
string theory.
The double-loop extension involves a piece of the universal enveloping of $\widehat G$,
which is responsible for Virasoro and $W$-constrains and, for quantization and for integrability.
The theory is non-trivial even at the level of ${\widehat{\widehat {gl}}}_1$,
which is also known as the family of $\hat W_\infty$ algebras.

One of the principal difficulties in working with $W_\infty$ is the lack of a distinguished basis.
There is no obvious {\it a priori} way to prescribe a generator to a particular point in the
half-plane, a ``natural" prescription depends on the context and on the representation.
We suggest to resolve this ambiguity by appealing to the inherent integrable structure.
Namely, as we demonstrate,  $W_\infty$ has a large set of commutative 1-loop subalgebras,
In a sense, they are built from an infinite set of ``ray" subalgebras,
and then there is a single ray passing through a given integer point on the half-plane.
Along a ray, there is a family of commuting Hamiltonians, and the suggestion is to ascribe
to a point exactly the Hamiltonian.
This provides an invariantly defined basis, often different from the ``naive" ones
considered for particular representations.

To illustrate the idea, we list some simple facts in the following table.
The rays are labeled by the points $(M,i)=(kp+1,\pm kq)$ with $p$ and $q$ co-prime, and
$k$ is the coordinate along the ray, which labels commuting Hamiltonians.
For the sake of simplicity, we restrict the table to one quarter-plane (by fixing the sign to be minus).

\be
\begin{array}{|c|c|c|}
\hline&&\\
{\rm representation} & {\rm ``natural" \ basis}
& {\rm commuting\ Hamiltonians}    \cr
&&\\
\hline \hline
&&\\
w_\infty &   \hat v_{M,i}
&  \hat h^{(p,q)}_{-k} = \hat v_{kp+1,-q} \\  &&\\
\hline
&&\\
z-{\rm rep\ of\ } W_\infty & z^{-m}\hat D^{n} & \hat H^{(p,q)}_{-k} = (z^{-q} \hat D^p)^k  \\
&&\\
\hline
&&\\
{\rm Matrix\ rep} & \Tr\Lambda^{-m}\hat{\cal D}^n & \Tr \Big(\Lambda^{-1}\hat {\cal D}^m\Big)^k     \\
&&\\
\hline
&&\\
{\rm Time\ rep} & \hbox{Eq.}(\ref{WH}) &
H^{(p,q)}_{-k} \stackrel{(\ref{eq:h-formula-explicit1})}{ =}  \ \sum_{l=0}^{\frac{pq+1}{2}}
\left(\sum_{a_1+a_{pk+1-2l} = -q k}  G_{p,q,k,l}\{a\} :p_{a_1} \cdots p_{a_{p k + 1 - 2 l}}:\right)
 \\
 &&\\
\hline
&&\\
{\rm Eigenvalue\ rep} &  & \hat H_{-k}^{(p,q)} = q^{pk}\sum_i \left(\lambda_i^{1-q\over 2}\cdot\lambda_i^{-1}\hat{\mathfrak{D}}_i^p\cdot\lambda_i^{1-q\over 2}
\right)^k     \\
&&\\
&&\\
\hline
&&\\
\ldots &&\\
&&
\nn
\end{array}
\ee

Our specification of basis is best seen
in the second line of this table ($z$-representation).

In the third, forth and fifth lines the reasons for commutativity are well hidden,
we did not yet find an obvious way to demonstrate it at conceptual level
(besides direct relation to the previous lines). Moreover, in the third line,
such a representation is currently known only for integer rays with $q=1$.

\section{Conclusion
\label{conc}}

To summarize, in this paper
we explained that the $W_{1+\infty}$ algebra
contains a vast variety of commutative subalgebras.
They give rise to integrable systems,
of which the standard Calogero model is just the simplest example.
They can be associated with straight rays, but rays can be a kind of
``thickened" and represented as ``cones" or ``bundles of rays".
A full description of the family involves an arbitrary number
of free parameters $\alpha_i$'s.

Further extensions to Yangian and DIM representations (and to the Ruijsenaars substitute of the Calogero integrable systems) are left for future, partly, to avoid overloading the text with too complicated formulas,
which can overshadow simple structures. In fact, the first steps of extension to the most
general case of these, to the DIM algebra are described in \cite{Ch3}. In this case, the problem is
reformulated as constructing Heisenberg subalgebras of the quantum toroidal algebra, so that commuting
Hamiltonians are just halves of these Heisenberg algebras.
These Heisenberg subalgebras are also associated with rational rays, or, better to say, lines \cite{Smirnov}, their existence being a direct consequence of the Miki $SL(2\mathbb{Z})$-automorphism \cite{Miki1,Miki}, however, the limit to
the $W_{1+\infty}$ algebra case from these rational lines is rather intricate \cite[Sec.2.5]{Ch3}.

There is also an intermediate case: the affine Yangian algebra \cite{Tsim,Prochazka}. This case is associated with the $\beta$-deformation,
which we discussed a little in this paper, and it is much simpler related to the $W_{1+\infty}$ algebra. Still, as we demonstrated it requires some care, and the results obtained for the $W_{1+\infty}$ algebra are not literally transferred to the affine Yangian case. We postpone a detailed description of this case to a separate publication \cite{MMMP2}.

In another direction, one needs to find a more formal explanation of the pattern
of commuting one-dimensional subalgebras.
As we saw, there is a rather drastic difference between integer and non-integer rays and cones.
Our expectation is that commutativity in the integer ray case is a direct consequence of {\it only} the
relations that can be associated with the Serre relations.
They do not depend on $\beta$-deformation (and more generally on the three parameters $h_i$) of the $W_{1+\infty}$ algebra
to the affine Yangian, and this can explain  the straightforwardness of this deformation in the integer ray case.
At the same time, for the rational rays the Serre relations are not enough,
which makes the $\beta$-deformation less straightforward, if at all preserving the commutativity
of Hamiltonians.
This conclusion should have interesting implications for integrable systems:
we remind that the Calogero model {\it per se} acquires a non-trivial interaction only for $\beta\neq 1$.
We will return to these questions in a separate more formal note on algebraic properties
of our construction \cite{MMMP2}.

\section*{Acknowledgements}

A.Mir. is grateful to A. Grigoriev-Savrasov for kind hospitality at late stages of the project. This work was supported by the Russian Science Foundation (Grant No.20-12-00195).

\section*{Appendix A. Obtaining general formulas in time variables from the second quantization}

Using the formulas for second quantized operators (\ref{Ofromz}), (\ref{bos}), one can construct general formulas for the generating operators $\hat F^{(q)}_k$, $\hat E^{(q)}_k$ and for the Calogero Hamiltonians $\hat H^{(1)}_k$ in time variables $p_k$, see sec.8.3-8.4 for the answers. Now we explain how to derive these formulas.

\subsection*{A.1 The Calogero series of Hamiltonians}

Let us construct the  generating function of the simplest (Calogero) series of one-body Hamiltonians $\hat H_{-k}^{(1)}$ \eqref{genformbinom}\cite{Ch2}:
\be
{\cal H}^{(1)}_-(\zeta)
=\sum_{k=0}^\infty \frac{\zeta^k}{k!}H_{-k}^{(1)}= \sum_{k=0}^\infty \frac{\zeta^k}{k!} \p_z^k
= \exp(\zeta \p_z)
\ee
Because of a simple exponential form of the operator, it is easier to apply the second quantization, which is given by
\be\label{secondquantizedhamiltonian}
W\left( {\cal H}^{(1)}_-(\zeta)\right) = \oint \frac{dz}{2 \pi i} \cdot
\frac{-1}{\zeta}
: \exp\left( \phi(z) - \phi(z+\zeta)\right) - 1 :
\ee
where we used
\begin{equation}
	e^{(\zeta \p_z)} e^{\phi(z)} = e^{\phi(z+\zeta)}
\end{equation}
Expanding the bosonic fields in $\xi$ and using the definition of the complete homogeneous symmetric polynomials, one obtains:
\begin{equation}
	W\left( {\cal H}^{(1)}_-(\zeta)\right) = - \oint \frac{dz}{2 \pi i} \cdot
	\frac{1}{\zeta} \sum_{r=0}^{\infty}: \zeta^r h_r \left( p_k = -\dfrac{\partial_z^k \phi}{(k-1)!} \right)  :
\end{equation}
Using the simple formula for the complete homogeneous symmetric polynomials, one gets
\begin{equation}\label{expofsymshur}
	h_r\left( p_k = \dfrac{\partial_z^k \phi}{(k-1)!} \right)=\sum_{n=1}^r \frac{1}{n!}
	\sum_{k_1+\dots+k_n=r}
	\left(\prod_{i=1}^{n} \frac{\left(-\partial_z^{k_i} \phi \right) }{k_i!}\right)
\end{equation}
Each free field is expanded according to \eqref{freeboson} (recall that, according to our notations, $a_n^\dag=-p_n,\ \ \hat a_n=-{\p\over\p p_n}$):
\begin{equation}\label{derofboson}
	\dfrac{1}{k_i!}\partial_z^{k_i} \phi =\sum_{a_i=-\infty}^{\infty}  z^{-a_i-k_i} \binom{-a_i}{k_i}  p_{a_i}
\end{equation}

Taking the residue enforces the sum of degrees of $z$ to be $-1$, hence when one plugs the expansions \eqref{derofboson} and \eqref{expofsymshur} into \eqref{secondquantizedhamiltonian}, one obtains

\begin{equation}\label{genformbinom1}
	\hat{H}^{(1)}_{-k} = - k! \cdot \sum_{n=1}^k \sum_{ \sum_{i=1}^n a_i = -k}
	:\prod_{j=1}^n \dfrac{p_{a_j}}{a_j} : \left( \sum_{\substack{ k_1, \ldots, k_n  > 0 \\ \sum_{i=1}^n k_i = k+1 } } \prod_{i=1}^{n} \binom{-a_i}{k_i} \right)
\end{equation}
where the sum runs over all tuples $a=\{a_1,a_2\ldots a_{n} \} \in \left(\mathbb{Z}/\left\{0\right\} \right)^n$ and $\{k_1,\ldots k_n \}$ with the respective conditions.

\subsection*{A.2 Generating operators}

First of all, let us construct a one-body (formal) generating function of the generating operators
(see Eqs.\eqref{Fnr},\eqref{Enq}):
\begin{align}
   {\cal F}^{(q)}(\zeta)
   = & \ \sum_{m=0}^\infty \frac{\zeta^m}{m!} \frac{1}{z^q} q^m \left(\hat{D} + \frac{(1-q)}{2}
   \right)^m
  = \frac{1}{z^q} \exp\left(\zeta \frac{q (1-q)}{2} \right ) \exp(\zeta q \hat D)\nn\\
 {\cal E}^{(q)}(\zeta)
= & \ \sum_{m=0}^\infty \frac{\zeta^m}{m!}  \left(\hat{D} + \frac{(1-q)}{2} \right)^m q^{m-1} z^q
=\frac{1}{q} \exp\left(\zeta q \left(\hat D+\frac{(1-q)}{2} \right)\right)z^q
= \frac{z^q}{q} \exp\left(\zeta \frac{q (1+q)}{2} \right )
\exp(\zeta q \hat D)
\end{align}
Then, using (\ref{Ofromz}), (\ref{bos}), one gets
\begin{align} \label{eq:w-for-hurwitz}
W\left({\cal F}^{(q)}(\zeta)\right)
= \oint \frac{dz}{2 \pi i} \cdot
  \frac{1}{z^q} \cdot e^{\zeta \frac{q(1-q)}{2}} \cdot \frac{1}{(1 - e^{\zeta q}) z}
  : \exp\left(\phi(z) - \phi(e^{\zeta q} z) \right) - 1 :\nn\\
W\left({\cal E}^{(q)}(\zeta)\right)
= \oint \frac{dz}{2 \pi i} \cdot
\frac{z^q}{q} \cdot e^{\zeta \frac{q(1+q)}{2}} \cdot \frac{1}{(1 - e^{\zeta q}) z}
: \exp\left(\phi(z) - \phi(e^{\zeta q} z) \right) - 1 :
\end{align}
from which one can readily obtain the explicit expressions for
$\hat F_m^{(q)}$, $\hat E_m^{(q)}$ in  time variables.

Now we gradually obtain the desired expansion. Firstly,
keeping in mind that in our representation $a^\dagger_n = - p_n$
and $a_n = -\frac{\partial}{\partial p_n}$, we have
\begin{equation}
	\begin{split}
	\phi(z)  - \phi(e^{\zeta q} z) = \sum_{k \in \mathbb{Z} \backslash \{0\}}
	\frac{{ p_k}}{kz^k}
	\Big(\!\!\!\!\!\!\!\!\!\! \underbrace{e^{-k\zeta q}-1}_{
		\sum_{n=1}^\infty \frac{(-1)^n}{n!} (\zeta q k)^n}\!\!\!\!\!\!\!\!\!\!\Big)
	&- \log \left ( \exp(\zeta q) \right ) p_0
	=
	\\&
	=\sum_{n=1}^\infty \frac{1}{n} \zeta^n q^n   \underbrace{\left(
		\frac{(-1)^{n}}{(n-1)!}
		\sum_{k \in \mathbb{Z}} k^{n-1} {{ p_k}\over z^k}\right)}_{p_n^\ast}
	=\sum_{n=1}^\infty \frac{1}{n} \zeta^n q^n p_n^\ast,
	\end{split}
\end{equation}
    where the summand with $p_0$ is incorporated into $p^\ast_n$
    in a uniform fashion. Secondly, we can express the exponential as follows:
    \begin{align}
    \left(:\exp \left ( \phi(z) - \phi(e^{\zeta q} z)  \right ) : - 1 \right)\ \ \
    = \ \ \ \sum_{m=1}^\infty \zeta^m q^m :h_m \{p^{\ast}   \} :
  \end{align}
  where $h_m\left\{p^{\ast} \right\}$
  is the complete homogeneous symmetric polynomial.

Taking into account that
  \begin{align}
    \frac{e^{\zeta \frac{q(1\pm q)}{2}}}{e^{\zeta q} - 1}
    = \sum_{l=0}^\infty \frac{(\zeta q)^{l-1}}{l!} \cdot B_l \left(\frac{1-q}{2}\right)
  \end{align}
  where $B_l(x)$ are the Bernoulli polynomials, one obtains from \eqref{eq:w-for-hurwitz}
  (for $E^{(q)}_m$, the treatment is analogous)
  \begin{align}
   W\left({\cal F}^{(q)}(\zeta)\right)
    = \sum_{m=0}^\infty \frac{\zeta^m}{m!} \hat{F}_m^{(q)}
    = \oint \frac{dz}{2 \pi i z^{q+1}} (-1) \sum_{m=0}^\infty \zeta^m q^m
    \sum_{l=0}^{m+1} \frac{B_l\left(\frac{1-q}{2}\right)}{l!}
    :S_{[m+1 - l]} \left \{ p^{\ast} \right\}:
  \end{align}

\noindent Using the manifest formula for the complete homogeneous symmetric polynomials
\begin{align}
    h_r\{p^{\ast}\} = & \ \sum_{j=1}^r \frac{1}{j!}
    \sum_{n_1+\dots+n_j=r}
    \frac{p^{\ast}_{n_1}}{n_1} \dots \frac{p^{\ast}_{n_j}}{n_j}
\ \ \ \   =  \ \ \ \ \sum_{j=1}^r \frac{1}{j!}
    \sum_{n_1+\dots+n_j=r}
    \left(\prod_{i=1}^{j} \frac{p^{\ast}_{n_i}}{n_i}\right)
\end{align}
one gets for $\hat{F}^{(q)}_m$:
\begin{align}
    \hat{F}_m^{(q)} = & \ m! q^m \sum_{l=0}^{m+1}
    \frac{(-1)B_{l}\left(\frac{1-q}{2}\right)}{l! } \oint \frac{dz}{2 \pi i z^{q+1}}
    \sum_{j=1}^{m+1-l} \frac{1}{j!}
    \sum_{n_1+\dots+n_j=m+1-l}
    \left( : \prod_{i=1}^{j}
    \frac{(-1)^{n_i}}{n_i!}
    \sum_{k_i \in \mathbb{Z}} \frac{k_i^{n_i-1}}{ z^{k_i}} { p_{k_i}}
    :
    \right)  = \nn \\
& = m! q^m \sum_{l=0}^{m+1}
    \frac{(-1) B_{l}\left(\frac{1-q}{2}\right)}{l! }
    \sum_{j=1}^{m+1-l} \frac{1}{j!}\!\!\!\!\!\!\!\!
     \sum_{\stackrel{k_1+\ldots+k_j=-q}{n_1+\dots+n_j=m+1-l}}
    \    \prod_{i=1}^{j}
    \frac{(-1)^{n_i}k_i^{n_i-1}}{n_i!}
   : \prod_{i=1}^j { p_{k_i}} :
\end{align}
where the residue in $z$ transforms into requirements that sums of the {\it integer} indices
$k_i$  in each sum are equal to $-q$, while the indices $n$ are all {\it positive}.

\section*{Appendix B. From matrices to eigenvalues}

This is a modernized version of old calculations in \cite{MMM,Mikh}.
Usually it was made like in perturbation theory in quantum mechanics,
the construction below can be considered as its non-perturbative substitute.

We begin
from expressing
the characteristic equation ${\rm det}(\Lambda-\lambda)=0$,
which defines eigenvalues $\lambda_i$ of the $N\times N$ matrix $\Lambda$,
through the elementary symmetric polynomials:
\be
{\rm det}\left(1-\frac{\Lambda}{\lambda}\right) =
\exp\left\{ \Tr \log \left(1-\frac{\Lambda}{\lambda}\right)\right\} =
\exp\left(-\sum_{k=1}^\infty \frac{p_k}{k\lambda^k}\right)
= \sum_{k=0}^N  (-)^k \frac{e_k\{p_k\}}{\lambda^k}  := D_N = 0
\ee
where $p_k:=\Tr\Lambda^k$. For finite $N$, the last sum is automatically cut at $k=N$.

We can differentiate this expression by $\Lambda$:
\be
\frac{\p \lambda}{\p \Lambda}\cdot
\sum_{k=1}^N (-)^{k+1}\frac{k e_k}{\lambda^{k+1}}  +
\sum_{k=1}^N \frac{(-)^k}{\lambda^k} \sum_{m=1}^k \frac{\p p_m}{\p \Lambda} \frac{\p e_k}{\p p_m} = 0
\label{lambdader00}
\ee
We use the standard definition for matrix derivatives with the transposition:
$\left(\frac{\p \lambda}{\p \Lambda}\right)_{ij} = \frac{\p \lambda}{\p \Lambda_{ji}}$,
so that $\frac{\p \Tr \Lambda^m}{\p \Lambda} = m\Lambda^{m-1}$ and
$\Tr \frac{\p^2 \Tr \Lambda^2}{\p\Lambda^2} = 2N^2$.

In the first sum in (\ref{lambdader00}),
one substitutes $e_N$ from the condition $D_N=0$ in order to return the powers of $\lambda$
into the range $\lambda^0,\ldots \lambda^{-N}$, after that it becomes
\be
\sum_{k=1}^N (-)^{k+1}\frac{k e_k}{\lambda^{k+1}}
= \sum_{k=0}^{N-1} (-)^k \frac{(N-k)e_k}{\lambda^{k+1}}
\label{lhs}
\ee
In the second sum, one expresses derivatives of the elementary symmetric polynomials (boxed):
\be
\sum_{k=1}^N \frac{(-)^k}{\lambda^k} \sum_{m=1}^k \frac{\p p_m}{\p \Lambda} \boxed{\frac{\p e_k}{\p p_m}}
= \sum_{k=1}^N \frac{(-)^k}{\lambda^k} \sum_{m=1}^k m \Lambda^{m-1} \boxed{\frac{(-)^{m+1}}{m} e_{k-m}}
= -\sum_{k=0}^{N-1} \frac{(-)^k}{\lambda^{k+1}} \sum_{m=0}^k   (-)^{m}\Lambda^{m} e_{k-m}
\ \ \
\label{rhs}
\ee
\i.e.
\be
\sum_{k=0}^{N-1} \frac{(-)^k}{\lambda^{k+1}}  \left( \frac{\p \lambda}{\p \Lambda} \cdot (N-k)e_k
-  \sum_{m=0}^k   (-)^{m}\Lambda^{m} e_{k-m}\right) = 0
\label{lambdader0}
\ee
The first sum at the l.h.s. has a simple interpretation in terms of eigenvalues:
\be
\sum_{k=0}^{N-1} \frac{(-)^k}{\lambda_i^{k+1}} (N-k)e_k =
\lambda_i^{-N}\prod_{j\neq i}^N (\lambda_i-\lambda_j)
\label{Vieta}
\ee
This is just the Vieta's formula for $D_N$.

For instance, for $N=4$
\be
(\lambda_1-\lambda_2)(\lambda_1-\lambda_3)(\lambda_1-\lambda_4) =
-\lambda_2\lambda_3\lambda_4 + \lambda_1(\lambda_2\lambda_3+\lambda_2\lambda_4+\lambda_3\lambda_4)
 - \lambda^2(\lambda_2+\lambda_3+\lambda_4)+ \lambda_1^3
= \nn \\
= - e_3 + 2\lambda_1(\lambda_2\lambda_3+\lambda_2\lambda_4+\lambda_3\lambda_4)
 - \lambda_1^2(\lambda_2+\lambda_3+\lambda_4)+ \lambda_1^3
= -e_3 + 2\lambda_1 e_2  - 3\lambda_1^2(\lambda_2+\lambda_3+\lambda_4)+ \lambda_1^3
=\nn \\
= -e_3 + 2\lambda e_2  - 3e_1+4\lambda_1^4
\nn
\ee

\noindent
Thus
\vspace{-0.5cm}
\be
\boxed{
\frac{\p\lambda}{\p \Lambda} = \frac{\lambda^N}{\prod_{j\neq i}^N (\lambda_i-\lambda_j)}
\sum_{k=0}^{N-1} \frac{(-)^k}{\lambda^{k+1}} \sum_{m=0}^k   (-)^{m}\Lambda^{m} e_{k-m}
}
\label{lambdader}
\ee
From this, one can deduce expressions for various $S_R\!\left\{\frac{\p}{\p \Lambda}\right\}\cdot\lambda$.

\subsection*{B.1 First derivative, $R=[1]$}

As the first example, one can find that
\be
\Tr\frac{\p \lambda}{\p \Lambda} =1
\label{firder}
\ee
Indeed, in this case, one needs to take a trace of (\ref{lambdader0}),
then the second sum at the r.h.s. becomes
\be
\sum_{m=0}^k   (-)^{m}\,\Tr \Lambda^{m} e_{k-m} =
Ne_k - \sum_{m=1}^k   (-)^{m}p_{m} e_{k-m}
=(N-k)e_k
\label{firderrel}
\ee
which is just the same as the first item in (\ref{lambdader0}).
Thus, (\ref{firder}) clearly is the answer.

Relation (\ref{firder}) holds separately for each eigenvalue $\lambda_i$ of $\Lambda$.
This implies that, for the action on functions which depend on times (and eigenvalues) only,
which we denote by $\cong$,
\be
\Tr \frac{\p }{\p \Lambda}  \cong \sum_{i=1}^N \Tr \frac{\p \lambda_i }{\p \Lambda}\frac{\p}{\p \lambda_i}
= \sum_{i=1}^N \frac{\p}{\p \lambda_i}
\ee

For the future use, we also need a generalization of (\ref{firder}) to
\be
{\rm for}\ n>0 \ \ \ \ \ \ \ \ \ \ \ \
\Tr \Lambda^n \frac{\p \lambda}{\p\Lambda}  = \frac{\lambda^N}{\prod_{j\neq i}^N (\lambda_i-\lambda_j)}
\sum_{k=0}^{N-1} \frac{(-)^k}{\lambda^{k+1}} \sum_{m=0}^k   (-)^{m}p_{m+n} e_{k-m}
= \nn \\
= \frac{\lambda^N}{\prod_{j\neq i}^N (\lambda_i-\lambda_j)}
\sum_{k=0}^{N-1} \frac{(-)^{k+n+1}}{\lambda^{k+1}}
\left((k+n)e_{k+n} +  \sum_{j=1}^{n-1} (-)^jp_j e_{k+n-j} \right)
\label{firderren}
\ee

\subsection*{B.2 Second derivative}

\be
\Tr \frac{\p^2}{\p \Lambda^2} \cong \sum_{i=1}^N \Tr \frac{\p^2 \lambda_i}{\p \Lambda^2} \cdot \frac{\p}{\p\lambda_i}
+ \sum_{i,j=1}^N \Tr \frac{\p \lambda_i}{\p \Lambda}\frac{\p \lambda_j}{\p \Lambda}
\cdot \frac{\p^2}{\p \lambda_i\p \lambda_j}
\ee
It follows from (\ref{lambdader}), that
\be
 \Tr \frac{\p \lambda_i}{\p \Lambda}\frac{\p \lambda_j}{\p \Lambda} = \delta_{ij}
\label{derderlambda}
\ee
and
\be
\Tr \frac{\p^2 \lambda_i}{\p \Lambda^2} =2 \frac{\p\log \Delta}{\p \lambda_i}
= \sum_{j\neq i} \frac{2}{\lambda_i-\lambda_j}
\label{der2lambda}
\ee
Therefore
\be
\Tr \frac{\p^2}{\p \Lambda^2} \cong
\sum_{i=1}^N \left(\frac{\p^2}{\p \lambda_i^2}
+ 2\sum_{j\neq i}\frac{1}{\lambda_i-\lambda_j}\frac{\p}{\p \lambda_i}
\right)
\ee

\subsubsection*{B.2.1 Derivation of (\ref{derderlambda})}

In order to derive (\ref{derderlambda}), we multiply two relations (\ref{lambdader0}):
\be
\frac{\p \lambda_i}{\p \Lambda}  \frac{\p \lambda_j}{\p \Lambda} \cdot
\sum_{k_1,k_2=0}^{N-1} \frac{(-)^{k_1+k_2}}{\lambda_i^{k_1+1}\lambda_j^{k_2+1}}
\cdot (N-k_1)(N-k_2)e_{k_1}e_{k_2}
= \nn \\
=\sum_{k_1,k_2=0}^{N-1} \frac{(-)^{k_1+k_2}}{\lambda_i^{k_1+1}\lambda_j^{k_2+1}}
\sum_{m_1=0}^{k_1}\sum_{m_2=0}^{k_2}   (-)^{m_1+m_2}\Lambda^{m_1+m_2} e_{k_1-m_1}e_{k_2-m_2}
\label{secder}
\ee
Taking a trace converts $\Lambda^{m_1+m_2}$ at the r.h.s. into $p_{m_1+m_2}$,
and we need the conditions $D_N=0$ and $e_n=0$ at $N+1\leq n\leq 2N$
to express all the extra time variables through $p_1,\ldots, p_{N-1}$
appearing at the l.h.s.

If $i=j$, this is simple:
\be
\sum_{k_1,k_2=0}^{N-1} \frac{(-)^{k_1+k_2}}{\lambda^{k_1+k_2+2}}\left(
\sum_{m_1=0}^{k_1}\sum_{m_2=0}^{k_2}   (-)^{m_1+m_2}p_{m_1+m_2} e_{k_1-m_1}e_{k_2-m_2}
-  (N-k_1)(N-k_2)e_{k_1}e_{k_2} \right)
= \nn \\
= \underline{D_N} \cdot\left(\sum_{k=0}^{N-2} \frac{(-)^{k+1}}{\lambda^{k+2}}(N-k)(N-k-1)e_k\right)
\ \  {\bf mod} \  e_n=0 \ {\rm for} \ n> N
\ \ \ \ \ \
\ee
or, in more detail,
\be
\sum_{k_1,k_2=0}^{N-1} \frac{(-)^{k_1+k_2}}{\lambda^{k_1+k_2+2}}\left(
\sum_{m_1=0}^{k_1}\sum_{m_2=0}^{k_2}   (-)^{m_1+m_2}p_{m_1+m_2} e_{k_1-m_1}e_{k_2-m_2}
-  (N-k_1)(N-k_2)e_{k_1}e_{k_2} \right)
=
\ee
\vspace{-0.3cm}
{\footnotesize
\be
=
\left(\sum_{k=0}^{N-2} \frac{(-)^{k+1}}{\lambda^{k+2}}(N-k)(N-k-1)e_k\right)
\cdot\underline{ \sum_{k=0}^N  (-)^k \frac{e_k }{\lambda^k}}
+\sum_{j=0}^N \frac{(-)^{N+j}}{\lambda^{N+j}}  \sum_{k=1}^{N-2-j} \frac{(-)^{k+1}}{\lambda^{k+2}}
(N-1-j-k)(N+k-j) e_j\underline{e_{N+k}}
\nn
\ee
}

\bigskip

For $i\neq j$, we need to prove that the trace of the r.h.s. of (\ref{secder}) is vanishing.
Indeed, it is decomposed as
\be
\sum_{k_1,k_2=0}^{N-1} \frac{(-)^{k_1+k_2}}{\lambda_i^{k_1+1}\lambda_j^{k_2+1}}
\sum_{m_1=0}^{k_1}\sum_{m_2=0}^{k_2}   (-)^{m_1+m_2}p_{m_1+m_2} e_{k_1-m_1}e_{k_2-m_2}
= \frac{D^{(1)}_{N-1}(\lambda_j)\underline{D_N(\lambda_i)} - D^{(1)}_{N-1}(\lambda_i)\underline{D_N(\lambda_j)}}
{\lambda_i-\lambda_j}
- \nn
\ee
\vspace{-0.4cm}
\be
- \frac{1}{(\lambda_i\lambda_j)^N}\sum_{k=N+1}^{2N-2}\underline{e_k}\cdot\left(
\sum_{l=0}^{2N-2-k} (-)^{k-l}(k-l)\frac{\lambda_i^{2N-1-k-l}-\lambda_j^{2N-1-k-l}}{\lambda_i-\lambda_j}e_l\right)
\ee
where $D^{(1)}_N(\lambda):= \frac{1}{\lambda^{N+1}}\frac{\p\left((\lambda^{N+1}D_N(\lambda)\right)}{\p\lambda}$.
The vanishing quantities at the r.h.s. are underlined, and there is one for each item at the r.h.s.

\subsubsection*{B.2.2 Derivation of (\ref{der2lambda})}

The last step in this subsection is the proof of (\ref{der2lambda}).
Conceptually, it is a simple exercise, in the spirit of (\ref{Vieta}).
However, technically it is a little lengthy.

Since $\Tr \frac{\p \Lambda^m}{\p \Lambda} = \sum_{a+b=m-1} p_ap_b$,
 we get from (\ref{lambdader}) and (\ref{Vieta}):
\be
\!\!\!
\left\{\Tr \frac{\p^2 \lambda_i}{\p \Lambda^2}
+\sum_{\j\neq i}^N \frac{1}{\lambda_i-\lambda_j}
\left(\Tr \frac{\p\lambda_i}{\p \Lambda}\frac{\p\lambda_i}{\p\Lambda}
- \Tr \frac{\p\lambda_i}{\p \Lambda}\frac{\p\lambda_j}{\p\Lambda}\right)
 \right\}  \prod_{j\neq i}^N(\lambda_i-\lambda_j) =
\ee
{\footnotesize
\be
 = \lambda_i^N \sum_{k=0}^{N-1} \frac{(-)^k}{\lambda_i^{k+1}}\left(
 \sum_{a,b=0}   (-)^{a+b+1}p_ap_b e_{k-a-b-1}
+ \frac{N-k-1}{\lambda_i} \sum_{m=0}^k   (-)^{m}
 e_{k-m}  \Tr \frac{\p\lambda_i}{\p \Lambda}
 \Lambda^{m}
+ \sum_{m=0}^k   (-)^{m}
 \sum_{n=1}^{k-m} \frac{\p e_{k-m}}{\p p_n}
 \Tr  \Lambda^{m}\frac{\p p_n}{\p \Lambda}\right) \ \ \
\nn
\ee
}

\bigskip

\noindent
Making the same substitution as in (\ref{rhs}) in the last term
and using (\ref{derderlambda}) at the l.h.s.,  we simplify the relation to
\be
\left(\Tr \frac{\p^2 \lambda_i}{\p \Lambda^2}  + \sum_{\j\neq i}^N \frac{1}{\lambda_i-\lambda_j}
 \right)  \prod_{j\neq i}^N(\lambda_i-\lambda_j)
 = \lambda_i^N\sum_{k=0}^{N-1} (N-k-1)\frac{(-)^k}{\lambda_i^{k+2}} \sum_{m=0}^k   (-)^{m}
  e_{k-m}  \cdot \Tr \Lambda^{m}\frac{\p\lambda_i}{\p \Lambda}
+ \!\!\!
\ee
\vspace{-0.5cm}
\be
\!\!+ \lambda_i^N \sum_{k=0}^{N-1} \frac{(-)^k}{\lambda_i^{k+1}}
\left(
\sum_{a,b=0}   (-)^{a+b+1}p_ap_b e_{k-a-b-1} +
 \sum_{m=0}^k   \sum_{n=1}^{k-m} (-)^{m+n-1} p_{m+n-1}  e_{k-m-n}\right)\!\!
 =\sum_{\j\neq i}^N \frac{3}{\lambda_i-\lambda_j}  \prod_{j\neq i}^N(\lambda_i-\lambda_j)
\nn
\ee
For the trace in the first term at the r.h.s. of the first line we use (\ref{firderren}),
and the final result is (\ref{der2lambda}).

\end{document}